\documentclass{jfm}
\usepackage{graphicx}
\usepackage{epstopdf, epsfig}

\usepackage{amssymb}
\usepackage{xspace}
\usepackage{amsmath}
\usepackage{subcaption}
\usepackage{makecell}
\usepackage{blkarray}
\usepackage{mathtools}
\usepackage{bm}
\newcommand{\dx}{\mathrm{d}x}
\newcommand{\dy}{\mathrm{d}y}

\newcommand{\dt}{\mathrm{d}t}
\newcommand{\dee}{\mathrm{d}}
\usepackage{xcolor}
\newcommand{\jl}[1]{\textcolor{black}{#1}}
\newcommand{\rmc}[1]{\textcolor{black}{#1}}
\newcommand{\oej}[1]{\textcolor{black}{#1}}

\shorttitle{Lagrangian surfactant dynamics in  confined domains}
\shortauthor{R. Mcnair, O. E. Jensen and J. R. Landel}

\title{Confinement-induced drift in Marangoni-driven transport of surfactant: \\ a Lagrangian perspective}

\author{Richard Mcnair\aff{1}
  \corresp{\email{richard.mcnair@manchester.ac.uk}},
  Oliver E. Jensen\corresp{\email{oliver.jensen@manchester.ac.uk}} \aff{1}
 \and Julien R. Landel\corresp{\email{julien.landel@univ-lyon1.fr}} \aff{1,2}
 }

\affiliation{\aff{1}Department of Mathematics, University of Manchester,
Oxford Road, M13 9PL, UK
\aff{2}Universite Claude Bernard Lyon 1, Laboratoire de Mecanique des Fluides et d'Acoustique (LMFA), UMR5509, CNRS, Ecole Centrale de Lyon, INSA Lyon, 69622
Villeurbanne, France
}

\begin{document}

\maketitle

\begin{abstract}
Successive drops of coloured ink mixed with surfactant are deposited onto a thin film of water to create marbling patterns in the Japanese art technique of Suminagashi. To understand the physics behind this and other applications where surfactant transports adsorbed passive matter at gas--liquid interfaces, we investigate the Lagrangian trajectories of material particles on the surface of a thin film of a confined viscous liquid under Marangoni-driven spreading by an insoluble surfactant. We study a model problem in which several deposits of exogenous surfactant simultaneously spread on a bounded rectangular surface containing a pre-existing endogenous surfactant. We derive Eulerian and Lagrangian formulations of the equations governing the Marangoni-driven surface flow. Both \oej{descriptions} show how confinement can induce drift and flow reversal during spreading. The Lagrangian formulation captures trajectories without the need to calculate surfactant concentrations; however, concentrations can still be inferred from the Jacobian of the map from initial to current particle position. We explore a link between thin-film surfactant dynamics and optimal transport theory to find the approximate equilibrium locations of material particles for any given initial condition by solving a Monge--Amp\`ere equation. We find that, as the endogenous surfactant concentration $\delta$ vanishes, the equilibrium shapes of deposits using the Monge--Amp\`ere approximation approach polygons with corners curving in a self-similar manner over lengths scaling as $\delta^{1/2}$. \jl{We explore how Suminagashi patterns may be produced by using computationally efficient successive solutions of the Monge--Amp\`ere equation}. 
\end{abstract}


\section{Introduction}
\label{sec:Intro}

\begin{figure}
    \centering
    \begin{subfigure}{0.422\textwidth}
\includegraphics[width=\textwidth]{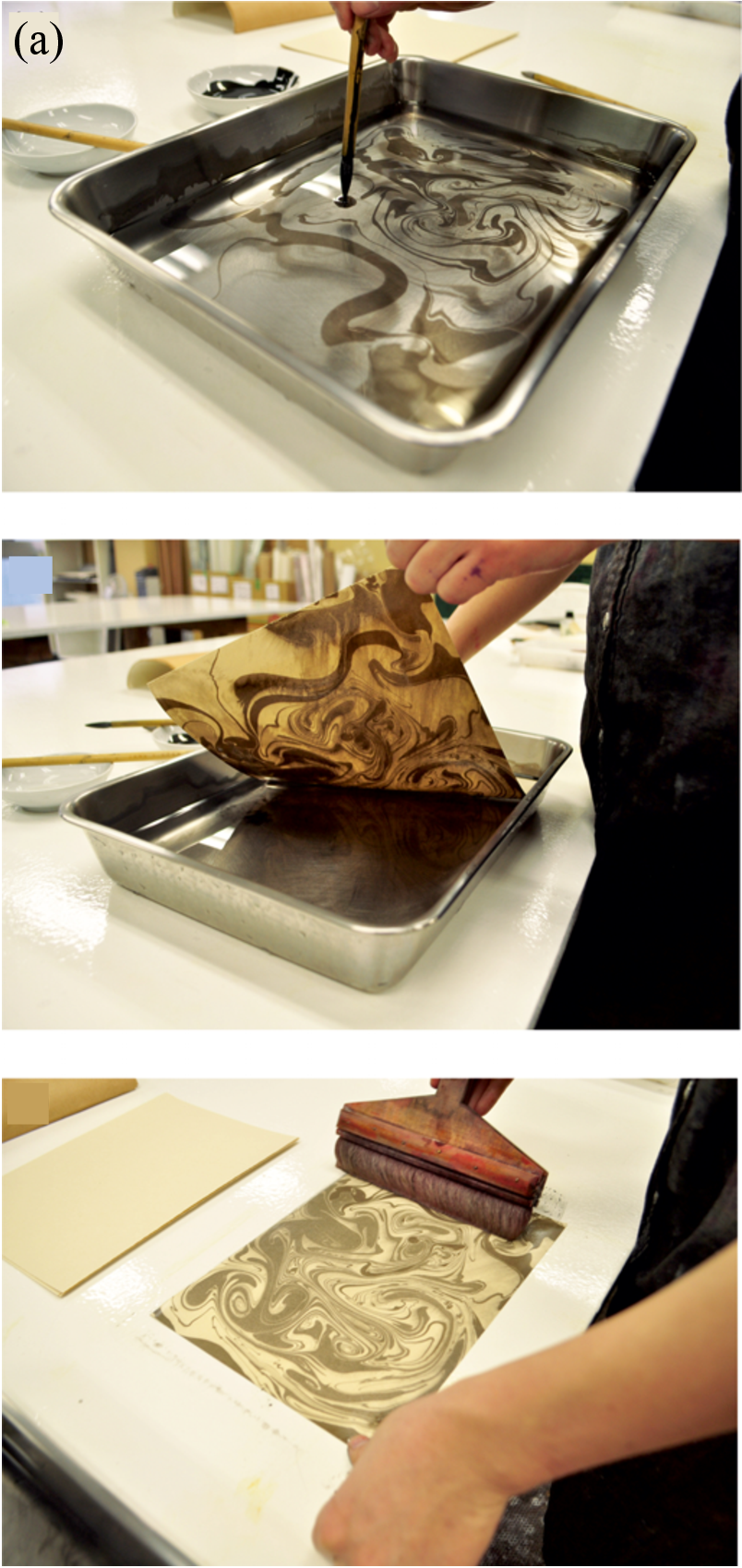}
\end{subfigure}
\begin{subfigure}{0.565\textwidth}
\centering
\begin{subfigure}{\textwidth}
\includegraphics[width=\textwidth]{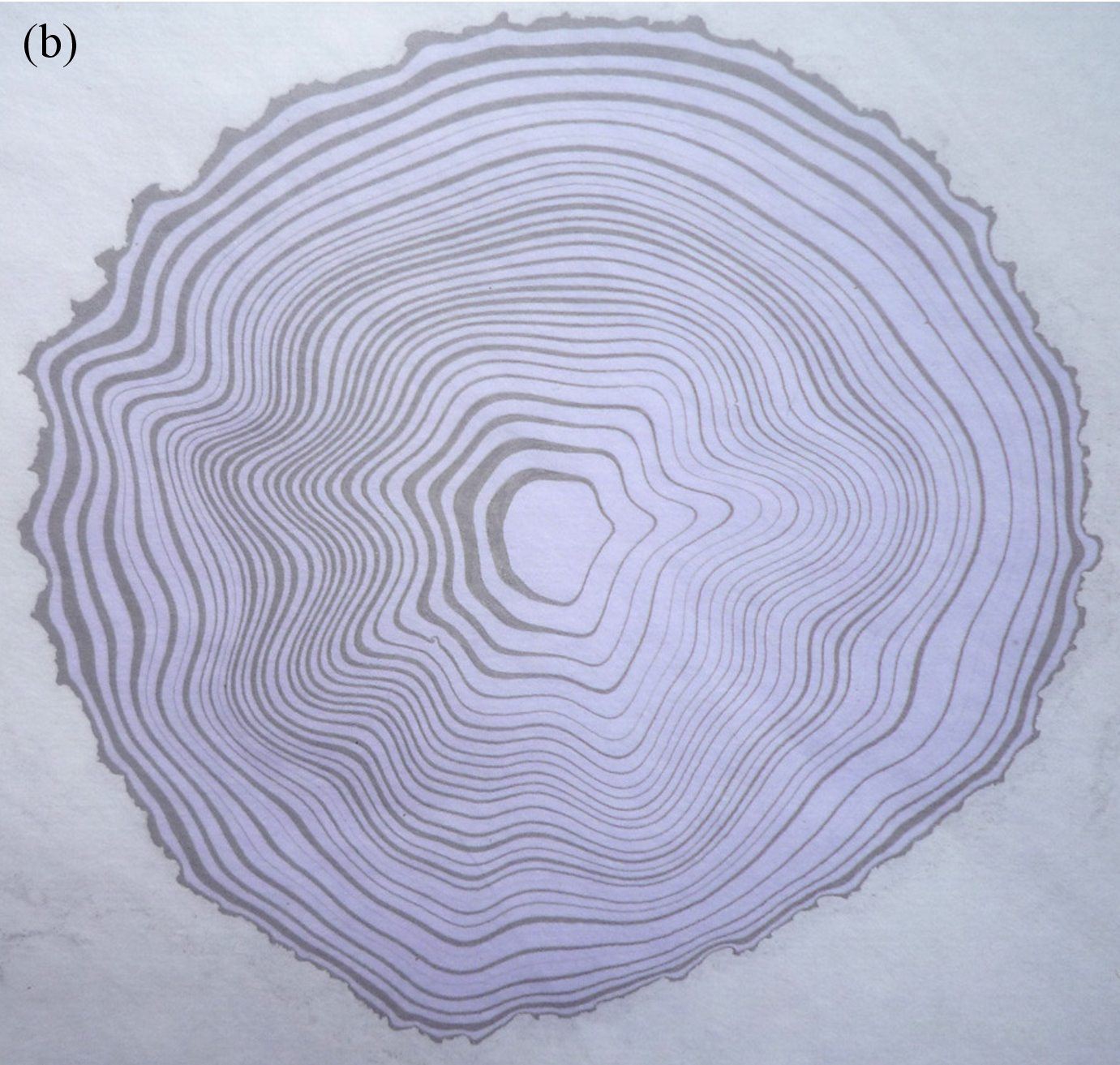}
\end{subfigure}
\begin{subfigure}{\textwidth}
\includegraphics[width=\textwidth]{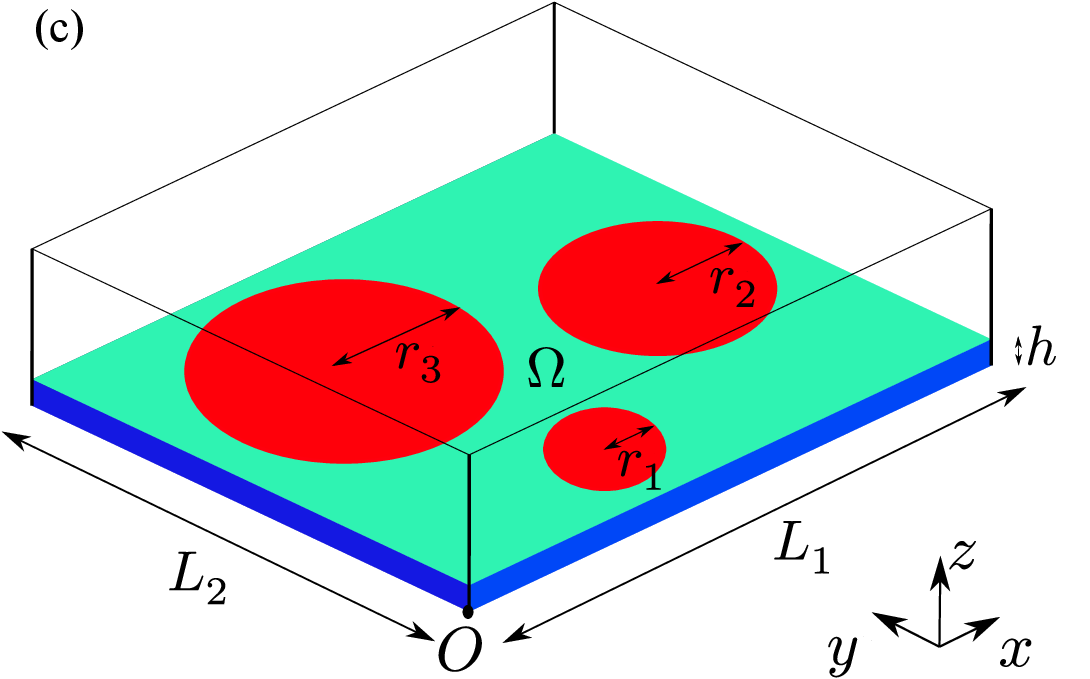}
\end{subfigure}
\end{subfigure}
    \caption{(a) Pictures showing the Japanese art technique of Suminagashi \citep{rouwet2017sedimentation}. Successive drops of a mixture of coloured ink and surfactant are deposited on the surface of a thin film of water to create a multicoloured pattern. Blowing on the surface then creates further intricate patterning. (b) Picture of a Suminagashi pattern of ink on water created by artist Bea \cite{beamehan}.
  (c) Schematic of the model problem. Circular deposits of insoluble exogenous surfactant (red) spread on the surface $\Omega$ of a thin layer of viscous liquid (blue) of mean height $h$ confined in a rectangular region of dimensions $L_1$ and $L_2$, where the surface contains an initially uniform endogenous surfactant (green). We assume that the ratio of vertical to horizontal length scales is small enough, and that the Bond number (ratio of gravitational to surface tension forces) is large enough, for height deflections caused by spreading to be negligible, confining spreading to the flat plane of the surface $\Omega$.}
    \label{fig:Schematic}
\end{figure}

Successive drops of coloured inks mixed with surfactant create intricate patterns by Marangoni spreading  in the Japanese art technique of Suminagashi (see figure \ref{fig:Schematic}\textit{a}). A surfactant--ink drop is gently deposited at the surface of a thin layer of water, which may have a small initial concentration of endogenous surfactant due to normal environmental contamination. It then spreads outwards and equilibrates before reaching the edges of the container. Then, successive drops deposited at different locations of the liquid surface form the intricate patterns. During pattern creation, the artist can blow on the surface with a straw after drop equilibrations to further deform the pattern. Eventually, the pattern is captured on pieces of paper placed onto the surface \citep{chambers1991suminagashi}. \cite{rouwet2017sedimentation} noticed similar patterns occurring in volcanic crater lakes, and hypothesised that similar physics were responsible: thermal gradients in the lake create Marangoni flows, and wind action creates a blowing effect resulting in marbling patterns of the adsorbed multi-coloured sediments. In this study, we seek to understand the Lagrangian trajectories of material points and curves on a surface during the spreading of surfactants on a confined surface, and thus the dynamics of any adsorbed passive tracer, similar to the advection of ink by surfactant in the Suminagashi technique.

In addition to the cultural importance of Suminagashi, which has been part of Japanese art since the 12th century, and similar practices in China for even longer \citep{ishii1989fractal}, understanding Marangoni-driven surface motion can help us better understand various industrial and biological applications involving surfactants carrying passive, adsorbed material. For example, \cite{deng2018surfactant} showed how small amounts of surfactant added to perovskite (a calcium titanium oxide mineral) can suppress the formation of islands during the drying phase of blade coating by creating Marangoni flows that keeps the solution coating even. Many other coating processes involve Marangoni flows induced by trace amounts of surfactants. Some methods of drug delivery in lungs  mix pharmaceutical substances with exogenous surfactant \citep{haitsma2001exogenous}, so that the surfactant acts as a carrier to spread the drug through the airways. \oej{In particular, surfactant replacement therapies have been used successfully in lungs of neonates affected with respiratory distress syndrome \cite[][]{avery59,jobe1993pulmonary,rodriguez2003management,halliday2008surfactants}.} The surfactant-driven spreading in the complex and confined tree-like geometry of the lungs acts against its natural endogenous surfactant \citep{espinosa1993spreading,jensen1994transport,grotberg1995interaction,halpern1998theoretical,temprano2018soap,mcnair2023exogenous}. These methods of delivery can help overcome difficulties such as poor solubility of the pharmaceuticals \citep{hidalgo2015barrier}. 


Molecules and substances which act as surfactants are ubiquitous in the environment. They can cause unexpected fluid flows which have confounded scientists and engineers, as described by \cite{manikantan2020surfactant} who discussed  the `hidden' variables related to surfactant dynamics in many fluid flows. The present study addresses insoluble surfactant spreading into pre-existing, endogenous surfactant on a thin film of a bounded Newtonian viscous liquid, allowing us to use lubrication theory to approximate the Stokes flow in the liquid film. Lubrication theory for insoluble surfactant-driven flows has its origins with the work of \cite{borgas1988monolayer} who derived coupled PDEs describing the leading-order evolution of the liquid film height and surfactant concentration. The work was then extended theoretically and experimentally by \cite{gaver1990dynamics,gaver1992droplet}. \oej{\cite{thess1997} and} \cite{jensen1998stress1} showed that the coupled equations in the limit of large Bond number could be combined into a single nonlinear diffusion (or `porous medium') equation governing surfactant concentration evolution as a function of space and time. The effect of gravity is to suppress deflections of the surface, removing the functional dependence of the spreading on the dynamic film height.

In this paper, we explore a link between the theory of surfactant spreading and the theory of optimal transport. This theory was initiated by \cite{monge1781memoire} who was trying to find the optimal way to transport mounds of soil under some cost function. The theory was extended into its modern formulation by \cite{kantorovich1942translocation,kantorovich2006translocation}. Most of its current uses are found in machine learning and image analysis \citep{kolouri2017optimal}. A powerful result, enabling significant simplification of optimal transport problems, occurs when the cost function takes a quadratic form, yielding the quadratic Monge--Kantorovich optimal transport problem (qMK). For such cost functions, solutions for the optimal map of material from initial to final location can be shown to be the gradient of a convex function which satisfies a so-called Monge--Amp\`ere equation. A variety of approaches have been taken to find solutions of this nonlinear equation \citep{froese2011fast,benamou2012viscosity,benamou2014numerical}. \cite{otto2001geometry}, building on work by \cite{jordan1998variational} and \cite{benamou2000computational}, showed that porous medium equations (a class of equations to which \cite{jensen1998stress1}'s surfactant equation that we use in this study belong) have the variational structure of a gradient flow on a Riemannian manifold measured by the quadratic Wasserstein distance. The square of the Wasserstein distance, which is defined as the minimiser of a functional, doubles as qMK, which suggests that  solutions to the surfactant-induced transport problem may be approximated by solving the Monge--Amp\`ere equation under certain conditions. We explore  these conditions in this paper, and consider whether the Monge--Amp\`ere equation could be an efficient tool to determine  equilibrium solutions to this complex confined transport problem.

The primary aim of this study is to understand the underlying physics behind surfactant-induced Marangoni dynamics in a confined environment when the surface contains an initial endogenous concentration of surfactant, which is the case for most environmental fluids. The spreading of multiple exogenous deposits is particularly considered; this was investigated experimentally and with COMSOL\textsuperscript{\textregistered} models recently by \cite{iasella2023interaction}, showing how adjacent droplets interact and deform.  A Lagrangian framework, which has been adopted in the analysis of other transport problems with nonlinear diffusive character \citep{meuirmanov1997evolution}, enables us to compute efficiently individual surface particle trajectories and equilibrium states as a function of initial distributions. Moreover, the Lagrangian framework reveals underpinning flow phenomena such as stretching, compression and rotational motion that govern the particle trajectories. 
While there have been limited investigations of Lagrangian surfactant dynamics in one spatial dimension \citep{grotberg1995interaction}, there is none (to our knowledge) in higher dimensions, despite the potential relevance to a variety of applications. 
Furthermore, while some authors have exploited the gradient-flow structure of thin-film evolution equations \citep{thiele2016gradient,henkel2021gradient}, we are not aware of prior studies linking thin-film flows to optimal transport. We show how we can exploit this link for practical purposes. In particular, we describe a procedure to reproduce the intricate patterns of  Suminagashi art, through  resolution of the Monge--Amp\`ere equation associated with the surfactant transport model. These results appear to capture, at least qualitatively, the dominant physics behind Suminagashi art, suggesting a powerful tool for other applications where surface transport is dominated by surfactants in confined environments.

In \S\ref{sec:Eul}, we use a two-dimensional extension of the model of \cite{jensen1998stress1} (derived in Appendix \ref{sec:appnonldiff}) to describe transport of material particles on a surface. We outline a physical problem in Eulerian coordinates in a confined rectangular domain, implementing initial conditions that represent multiple deposits of exogenous surfactant spreading on a surface with an initially uniform endogenous surfactant concentration. We solve the particle-tracking problem using a finite-difference method by first solving for the evolution of surfactant concentration, and then interpolating the gradient of this solution onto a second Lagrangian grid where we integrate the surface velocity to find the trajectories of surface particles initially located at each grid point. In \S\ref{sec:MethodsLag} we reformulate the problem in Lagrangian coordinates and show how the problem can be reduced from three to two scalar PDEs, enabling the same calculation without the intermediate step of finding the evolution of the surfactant concentration, and without the need to interpolate concentration gradients from an Eulerian to a Lagrangian grid. We solve the resulting scheme using a finite-element method. In \S\ref{sec:MethodsSteadyState} we show how to approximate the equilibrium locations of surface particles as a function of their initial locations via a Monge--Amp\`ere equation, without having to compute their intermediate trajectories. In \S\ref{sec:ResEul} we show consistency between the Eulerian and Lagrangian methods, and describe dynamical phenomena not normally associated with spreading surfactants, such as drift and flow reversals due to confinement. In \S\ref{sec:ResSteadyState} we show that solutions of the Monge--Amp\`ere equation approximate the equilibrium solution well when the endogenous and exogenous concentrations are of comparable magnitude, and also provide a credible approximation when the endogenous concentration is much smaller. We show how, in the limit of small endogenous concentration, the boundaries of the deposits become almost polygonal with self-similar structures at the corners, resembling a two-dimensional foam. We discuss subtle discrepancies between the Monge--Amp\`ere solution, and a solution computed with the Eulerian particle-tracking method, indicating that surfactant transport can be considered almost, but not exactly, optimal. We analyse the two-dimensional mapping between the initial surfactant distribution and its equilibrium distribution and discuss how the divergence and curl of the mapping can reveal regions of stretching, compression and rotational motion. Finally, we show that successive solutions of the Monge--Amp\`ere equation, combined with divergence-free maps to mimic blowing, can be used to create a computational Suminagashi marbling pattern, illustrating the power of the optimal-transport approximation. Additional results are shown in supplementary material to provide further evidence supporting the main findings and discussion presented in this paper.

\section{Model and methods}
\label{sec:ModelandMethods}

\subsection{The Eulerian particle-tracking problem}
\label{sec:Eul}

\subsubsection{The problem and derivation of the model}

We investigate the trajectories of particles on the surface of a viscous Newtonian liquid advected by surface-tension gradients caused by a non-uniform concentration profile of insoluble surfactant, {\oej{which is assumed to have negligible molecular diffusivity}}. Concentration gradients are caused by deposits of exogenous surfactant added to a uniform concentration field of endogenous surfactant. We assume both species of surfactant have the same material properties, which combine to create a single concentration field which has a linear relationship with surface tension. A typical length scale is found from the initial size of an exogenous deposit, which is much greater than the initial height of the film. The thickness of the film is assumed to remain approximately uniform during the spreading, as we assume that any large vertical deflections are suppressed by gravity (in a large-Bond-number limit). The spreading takes place in a closed region with rectangular horizontal cross-section $\Omega$, given in non-dimensional Cartesian coordinates as $0\leq x\leq L_1$, $0 \leq y\leq L_2$ confined by impermeable walls. Surfactant concentrations are scaled by the maximum initial concentration of one of the deposits. \oej{As explained in Appendix~\ref{sec:appnonldiff}}, the surfactant is transported from its initial profile to its final equilibrium state via the nonlinear diffusion equation, which describes the evolution of the surfactant concentration as a function of space and time,
\begin{equation}\label{eq:nonlineardiff}
    \Gamma_t = \tfrac{1}{4}\boldsymbol{\nabla}_{\mathbf{x}}\cdot (\Gamma \boldsymbol{\nabla}_{\mathbf{x}}\Gamma),
\end{equation} 
where $\bm{\nabla}_{\mathbf{x}}$ is the gradient operator in the $\mathbf{x}= (x,y)$ plane of the Eulerian coordinates, and $\Gamma_t$ is the derivative of surfactant concentration with respect to non-dimensional time $t$; here, time is scaled by the ratio of liquid viscosity to maximum surface tension gradient (Appendix \ref{sec:appnonldiff}).
We impose a no-flux boundary condition at the periphery of the domain
\begin{equation}\label{eq:EulBoundConds}
    \boldsymbol{\nabla}_{\mathbf{x}}\Gamma \cdot \mathbf{n}_b = 0 \qquad \text{ on } \partial \Omega,
\end{equation} 
where $\partial \Omega$ is the boundary of the domain $\Omega$, and $\mathbf{n}_b$ is a unit normal vector to the boundary of the domain. Comparison of \eqref{eq:nonlineardiff} with the non-dimensional surface transport equation $\Gamma_t+\bm{\nabla}_{\mathbf{x}}\cdot(\mathbf{u}_s\Gamma)=0$ for a flat surface and non-diffusive surfactant shows that the surface velocity is
\begin{equation}\label{eq:velocityexp}
    \mathbf{u}_s(\mathbf{x},t) = -\tfrac{1}{4}\boldsymbol{\nabla}_{\mathbf{x}} \Gamma.
\end{equation} 
\oej{}
\rmc{
Since we impose no flux of surfactant at the boundaries of $\Omega$, as time goes to infinity concentration gradients vanish to reach an equilibrium or steady-state, and so the initial concentration profile of surfactant $\Gamma_0(x,y)$ spreads to a uniform state with concentration $\bar{\Gamma}>\delta>0$, where $\delta$ is the initial endogenous concentration. We do not consider the singular limit $\delta=0$, which is beyond the scope of this study. In that case spreading  at the edges of the deposits would continue until the edges meet a solid boundary or the edges of another deposit. The final concentration relates to the initial concentration profile by
\begin{equation}\label{eq:GammaBar}
    \bar{\Gamma}=\frac{\int_{\Omega}\Gamma_0(x,y) \dx \dy}{\int_{\Omega}  \dx \dy} = \frac{1}{L_1L_2}\int_{\Omega} \Gamma_0(x,y) \dx \dy .
\end{equation} }

\oej{Equation~(\ref{eq:nonlineardiff}) represents a natural generalisation of the spatially one-dimensional nonlinear diffusion equation derived in \cite{jensen1998stress1}, and aligns with the two-dimensional formulation of \cite{thess1997}. In stepping from one to two dimensions, an extra degree of freedom must be considered: any surface velocity field for which $\mathbf{u}_s\Gamma$ has zero divergence will not change surface concentrations but will nevertheless transport surface particles.  This is illustrated in Appendix~\ref{sec:appnonldiff} by considering the influence of an imposed surface stress, as might arise from external blowing on the liquid film.  For a monolayer close to equilibrium, the divergence of the stress field is area-changing; this is resisted by Marangoni effects (\ref{eq:lin}).  However, the curl of the stress field in this simple model generates a flow that can redistribute surfactant (i.e. surface material elements carrying either endogenous or exogenous surfactant) without inducing surface tension gradients (\ref{eq:rotationalvelocity}).  As well as being exploited by Suminagashi artists, this feature highlights a potential degeneracy in (\ref{eq:nonlineardiff}): namely, that the energetic cost of any flow that preserves concentrations of surface material elements is not captured by the evolution equation.}

We now introduce a Lagrangian coordinate system $(x_0,y_0,\tau)$ to complement the Eulerian system $(x,y,t)$. We define a mapping $\mathbf{X}=(X,Y)$ between them, such that particles starting on the interface at $\mathbf{x}_0=(x_0,y_0)\in\Omega$ at $t=0$ are advected at time $t=\tau$ to
\begin{equation}\label{eq:pushforward}
	x = X(x_0,y_0,\tau), \qquad y = Y(x_0,y_0,\tau).
\end{equation} 
Since surfactant transport is purely advective under (\ref{eq:nonlineardiff}), the mapping satisfies
\begin{equation}\label{eq:particlevelocity}
    \frac{\partial \mathbf{X}(x_0,y_0,\tau)}{\partial \tau} = -\frac{1}{4}\boldsymbol{\nabla}_{\mathbf{x}} \Gamma(\mathbf{X}(x_0,y_0,\tau),\tau).
\end{equation} 
The mapping function $\mathbf{X}(x_0,y_0,\tau)$ from initial to current particle location is the main quantity we seek throughout this study. \rmc{The initial conditions for each simulation we perform in this study will be of the form
\begin{equation}\label{eq:InitCondsGenNonD}
      \Gamma_0(x_0,y_0) =   \begin{cases}
		\delta + \mathcal{F}(x_0,y_0) \qquad & \text{in } \Omega',
		\\
		\delta \qquad & \text{in } \Omega -\Omega',
	\end{cases}
\end{equation}
where $\delta=\min{\Gamma_0(x_0,y_0)}$ for all $(x_0,y_0)$ in $\Omega$ represents the initially uniform endogenous surfactant, and $\mathcal{F}$ is a function describing the initial distribution of exogenous surfactant deposited in  $\Omega'$, a region of $\Omega$. In this study, we only consider non-overlapping depositions of exogenous surfactant which are   axisymmetric  about their own centre, and with a radially decreasing concentration profile. Although we have studied various initial distributions for the exogenous surfactant deposits (see supplementary material), we focus on quadratic distributions, which  we denote as 
\begin{equation}\label{eq:circlenotation}
 \mathcal{C}_q(\mathbf{x}_0;\mathbf{x}_c,r,\Gamma_{0,c}-\delta)  = \begin{cases}
   (\Gamma_{0,c}-\delta)\left(1- \frac{|\mathbf{x}_0-\mathbf{x}_c|^2}{r^2}\right)&\quad |\mathbf{x}_0-\mathbf{x}_c|\leq r \\
0 &\quad |\mathbf{x}_0-\mathbf{x}_c|> r,\end{cases}
\end{equation} 
which is centered at $\mathbf{x}_c=(x_c,y_c)$, where the initial concentration has a local maximum $\Gamma_{0,c}$,  with deposit radius $r$. The concentration profile $\Gamma_0$ is continuous when added to the endogenous field, and the Euclidean distance is given by $|\mathbf{x}_0-\mathbf{x}_c| \equiv \sqrt{(x_0-x_c)^2+(y_0-y_c)^2}$.
The subscript $q$ in \eqref{eq:circlenotation} refers to the quadratic nature of the initial concentration profile. In Appendix \ref{sec:Differencedelta} and in \S{S5} of the supplementry material we consider circular concentration profiles with other functional forms.}

\subsubsection{Scenarios studied}
\label{sec:3blobsoutline}

\jl{We have investigated scenarios involving one, two or three distinct deposits (i.e. $\Omega'$ is constituted of one, two or three disconnected regions in $\Omega$). The different configurations studied for the one- and two-deposit problems are presented in the supplementary material (see table S1). These two problems are helpful to understand basic dynamical features and the impact of the relevant non-dimensional parameters, as we will discuss briefly in \S\ref{sec:Results}. However, the one- and two-deposit problems miss topological features that appear only with three or more exogenous deposits, such as internal corners where the edges of the deposits meet away from the domain boundaries. As we will discuss in \S\ref{sec:Results}, internal corners display self-similar features. For the sake of simplicity and to enable analytical progress, we focus mainly on the three-deposit problem for the rest of this paper. Nevertheless, we anticipate that many of the results found with the three-deposit problem will also apply to problems involving more deposits.}
%
Therefore, we devise a model problem where \rmc{$\mathcal{F}(x_0,y_0)$} consists of three circular regions of different radii ($r_1=1,r_2$ and $r_3$ in non-dimensional variables, see figure \ref{fig:Schematic}\textit{c}) containing exogenous surfactant with quadratic concentration profiles, with differing non-dimensional maximum values $1,\Gamma_2$ and $\Gamma_3$ in the different regions (the number in the subscript corresponds to the region). Deposit 1, the smallest, is centred at $(x_1,y_1)$, the second largest circular deposit is centred at $(x_2,y_2)$, the largest is centred at $(x_3,y_3)$. Therefore, \rmc{using our notation for circular deposits \eqref{eq:circlenotation}},
\rmc{\begin{equation}\label{eq:InitCondsGenNonD2}
 \mathcal{F}(x_0,y_0) = \mathcal{C}_q(x_1,y_1,1,1-\delta) + \mathcal{C}_q(x_2,y_2,r_2,\Gamma_2-\delta) +  \mathcal{C}_q(x_3,y_3,r_3,\Gamma_3-\delta).
\end{equation}For the three-deposit problem, we choose $r_2=2$, $r_3=3$, $\Gamma_2=1$ and $\Gamma_3=2$. For every problem tackled in this paper and in the supplement, we choose $L_1=13$ and $L_2=11$. The centres of the deposits are chosen to be $(x_1,y_1) = (6,2)$, $(x_2,y_2) = (10,5)$, and $(x_3,y_3)=(4,7)$ for most of the solutions presented, unless otherwise stated.}

\subsubsection{Numerical scheme for the Eulerian particle-tracking problem}
\label{sec:EulNum}

 A finite-difference approximation of (\ref{eq:nonlineardiff}, \ref{eq:particlevelocity}) is calculated using two rectangular grids. The first grid is used to solve for an approximation of \eqref{eq:nonlineardiff} subject to boundary conditions \eqref{eq:EulBoundConds} and initial conditions \rmc{(\ref{eq:InitCondsGenNonD}, \ref{eq:InitCondsGenNonD2})} in an Eulerian reference frame, which is accomplished using a second-order central differencing system in space, and a first-order forward Euler method in time (choosing a sufficiently small time-step to ensure stability). This is solved simultaneously with a forward-Euler approximation of \eqref{eq:particlevelocity} for the dynamics of the particle paths on a second grid in the Lagrangian reference frame. At each time step, the concentration gradient is approximated on the Eulerian grid, and interpolated onto the Lagrangian grid at the current particle locations using a linear interpolation method, \rmc{meaning the method as a whole is first-order in space and time}.

The simulation is computed from $t=0$ to a large time $t=t_f$ when the solution approximates the steady-state. The value of $t_f$ is found by considering the analysis in Appendix \ref{sec:AppX_cr}, which shows how to ensure that the map is within a small tolerance vector $[X_{tol},Y_{tol}]^T$ of the steady state everywhere (we set $[X_{tol},Y_{tol}]^T=[10^{-3},10^{-3}]^T$).


\subsection{The Lagrangian particle-tracking problem}
\label{sec:MethodsLag}

\subsubsection{Derivation of the Lagrangian method}\label{sec:DerLag}

Rather than solving the three scalar PDEs in (\ref{eq:nonlineardiff}, \ref{eq:particlevelocity}) in an Eulerian framework, it is sufficient to solve only two PDEs by adopting a Lagrangian framework, as we now demonstrate, by calculating $\mathbf{X}(x_0,y_0,\tau)$ without the intermediate step of determining surfactant concentrations. We present a Lagrangian scheme reminiscent of that presented by \cite{carrillo2021lagrangian} for a general Wasserstein gradient flow. The chain rule combined with \eqref{eq:pushforward} yields the material derivative $
	\partial/\partial \tau|_{x_0,y_0} =  \partial/\partial t |_{x,y}+ \mathbf{u}_s\cdot \boldsymbol{\nabla}_{\mathbf{x}}$,  where $\mathbf{u}_s = \mathbf{X}_{\tau}$, where the $\tau$ subscript means the partial derivative with respect to $\tau$. It is also the case that
\begin{equation}\label{eq:LtoEGrad}
	\begin{pmatrix} \frac{\partial}{\partial x_0} \\ 	\frac{\partial}{\partial y_0}
	\end{pmatrix} = 
	\begin{pmatrix}
		X_{x_0} & Y_{x_0} \\ X_{y_0} & Y_{y_0}
	\end{pmatrix}
	\begin{pmatrix}
		\frac{\partial}{\partial x} \\ \frac{\partial}{\partial y}
	\end{pmatrix},
	\quad \text{or} \quad
	\boldsymbol{\nabla}_{\boldsymbol{x_0}} = 
	(\boldsymbol{\nabla}_{\boldsymbol{x_0}} \mathbf{X} )^T \boldsymbol{\nabla}_{\mathbf{x}}.
 \end{equation} 
 We define tensor calculus operators as
\begin{equation}
	\boldsymbol{\nabla}_{\boldsymbol{x_0}} \begin{pmatrix}
		a_1 \\ a_2
	\end{pmatrix} = \begin{pmatrix}
		a_{1x_0} & a_{1y_0}  \\ a_{2x_0} & a_{2y_0} 
	\end{pmatrix},
	\qquad 
	\boldsymbol{\nabla}_{\boldsymbol{x_0}} \cdot \begin{pmatrix}
		a_1 & a_2 \\ a_3 & a_4
	\end{pmatrix} = \begin{pmatrix}
		a_{1x_0} + a_{3y_0}  & a_{2x_0} +a_{4y_0} 
	\end{pmatrix}.
\end{equation} 
The Jacobian of the mapping \eqref{eq:pushforward},
\begin{equation}\label{eq:alphadef}
    \alpha \equiv \det(\boldsymbol{\nabla}_{\boldsymbol{x_0}}\mathbf{X}) = X_{x_0}Y_{y_0}-X_{y_0}Y_{x_0},
\end{equation} quantifies how area elements are deformed by the map between initial and current particle positions, such that area elements $\dee A_{\boldsymbol{x_0}}$ and $\dee A_{\mathbf{x}}$ are related by $\dee A_{\mathbf{x}}  =  \alpha \dee A_{\boldsymbol{x_0}}$. By conservation of mass, we can equate integrals of the surfactant concentration over the Lagrangian and Eulerian domains, respectively,
\begin{equation}\label{eq:consmass}
	\int_{\mathbf{X}^{-1}(\Delta \Omega)} \Gamma_0(x_0,y_0) 
	\ \dee A_{\boldsymbol{x_0}} 
	= \int_{\Delta \Omega} \Gamma(\mathbf{X},t) \ \dee A_{\mathbf{x}}  ,
\end{equation} where $\mathbf{X}^{-1}(\Delta \Omega)$ is the pre-image of any subset $\Delta \Omega$ of the Eulerian domain $\Omega$, and there is a one-to-one mapping between the domains. Using the Jacobian of the mapping, we can change variables on the right hand side of \eqref{eq:consmass} to give
\begin{equation}
	\int_{\mathbf{X}^{-1}(\Delta \Omega)} \Gamma_0(x_0,y_0) \dee A_{\boldsymbol{x_0}}  = \int_{\mathbf{X}^{-1}(\Delta \Omega)} \Gamma(\mathbf{X}(x_0,y_0,\tau),\tau)  \alpha(\mathbf{X}(x_0,y_0,\tau),\tau)  \dee A_{\boldsymbol{x_0}}  .
\end{equation} We are now integrating over the same space with respect to the same variables, and as $\Delta \Omega$ is arbitrary the integrands must be equal, yielding
\begin{equation}\label{eq:LagMassConserv}
	 \Gamma(\mathbf{X}(x_0,y_0,\tau),\tau)  \alpha(\mathbf{X}(x_0,y_0,\tau),\tau) =  \Gamma_0(x_0,y_0).
\end{equation} This is the main statement of mass conservation in $\Omega$, valid for any $\tau\geq0$, and is key for our analysis in this subsection and the next. 

\rmc{The choice of Lagrangian coordinate system is arbitrary, and in the rest of this subsection we choose a spatially non-uniform coordinate system $(\xi, \eta)$. This coordinate system, non-uniform in $\Omega$, also defines a geometric transformation of the domain $\Omega$, which is achieved by deforming $\Omega$ such that $(\xi, \eta)$ become regularly spaced Cartesian coordinates. We call this new domain the deformed Lagrangian domain, with coordinates $(\xi,\eta)$ replacing $(x_0,y_0)$. In \S\ref{sec:MethodsSteadyState} below we will revert back to $(x_0,y_0)$ which there will refer to regular Cartesian coordinates in an undeformed copy of the Eulerian domain such that $(x,y)=(x_0,y_0)$ at $\tau=0$ (these two domains will be referred to as the deformed and undeformed Lagrangian domains, respectively). For now, however, we choose a coordinate system $(\xi,\eta)$ such that the initial surfactant concentration is uniform in the deformed domain, with $\Gamma_0(\xi,\eta)=1$ everywhere.
 This new coordinate system $(\xi, \eta)$ defines a geometric transformation of the rectangular domain $\Omega$, such that surface areas are stretched or compressed until the concentration per unit area in the deformed system is $1$ everywhere. 
To illustrate, if a region of unit area has an initial uniform concentration of $0.25$ in the undeformed domain, in the deformed domain it would have an area of $0.25$ and therefore an initial uniform concentration of $1$. In the coordinate system of the deformed domain }\eqref{eq:LagMassConserv} becomes
\begin{equation}\label{eq:alphgam1}
\alpha(\mathbf{X}(\xi,\eta,\tau),\tau) \Gamma(\mathbf{X}(\xi,\eta,\tau),\tau)  = 1.
\end{equation}  With this choice, and using \eqref{eq:pushforward}, it follows that $\boldsymbol{\nabla}_{\mathbf{x}} (\alpha \Gamma) = \alpha \boldsymbol{\nabla}_{\mathbf{x}} \Gamma + \Gamma \boldsymbol{\nabla}_{\mathbf{x}}\alpha = 0$, and so
\begin{equation}\label{eq:**}
	\alpha \boldsymbol{\nabla}_{\mathbf{x}} \Gamma = -\Gamma \boldsymbol{\nabla}_{\mathbf{x}}\alpha = -\Gamma \left(\boldsymbol{\nabla}_{\boldsymbol{\xi}} \mathbf{X}\right)^{-T}
	\boldsymbol{\nabla}_{\boldsymbol{\xi}}\alpha,
\end{equation} where $\alpha = \det(\bm{\nabla}_{\bm{\xi}}\mathbf{X})$ and $\bm{\nabla}_{\bm{\xi}} = [ \partial/\partial \xi, \partial/\partial \eta]^T$. The particle velocity \eqref{eq:particlevelocity} is given by $\mathbf{X}_{\tau} = - \boldsymbol{\nabla}_{\mathbf{x}} \Gamma/4$, so (\ref{eq:alphadef}, \ref{eq:alphgam1}, \ref{eq:**}) give
\begin{equation}\label{eq:KeyEqn1stForm}
	4\alpha^2\left(\boldsymbol{\nabla}_{\boldsymbol{\xi}} \mathbf{X}\right)^T \mathbf{X}_{\tau} = \boldsymbol{\nabla}_{\boldsymbol{\xi}}\alpha. 
\end{equation} This expresses the time-evolution of material particle locations in Eulerian coordinates as a function of the deformed Lagrangian coordinates. We can expand \eqref{eq:KeyEqn1stForm} as the system
\begin{subequations}\label{eq:KeyEqn4thForm}
\begin{eqnarray}
	X_{\tau} = \frac{1}{4 \alpha^3} \left(\alpha_{\xi}Y_{\eta} - \alpha_{\eta}Y_{\xi}\right),
 \\
	Y_{\tau} = \frac{1}{4 \alpha^3} \left(\alpha_{\eta} X_{\xi} - \alpha_{\xi}X_{\eta}\right),
 \end{eqnarray} 
\end{subequations} with $\alpha = X_{\xi} Y_{\eta}-X_{\eta}Y_{\xi}$. In turn, \eqref{eq:KeyEqn4thForm} can be rewritten as
\begin{equation}\label{eq:KeyEqn2ndForm}
	\mathbf{X}^T_{\tau} 
	=
	-\frac{1}{8}\boldsymbol{\nabla}_{\boldsymbol{\xi}}\cdot \left(\frac{1}{\alpha^2}
	\begin{pmatrix}
		Y_{\eta} & -X_{\eta}
		\\
		-Y_{\xi} & X_{\xi}
	\end{pmatrix}
	\right), 
\end{equation} which is an equation in divergence form that is easier to solve than \eqref{eq:KeyEqn1stForm} or \eqref{eq:KeyEqn4thForm} when using a finite-element method.
Initial conditions are imposed via \eqref{eq:alphgam1}, so
\begin{equation}\label{eq:DynInitConds}
\frac{1}{\Gamma(\mathbf{X}(\xi,\eta,0),0)} = X_{\xi}(\xi,\eta,0)Y_{\eta}(\xi,\eta,0) - X_{\eta}(\xi,\eta,0)Y_{\xi}(\xi,\eta,0) . 
\end{equation} 
We choose $Y_{\eta}(\xi,\eta, 0 )=1$, and $Y_{\xi}(\xi,\eta, 0 )=0$, so that $X_{\xi}(\xi,\eta,0)=  1/\Gamma(\mathbf{X}(\xi,\eta,0),0)$. \rmc{This yields a purely one-dimensional transformation, as illustrated in figure~\ref{fig:LagrangianDomain}, from the undeformed to the deformed Lagrangian domain, simplifying the calculation of the deformed geometry}. The initial conditions for $\xi$ are therefore obtained through
\begin{equation}\label{eq:xiintegral}
	\int_0^x \Gamma_0(x',y) \mathrm{d}x' +C(y) = \xi(x,y,0), \qquad \xi(0,y,0)=0.
\end{equation} Here, $C(y)$ is an arbitrary piecewise function chosen such that $\xi$ is continuous, and $1/X_{\xi}  =  \xi_x$ because we have fixed $Y$ and $t$. After finding the indefinite partial integral \eqref{eq:xiintegral}, we substitute $y=\eta$, and $x=X$ and invert \eqref{eq:xiintegral} (numerically if needed) to find $X$ as an explicit function of $(\xi,\eta)$. Calling this solution $G(\xi,\eta)$, the initial conditions can be summarised as
\begin{equation}\label{eq:InitConds}
    Y(\xi,\eta,0) = \eta, \qquad X(\xi,\eta,0) = G(\xi,\eta).
\end{equation} 

\subsubsection{The three-deposits problem}

\begin{figure}
	\centering
	\begin{subfigure}{0.59\textwidth}
		\includegraphics[width=\textwidth]{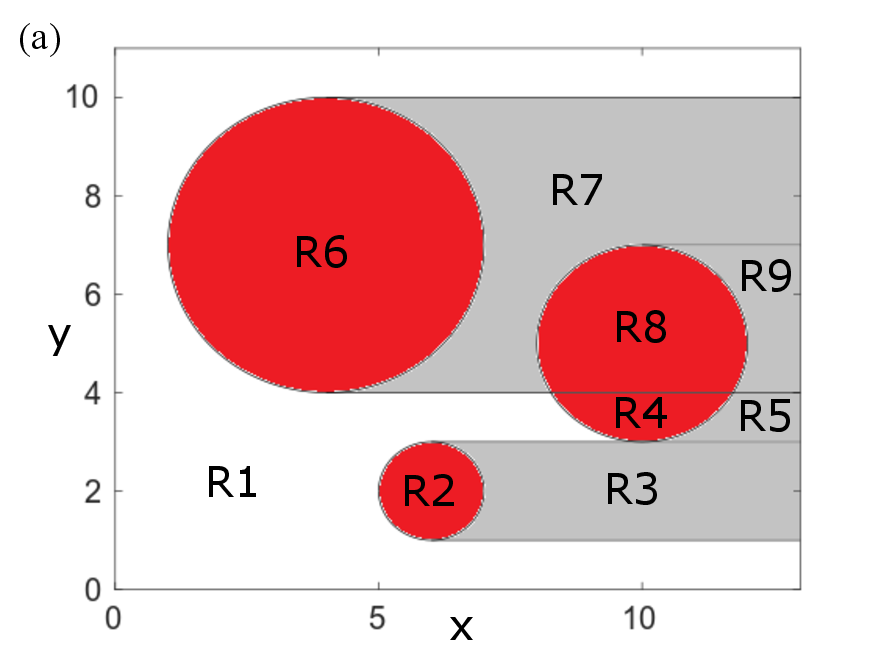}
	\end{subfigure}
	\begin{subfigure}{0.40\textwidth}
		\includegraphics[width=\textwidth]{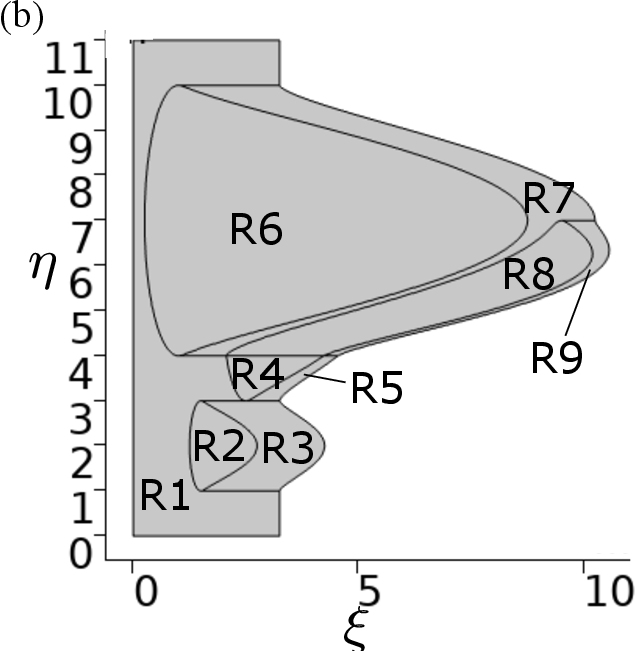}
	\end{subfigure}
	\caption{ (a) The Eulerian domain of the dynamic Lagrangian problem presented in \S\ref{sec:MethodsLag}. This domain is broken into $9$ different regions (denoted R1 to R9) to compute the piecewise continuous definition of $\xi$, given in \rmc{supplementary material, \S{S1}}. The red circles are the locations of the initial deposits of exogenous surfactant. (b) The deformed Lagrangian domain, calculated such that \eqref{eq:alphgam1} holds for the Eulerian initial conditions \rmc{(\ref{eq:InitCondsGenNonD}, \ref{eq:InitCondsGenNonD2})} with the same parameter choices taken in \S\ref{sec:EulNum}. This is the domain in which we compute the numerical solution of \eqref{eq:KeyEqn2ndForm} with boundary conditions \eqref{eq:DynBoundConds}.}
	\label{fig:LagrangianDomain}
\end{figure}

We illustrate the Lagrangian method introduced in \S\ref{sec:DerLag} by solving the model problem with the parameters outlined in \S\ref{sec:3blobsoutline}. For the initial conditions \rmc{(\ref{eq:InitCondsGenNonD}, \ref{eq:InitCondsGenNonD2})} the solution of \eqref{eq:xiintegral} is 
\begin{equation}\label{eq:piecewisexi3blobs2}
	\xi = \begin{cases}
		 x- \left(\frac{(x-x_1)^3}{3}+x(y-y_1)^2\right)\left(1-\delta\right) +C_1(y) \qquad & |\mathbf{x}-\mathbf{x}_1|  \leq 1,
		\\
		\Gamma_2x- \frac{1}{r_2^2}\left(\frac{(x-x_2)^3}{3}+x(y-y_2)^2\right)\left(\Gamma_2-\delta\right) +C_2(y) \qquad & |\mathbf{x}-\mathbf{x}_2| \leq r_2,
		\\
		\Gamma_3 x- \frac{1}{r_3^2}\left(\frac{(x-x_3)^3}{3}+x(y-y_3)^2\right)\left(\Gamma_3-\delta\right) +C_3(y) \qquad & |\mathbf{x}-\mathbf{x}_3|  \leq r_3,
		\\
		\delta x + C_4(y)  \qquad & \text{everywhere else in } \Omega.
	\end{cases}
\end{equation} 
Here, $C_1(y),C_2(y),C_3(y)$ and $C_4(y)$ are determined for the choice of $\Gamma_2=1$, $\Gamma_3=2$, $(x_1,y_1) = (6,2)$, $(x_2,y_2) = (10,5)$, and $(x_3,y_3)=(4,7)$ \rmc{in \S{S1} of the supplementary material}  along with the definition of the Lagrangian coordinates of the three circles. Finding this initial condition involves breaking the Lagrangian domain into $9$ regions, as shown in figure \ref{fig:LagrangianDomain}. By imposing $Y=\eta$, and imposing that the line $X=0$ corresponds to $\xi=0$, only the right-hand side of the Lagrangian domain, which we call $\partial \Omega_R$ (defined for this problem in \rmc{ equation (S1.2) of the supplementary material}), is not a straight line. We substitute $\eta=y$ into \eqref{eq:piecewisexi3blobs2} and then invert \eqref{eq:piecewisexi3blobs2} numerically to find the initial expression for $X$ as an explicit function of $\xi$ and $\eta$.

The boundary conditions for \eqref{eq:KeyEqn2ndForm}, and for the steady-state problem presented in the next subsection, are derived from the dynamic boundary condition \eqref{eq:EulBoundConds}. Analysis in Appendix \ref{sec:Appsharpcorner} reveals that 
for corner angles less than $\pi$, such as we have in the domain we consider, a particle that begins on one of the four edges of the rectangle must stay on that edge for all time, and the appropriate boundary conditions accompanying \eqref{eq:SSproblemwithoutphi} in the undeformed Lagrangian domain are the Dirichlet conditions
\begin{equation}\label{eq:BoundCondsinxy} 
    X = 0, L_1 \text{ on } x_0=0,L_1 \qquad \text{and} \qquad Y=0, L_2 \text{ on } y_0=0,L_2,
 \end{equation} 
which ensures that \eqref{eq:EulBoundConds} is satisfied. This means that in the deformed Lagrangian domain
\begin{equation}\label{eq:DynBoundConds}
 X(0,\eta,\tau) = 0, \qquad  X(\xi,\eta,\tau) = L_1 \text{ on } \partial \Omega_R,   \qquad Y(\xi,0,\tau) = 0, \qquad Y(\xi,L_2,\tau) =   L_2.
\end{equation}

\subsubsection{Numerical solution}
\label{sec:LagNum}

Having inverted \eqref{eq:piecewisexi3blobs2} numerically to find the initial conditions \eqref{eq:InitConds}, we use these initial conditions to solve \eqref{eq:KeyEqn2ndForm} subject to boundary conditions \eqref{eq:DynBoundConds}, from $\tau=0$ to a final time taken to approximate the steady-state $\tau=t_f$, in the Lagrangian domain shown in figure \ref{fig:LagrangianDomain}(\textit{b}) using COMSOL\textsuperscript{\textregistered}. For reproducibility purposes we provide the details of the COMSOL\textsuperscript{\textregistered} settings chosen: we use the Mathematics suite, using the coefficient form PDE setup which is designed to handle PDEs in divergence form such as \eqref{eq:KeyEqn2ndForm}. We discretise using COMSOL\textsuperscript{\textregistered}'s standard triangulation method, and we use quadratic Lagrange basis functions with $314198$ degrees of freedom plus $16578$ internal degrees of freedom, and set the relative tolerance to $10^{-9}$. We store the solution at every $2$ time units. 

\subsection{The steady-state problem}
\label{sec:MethodsSteadyState}

\subsubsection{Formulation of the problem}

We now consider the problem of approximating the equilibrium locations of surface particles (their locations as $t\to\infty$) as a function of their initial locations directly, i.e. without any intermediate calculation of surfactant concentrations, or intermediate calculation of particle trajectories. We return to the coordinate systems used in \S\ref{sec:MethodsLag}; however, here we revert to calling the Lagrangian coordinates $(x_0,y_0)$ to indicate that the Lagrangian domain is a copy of the Eulerian domain, defined as $0\leq x_0\leq L_1$ and $0\leq y_0\leq L_2$ and $(X,Y) = (x_0,y_0)$ at $t=\tau=0$. Using these variables, at steady state, \eqref{eq:LagMassConserv} becomes
\begin{equation}\label{eq:SSproblemwithoutphi}
X_{x_0}Y_{y_0} - X_{y_0}Y_{x_0}  = \frac{\Gamma_0(x_0,y_0)}{\bar{\Gamma}},
\end{equation} 
which is a partial differential equation describing the mapping function $(X,Y)$ to the spatial coordinates $(x,y)$ for particles starting at $(x_0,y_0)$, in the limit $t\to\infty$. Equation \eqref{eq:SSproblemwithoutphi} needs to be solved subject to boundary conditions \eqref{eq:BoundCondsinxy}.

\oej{For one-dimensional problems, \hbox{e.g.} $\Gamma_0=\Gamma_0(x_0)$,} \rmc{we can impose $Y=y_0$ and} \oej{(\ref{eq:SSproblemwithoutphi}) has a unique solution.  However, in two dimensions,}  \eqref{eq:SSproblemwithoutphi} constitutes only one equation for the two unknowns $(X,Y)$, and therefore does not have a unique solution, so we turn to the Helmholtz decomposition theorem to make progress. By this theorem, we know that we can write the map $\mathbf{X}=[X,Y]^T$ in terms of two scalar potentials $\phi(x_0,y_0)$ and $\psi(x_0,y_0)$ such that
\begin{equation}\label{eq:HelmholtzDecomp}
[X,Y]^T = \bm{\nabla}_{\mathbf{x}_0}\phi + \bm{\nabla}_{\mathbf{x}_0}\times\bm{\psi},
\end{equation}
where $\bm{\psi}$ is a vector of magnitude $\psi$ pointing out of the plane (in the $z$-direction), with $\bm{\nabla}_{\mathbf{x}_0}\cdot \mathbf{X}=\nabla_{\mathbf{x}_0}^2\phi$ and $\bm{\nabla}_{\mathbf{x}_0}\times\mathbf{X} = - \nabla_{\mathbf{x}_0}^2\bm{\psi}$. To make this Helmholtz decomposition unique up to constants, we impose the boundary conditions
\begin{equation}\label{eq:HelmholtzBoundConds}
  \bm{\nabla}_{\mathbf{x}_0}\phi \cdot \mathbf{n}_b = [x_0,y_0]^T\cdot \mathbf{n}_b, \qquad \bm{\nabla}_{\mathbf{x}_0}\times \bm{\psi}\cdot \mathbf{n}_b = 0 \qquad \text{on } \partial \Omega ,
\end{equation}
which satisfies \eqref{eq:BoundCondsinxy}. 

The map at time $t$ is generated by \eqref{eq:particlevelocity}, \rmc{the right-hand-side of which is an Eulerian} gradient of the instantaneous surfactant concentration. Thus, the map remains irrotational with respect to the Eulerian coordinates. Now, we investigate whether the map at time $t$ can be approximated by a map that is irrotational with respect to the Lagrangian coordinates, as this would allow us to remove the indeterminacy in \eqref{eq:SSproblemwithoutphi}, since $\bm{\nabla}_{\mathbf{x}_0}\times [X,Y]^T =\bm{0}$ yields $\psi$ equal to a constant, reducing the problem \eqref{eq:SSproblemwithoutphi} to finding a solution for a single scalar potential $\phi$. 
We summarise \rmc{the statement that we want to test as that}, for all time $t$,
\begin{equation}\label{eq:mapassumption}
|\bm{\nabla}_{\mathbf{x}_0}\phi| \gg |\bm{\nabla}_{\mathbf{x}_0}\times\bm{\psi}|.
\end{equation} 
In effect, we test the idea that because the Eulerian curl of $\mathbf{u}_s$ is zero, and that material particles on boundaries are not allowed to traverse corners, \eqref{eq:mapassumption} might hold for all time, at least when the rearrangement of the surface is small. We will test this hypothesis \textit{a-posteriori} in \S\ref{sec:Results}. 

Assuming the map \eqref{eq:HelmholtzDecomp} is given by $[X,Y]^T = \bm{\nabla}_{\mathbf{x}_0}\phi$, equation \eqref{eq:SSproblemwithoutphi} and boundary conditions \eqref{eq:BoundCondsinxy} reduce to the Monge--Amp\`ere equation
\begin{equation}\label{eq:MongeAmpere}
 \phi_{x_0 x_0}\phi_{y_0 y_0} - \phi_{x_0 y_0}^2 = \frac{\Gamma_0(x_0, y_0)}{\bar{\Gamma}} \qquad \text{on } 0\leq x_0 \leq L_1, \quad 0\leq y_0 \leq L_2 ,
\end{equation} subject to
\begin{equation}\label{eq:MABoundConds}
\phi_{x_0} = x_0 \quad \text{on } x_0 = 0,L_1, \qquad \phi_{y_0}=y_0 \quad \text{ on } y_0=0,L_2,  \qquad \phi(0,0)=0.
\end{equation} \rmc{The last boundary condition is necessary to close the problem, as $\phi$ is unique only up to a constant.} The Monge--Amp\`ere equation arises often in the theory of optimal transport, a connection we will further discuss in \S\ref{sec:Discussion}.

\subsubsection{Numerical method}
\label{sec:MethodsSteadyNum}

We solve \eqref{eq:MongeAmpere} subject to the boundary conditions \eqref{eq:MABoundConds} for the initial concentration profile of surfactant \rmc{(\ref{eq:InitCondsGenNonD}, \ref{eq:InitCondsGenNonD2})} using an iterative Newton--Raphson scheme for a finite-difference approximation of the solution, the full details of which are in \jl{Appendix \ref{sec:AppNumScheme}}. The Newton--Raphson scheme converges to the desired solution only if the initial guess is in the basin of attraction of the desired solution, which for a nonlinear problem such as (\ref{eq:MongeAmpere}, \ref{eq:MABoundConds}) is difficult to determine \textit{a priori}. We surmount this problem with the following continuation scheme: using a parameter $\beta_j \in [0,1]$, we take advantage of the fact that the PDE
\begin{equation}\label{eq:MongeAmperebeta}
 \phi_{x_0 x_0}\phi_{y_0 y_0} - \phi_{x_0 y_0}^2 - 1 + \beta_j \left(1 -  \frac{\Gamma_0(x_0,y_0)}{\bar{\Gamma}} \right) = 0,
\end{equation} subject to boundary conditions \eqref{eq:MABoundConds}, has a known solution when $\beta_j = 0$, namely $\phi = x_0^2/2 + y_0^2/2$; when $\beta_j=1$, we have the desired solution to (\ref{eq:MongeAmpere}, \ref{eq:MABoundConds}). We step from $\beta_0=0$ to $\beta_J=1$, in steps of some fixed quantity $\Delta \beta=1/J$ where $J$ is an integer, solving (\ref{eq:MongeAmperebeta}, \ref{eq:MABoundConds}) each time. Starting from $\beta_0=0$ and $\phi_0=x_0^2/2+y_0^2/2$, we find $\phi_{j+1}$ by using $\phi_j$ as a guess solution for (\ref{eq:MongeAmperebeta}, \ref{eq:MABoundConds}) where $\beta_{j+1}=\beta_j +\Delta \beta$. If we choose $\Delta \beta$ to be small enough we ensure that we stay inside the basin of attraction of solutions, finding the desired solution to (\ref{eq:MongeAmpere}, \ref{eq:MABoundConds}) when $j=J$. 

We use this process to solve (\ref{eq:MongeAmpere}, \ref{eq:MABoundConds}) for intermediate and low values of endogenous surfactant, $\delta=0.25$ and $\delta = 0.002$. We solve for the larger value of $\delta$ using a grid with gridpoints uniformly spaced 0.05 units apart in MATLAB\textsuperscript{\textregistered}, using the software's `sparse' variable type to handle the large sparse matrices, and its efficient algorithms for finding solutions to linear systems such as \eqref{eq:NewtRaph} \oej{with a direct LU factorisation scheme}. This solution is obtained by using $\Delta \beta=0.1$. For the solution with the smaller value of $\delta$ we use grid-points evenly spaced 0.05 units apart. We need $\Delta \beta=0.0025$ for this second solution, which means the computational cost is increased. The convergence of the numerical scheme is presented in \jl{Appendix \ref{sec:AppConv}}. In addition, we present a method for creating a computational Suminagashi picture in \jl{Appendix \ref{sec:appSumiMethod}}. 

\rmc{To quantify how well the Monge--Amp\`ere method approximates the  solution  found by the Eulerian particle-tracking method at $t=t_f$ (assumed to be an accurate solution of the steady state), we define metrics that characterize the difference between solutions found using the two methods for the same initial conditions. We define the Euclidean distance between final particle locations $X_{EU}$ and $X_{MA}$ predicted by both methods and normalised by the longest side of the domain
\begin{equation}\label{eq:EucDist}
\frac{1}{L_1}|{X}_{EU}-{X}_{MA}| \equiv \frac{1}{L_1}\sqrt{(X_{EU}-X_{MA})^2+(Y_{EU}-Y_{MA})^2},
\end{equation}
which we call the normalised absolute error between the two methods for a given initial particle location. Statistics of the error are then obtained by analyzing distributions for a large number of the initial particle locations, particularly the median, the upper quartile, the 90th percentile and the maximum values of \eqref{eq:EucDist}}.

\section{Results}
\label{sec:Results}

Table \ref{tab:SimulationSummary} summarises all of the simulations and their parameters that are presented in the results section, with a key with which we refer to each simulation.

\begin{table}
\centering
\begin{tabular}{c c c c c }
  Method   & $\delta$ & Centre deposits $1/2/3$ & $t_f$ &  Key  \\ \hline
   Eulerian particle tracking   & $0.25$ & (6,2)/(10,5)/(4,7) & 1047.8 & Eul[0.25] \\ 
   Lagrangian particle tracking   & $0.25$ & (6,2)/(10,5)/(4,7) & 1047.8 & Lag[0.25] \\ 
   Monge--Amp\`ere   & $0.25$ & (6,2)/(10,5)/(4,7) & N/A & MA[0.25] 
   \\
   Eulerian particle tracking   & $0.002$ & (6,2)/(10,5)/(4,7) & 2094.1 &  Eul[0.002] \\ 
   Monge--Amp\`ere   & $0.002$ & (6,2)/(10,5)/(4,7) &  N/A & MA[0.002] 
   \\ 
   Monge--Amp\`ere   & $0.04$ & (6,2)/(10,5)/(4,7) &  N/A & MA[0.04]
   \\ 
   Monge--Amp\`ere   & $0.005$ & (6,2)/(10,2)/(4,7) &  N/A & MA[0.005]Alt1
   \\ 
   Monge--Amp\`ere   & $0.005$ & (6,2)/(10,8.5)/(4,7) &  N/A & MA[0.005]Alt2
   \\ 
   Monge--Amp\`ere   & $0.005$ & (3,2)/(10,5)/(4,7) &  N/A & MA[0.005]Alt3
      \\ 
   Monge--Amp\`ere   & $0.005$ & (8,2)/(10,5)/(4,7) &  N/A & MA[0.005]Alt4
   \\ 
   Monge--Amp\`ere   & $0.005$ & (9,2)/(10,5)/(4,7) &  N/A & MA[0.005]Alt5
   \\ 
   Monge--Amp\`ere   & $0.005$ & (11,2)/(10,5)/(4,7) &  N/A & MA[0.005]Alt6
\end{tabular}
\caption{A table presenting a summary of the simulations presented in \S\ref{sec:Results}, together with parameters used, and a key with which we refer to each simulation. The methods used are the Eulerian particle-tracking method (\ref{eq:nonlineardiff}, \ref{eq:particlevelocity}), the Lagrangian particle-tracking method \eqref{eq:KeyEqn2ndForm}, and the Monge--Amp\`ere method \eqref{eq:MongeAmpere}. For all these simulations we choose $r_2=2$, $r_3=3$, $\Gamma_2=1$ and $\Gamma_3=2$.}
\label{tab:SimulationSummary}
\end{table}

\begin{figure}
	\centering
    \includegraphics[keepaspectratio=true,width=\textwidth]{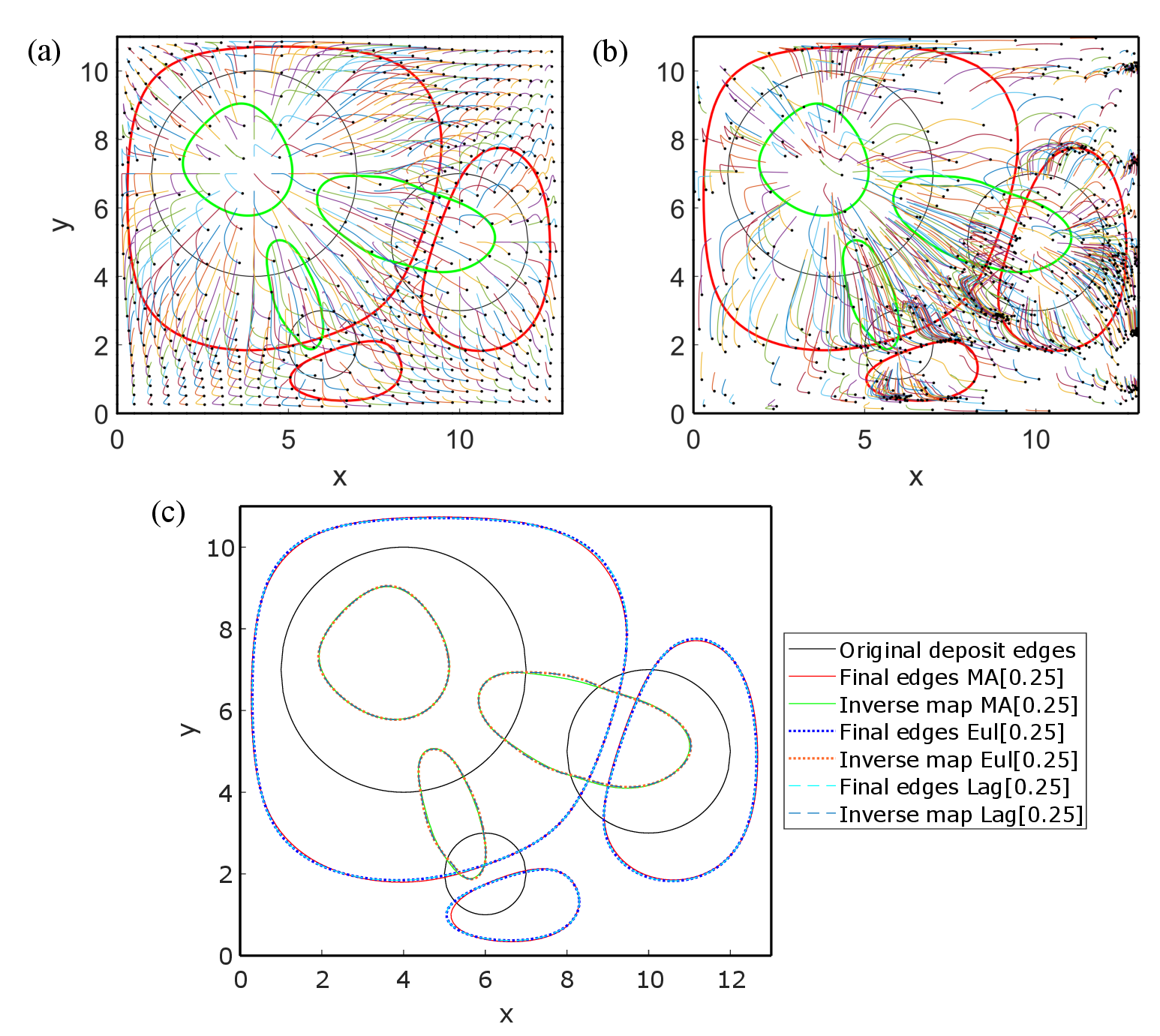}
	\caption{Solution to the example problem of three circular deposits of exogenous surfactant spreading together with $\delta=0.25$. (a) The results of Eul[0.25] (see movie 1 of the supplementary material). The initial boundaries of the exogenous surfactant circular deposits are the black circles, and the final locations are the thick red lines. The points described by the green curves map to the black circles at $t=t_f$. Individual particle trajectories are plotted using thin coloured lines terminating at black points. (b) The results of Lag[0.25] with the same colour scheme as in (a) (see movie 2 of the supplementary material). The particles represent $1/50$ of all the particle trajectories calculated which are chosen at random so the density of particles shown is not significant. (c) Graph showing the results of MA[0.25] overlaid onto Eul[0.25] and Lag[0.25]. The steady-state boundaries of the three deposits and the curves found by the inverse map (which spread from and to the black circles in the steady state respectively) are given by the colour scheme shown in the figure legend. 
 }
	\label{fig:ThreeBlobs}
\end{figure}  

\subsection{Particle-tracking solutions}
\label{sec:ResEul} 

The results for the Eulerian (Eul[0.25]) and Lagrangian (Lag[0.25]) particle-tracking methods (presented in \S\ref{sec:EulNum} and \S\ref{sec:LagNum}) with $\delta=0.25$ are shown in figures \ref{fig:ThreeBlobs}(\textit{a}) and \ref{fig:ThreeBlobs}(\textit{b}) respectively and also as movie 1 and movie 2 of the supplementary material, where each thin coloured line represents a particle trajectory, terminating at a black dot at $t=t_f$. The trajectories shown in figure \ref{fig:ThreeBlobs}(\textit{a}) represent 1 in every 225 trajectories calculated, selected such that their initial locations are evenly spaced. The data obtained by the solution for the Lagrangian method presented in figure \ref{fig:LagrangianDomain}(\textit{b}) are irregularly spaced, with each data point corresponding to a node of the mesh used in COMSOL\textsuperscript{\textregistered} to discretise the deformed Lagrangian domain; we display 1 in every 50 particles from the data list obtained from the simulation, so the density of particles shown is not significant.

In figures \ref{fig:ThreeBlobs}(\textit{a}) and \ref{fig:ThreeBlobs}(\textit{b}), the largest deposit spreads out through Marangoni stresses, and compresses the other two deposits. Flow reversals (sharp turns of particle trajectories of more than $90^{\circ}$) arise in several areas for two reasons. First, reversals in the top left-hand corner of the figure are due to confinement. Early outward spreading is into a region containing endogenous surfactant at low concentration $\delta$; later reversals arise once the surfactant concentration in this region is much larger due to non-local compression of the endogenous material. Second, points which begin on the edges of the smaller two deposits nearest the centre of the domain first spread into the centre, but soon the effect of the largest deposit spreading is felt and these points reverse their trajectories. The final shapes of the smallest deposits are non-trivial oval shapes, the centres of which are shifted away from their initial locations. Some particles to the top-left of the centre of the largest deposit traverse distances close to $1$ unit in length and then move approximately the same distance back, close to where the particles started. Particles compress into the top- and bottom-right-hand corner. A variety of trajectories are evident: for example, particles in the top and left of the figure have trajectories which involve straight lines and sharp turns, whereas particles towards the bottom right describe gentle arcs. \rmc{In figure S1 in \S{S2} of the supplementary material we present an overlay of the contour plots of $X(x_0,y_0)$ and $Y(x_0,y_0)$ for the solutions at $t=t_f$ obtained from the Eulerian and Lagrangian particle-tracking methods, respectively. The methods find the same particle locations to within a distance of $0.05$ almost everywhere, apart from the locations of small oscillations in the Lagrangian solution which appear to be an artefact of the domain deformation as discussed in section S2, and much closer than that in most places. Some of the small discrepancies that do exist can be partly explained by the fact that small errors arise by interpolating the Lagrangian solution onto a regular, rectangular grid to make the comparison, and errors occur in the Eulerian solution by the interpolation of the gradient of the evolving concentration shown in \S{S3} (figure~S2) of the supplementary material at every time-step in that solution.}

\subsection{The steady-state solution}
\label{sec:ResSteadyState}

\rmc{We investigated the steady-state solution of a variety of configurations involving one and two deposits (see table S1 in the supplementary material). We compare the results obtained between Eulerian (EU) particle-tracking and the Monge--Amp\`ere (MA) method in \S{S5} of the supplementary material. We tested how the final equilibrium  shape of the deposits is influenced by the proximity of the domain boundaries. In the case of a single initial deposit, we find only small differences in the discrepancy between the EU and MA results, quantified using \eqref{eq:EucDist} for  deposit locations at various distances from the domain boundaries. The median normalised absolute error is approximately $10^{-4}$ and the maximum error is bounded by $2\times 10^{-3}$. In the case of two initial deposits, the median normalised absolute error is about $5\times10^{-4}$ and the maximum error is bounded by $5\times 10^{-3}$. The median discrepancy between the two methods tends to be inversely correlated with the symmetry of the initial configuration, whereas the upper quartile, 90th percentile and maxiumum discrepancy are much noisier for both the one-deposit and two-deposit problems studied. Discrepancies between the EU and MA methods increase with an increase in the number of deposits, and with a decrease in $\delta$ (the normalized initial endogenous surfactant concentration), as shown in Appendix \ref{sec:Differencedelta}.  As stated previously, we choose to focus on the three-deposit case.}

\subsubsection{The three-deposits problem with $\delta=0.25$}

The solution for the approximation of the edges of the three deposits in the the steady state (MA[0.25]) found by the Monge--Amp\`ere method (outlined in \S\ref{sec:MethodsSteadyNum}) is presented in figure \ref{fig:ThreeBlobs}(\textit{c}) for $\delta = 0.25$, with the approximations of the steady state found from the Eulerian (Eul[0.25]) and Lagrangian (Lag[0.25]) particle-tracking solutions overlaid. The final edges of the deposits predicted by MA[0.25] are almost indistinguishable except for a few places, which supports assumption \eqref{eq:mapassumption}. \rmc{Predictions of Eul[0.25] and Lag[0.25] are indistinguishable to the naked eye in figure S1 of the supplement, providing a reliable benchmark against which to test the prediction of MA[0.25].} We also use the inverse maps to calculate the contours which map to the initial drop boundaries under the spreading in figure \ref{fig:ThreeBlobs}; again, only very small discrepancies between MA[0.25] and Eul[0.25] are evident.

\begin{figure}
	\centering
 \begin{subfigure}{0.49\textwidth}\centering
		\includegraphics[width=1.1\textwidth]{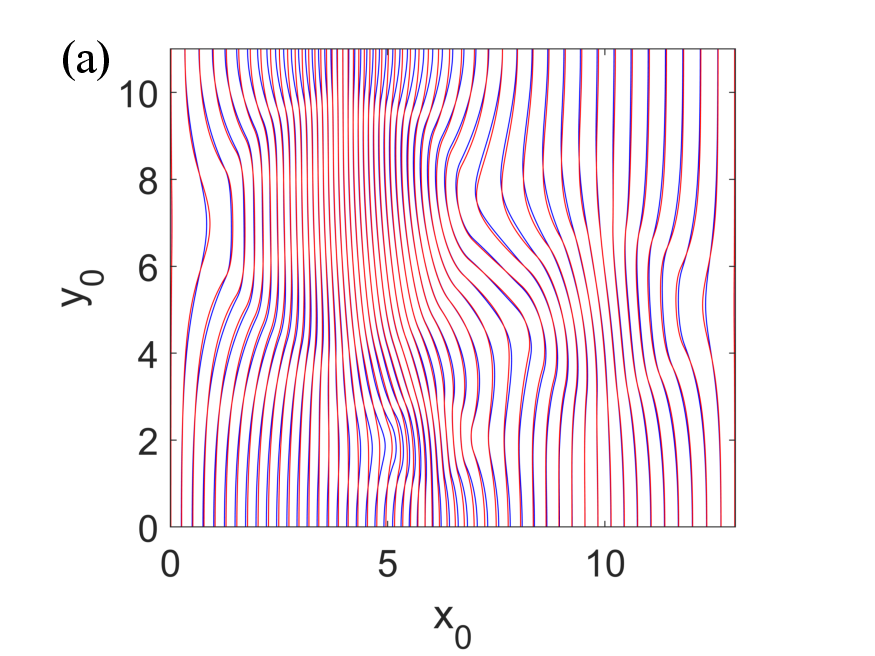}
	\end{subfigure}
 \begin{subfigure}{0.49\textwidth}
		\includegraphics[width=1.1\textwidth]{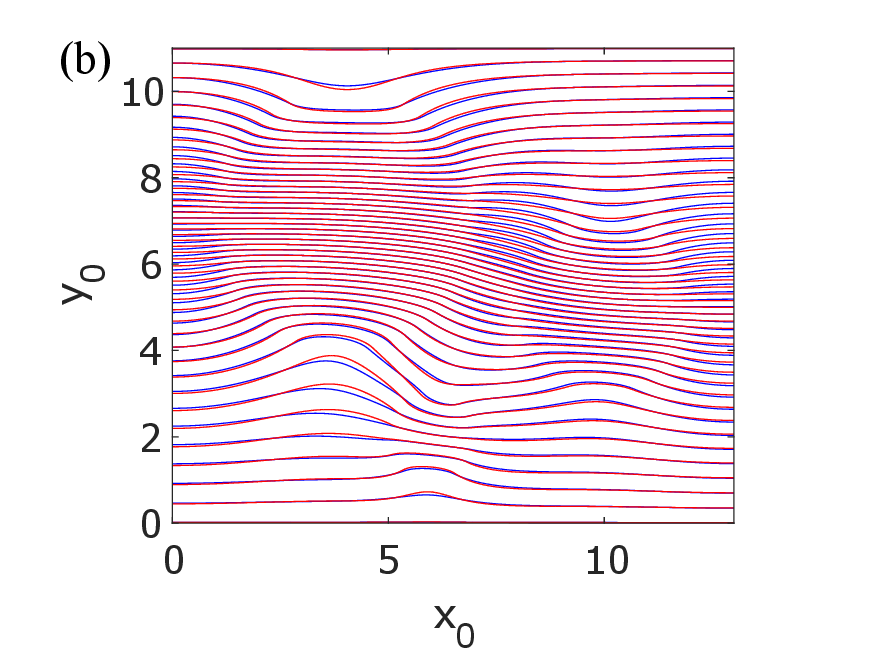}
	\end{subfigure}
 \begin{subfigure}{0.49\textwidth}
    \centering
\includegraphics[keepaspectratio=true,width=1.1\textwidth]{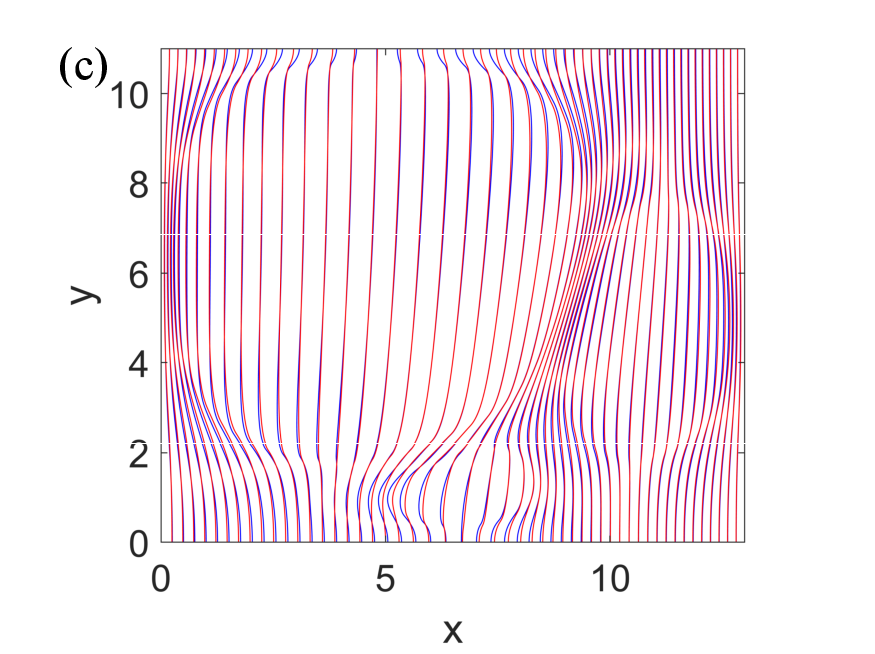}
    \end{subfigure}
     \begin{subfigure}{0.49\textwidth}
    \centering
\includegraphics[keepaspectratio=true,width=1.1\textwidth]{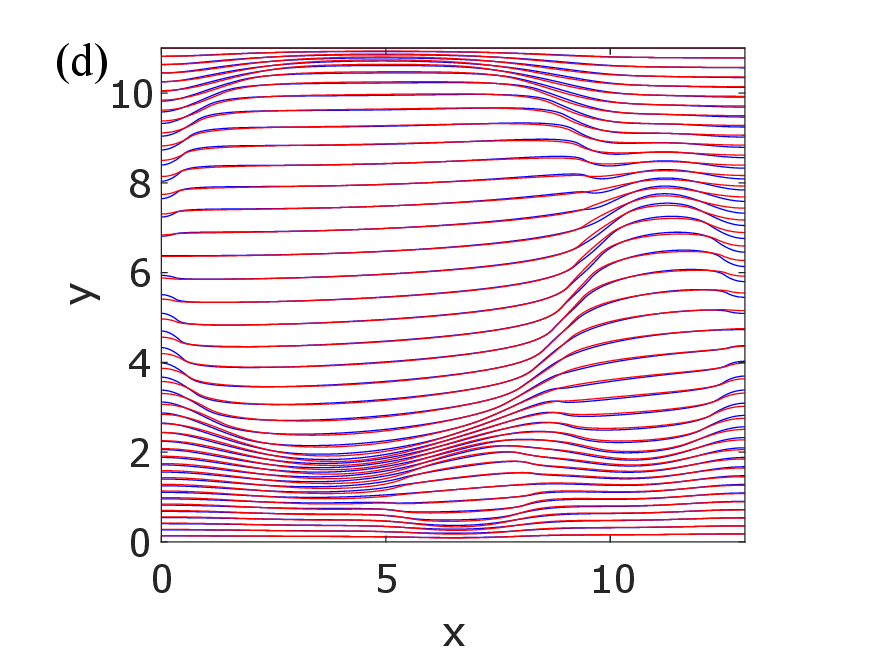}
    \end{subfigure}
	\caption{Contour plots of solutions for the map from initial configuration to steady-state found from MA[0.25] and Eul[0.25]. (a) 25 evenly spaced contours of $X_{MA}$ taken from MA[0.25] (red) overlaid with the same valued contours of $X_{EU}$ taken from Eul[0.25] (blue). (b) 25 contours of $Y_{MA}$ (red) overlaid with $Y_{EU}$ (blue). (c) The inverse map $X_{MA}^{-1}$ (red) overlaid with $X_{EU}^{-1}$ (blue), with the same contour scheme. (d) The same for $Y_{MA}^{-1}$ (red) overlaid with $Y_{EU}^{-1}$ (blue). }
	\label{fig:OverlayContourMapsdelta0p25}
\end{figure}

A comparison of the global behaviour of the Monge--Amp\`ere approximation of the map from initial to final particle configuration (MA[0.25]) with the map calculated from the particle-tracking solution (Eul[0.25]) is given by contour plots in figure \ref{fig:OverlayContourMapsdelta0p25}  \rmc{(see also a colour map of the absolute error between the two predictions for the final particle location in \S{S4}, figure S3, of the supplementary material)}. Dense contours in figures \ref{fig:OverlayContourMapsdelta0p25}(\textit{a}) and \ref{fig:OverlayContourMapsdelta0p25}(\textit{b}) indicate that surface areas starting at these locations are stretched by the mapping, and similarly large gaps between contours indicate that the map compresses the surface. Conversely, in figures \ref{fig:OverlayContourMapsdelta0p25}(\textit{c}) and \ref{fig:OverlayContourMapsdelta0p25}(\textit{d}) dense contours of the inverse maps indicate that surface areas finishing at these locations have been compressed by the spreading, and large gaps between contours indicate that the spreading has stretched the surface. The Monge--Amp\`ere approximation agrees with the particle-tracking solution in most places, although some noticeable discrepancies exist, such as in the left half of the smallest deposit most notably. \oej{The median error across the solution is approximately 0.25\% of the domain length, and the error for every particle is within $1.5$\% of the domain length as shown in Figure S7 of the supplementary material.} 


\begin{figure}
	\centering
 \begin{subfigure}{0.48\textwidth}\centering
		\includegraphics[width=\textwidth]{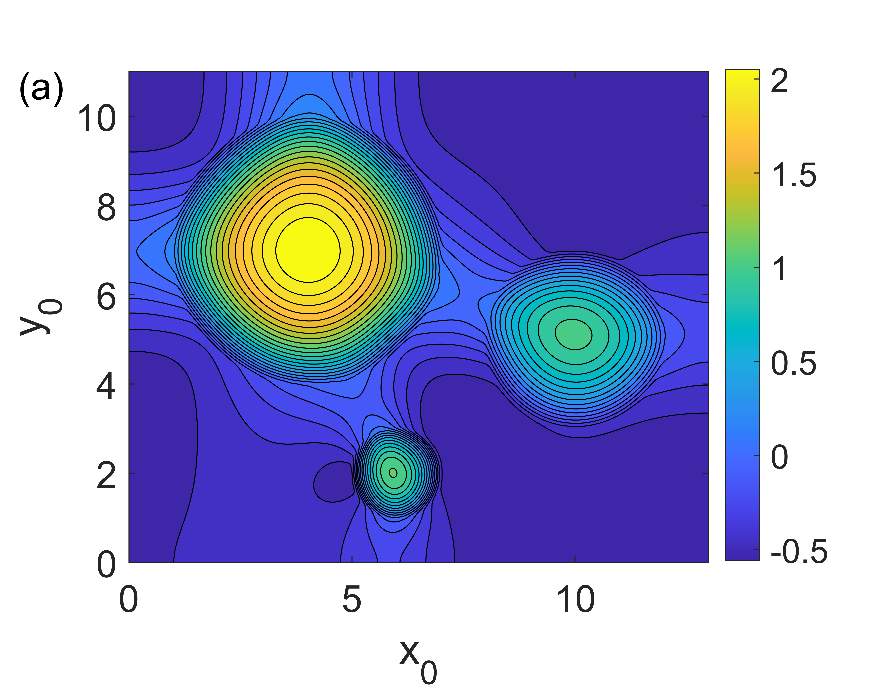}
	\end{subfigure}
 \begin{subfigure}{0.48\textwidth}
		\includegraphics[width=\textwidth]{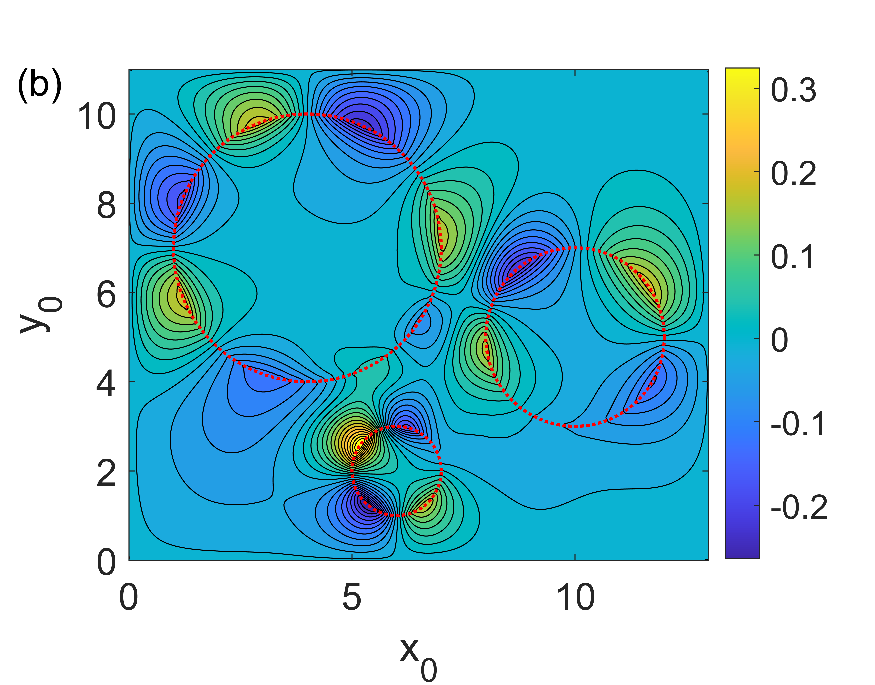}
	\end{subfigure}
 \begin{subfigure}{0.48\textwidth}
    \centering
    \includegraphics[keepaspectratio=true,width=\textwidth]{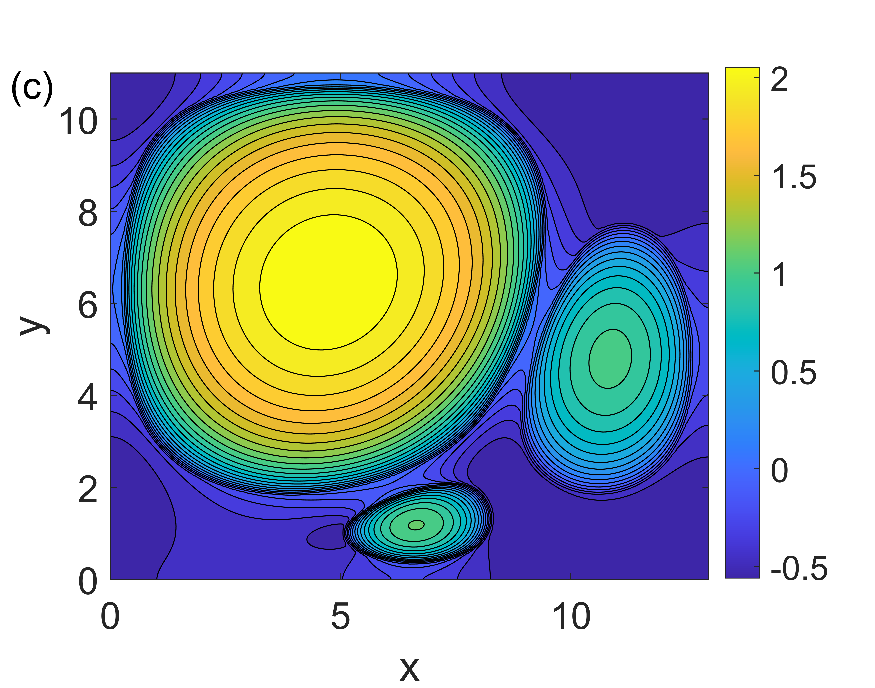}
    \end{subfigure}
     \begin{subfigure}{0.48\textwidth}
    \centering
    \includegraphics[keepaspectratio=true,width=\textwidth]{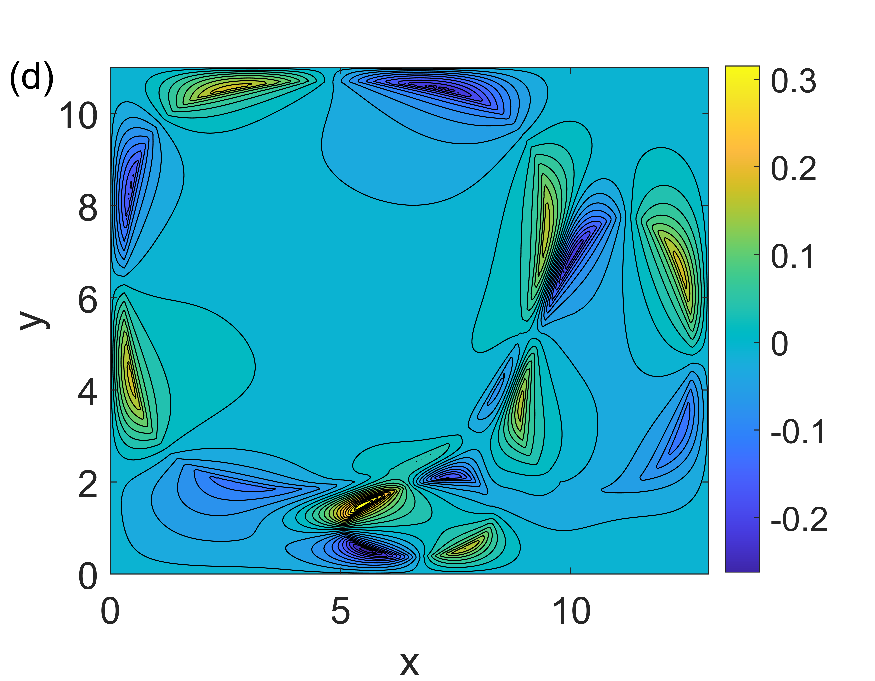}
    \end{subfigure}
	\caption{Contour plots showing the divergence and curl of the vector field from initial to final particle location taken from Eul[0.25]. (a) $\bm{\nabla}_{\mathbf{x}_0}\cdot(\mathbf{X}-\mathbf{x}_0)= \nabla_{\mathbf{x}_0}^2\phi-2$ is plotted in the Lagrangian coordinates. (b) $(\bm{\nabla}_{\mathbf{x}_0}\times(\mathbf{X}-\mathbf{x}_0))_{\perp} = -\nabla_{\mathbf{x}_0}^2\psi$ is plotted in the Lagrangian coordinates. (c) $\bm{\nabla}_{\mathbf{x}_0}\cdot(\mathbf{X}-\mathbf{x}_0)= \nabla_{\mathbf{x}_0}^2\phi-2$ in Eulerian coordinates. (d) $(\bm{\nabla}_{\mathbf{x}_0}\times(\mathbf{X}-\mathbf{x}_0))_{\perp} = -\nabla_{\mathbf{x}_0}^2\psi$ in Eulerian coordinates.}
	\label{fig:DivCurldelta0p25}
\end{figure}

To illustrate where and how the discrepancies arise, figure \ref{fig:DivCurldelta0p25} shows the divergence and the curl of the map found from Eul[0.25], shown in both Lagrangian and Eulerian coordinate systems. From the Helmholtz decomposition \eqref{eq:HelmholtzDecomp}, figures \ref{fig:DivCurldelta0p25}(\textit{a}) and \ref{fig:DivCurldelta0p25}(\textit{c}) show $\bm{\nabla}_{\mathbf{x}_0}\cdot (\mathbf{X}-\mathbf{x}_0)=\nabla_{\mathbf{x}_0}^2\phi-2$ and figures \ref{fig:DivCurldelta0p25}(\textit{b}) and \ref{fig:DivCurldelta0p25}(\textit{d}) show $(\bm{\nabla}_{\mathbf{x}_0}\times(\mathbf{X}-\mathbf{x}_0))_{\perp} = -\nabla_{\mathbf{x}_0}^2\psi $, where $(.)_{\perp}$ means the z-component, perpendicular to the plane of the solution. The vector field $\mathbf{X}-\mathbf{x}_0$, which points from initial to final particle locations, is an easier quantity to interpret physically than $\mathbf{X}$ itself. Given boundary conditions \eqref{eq:HelmholtzBoundConds} (the boundary condition for $\psi$ can be taken to be equivalent to the Dirichlet condition $\psi=0$ on all four boundaries), the fact that $|\nabla_{\mathbf{x}_0}^2\phi|$ is an order of magnitude greater than $| \nabla_{\mathbf{x}_0}^2\psi |$ almost everywhere is further evidence justifying our assumption \eqref{eq:mapassumption} (\rmc{the ratio $\nabla^2_{x_0}\psi/\nabla^2_{x_0}\phi$ is plotted in \S{S4}, figure S4, of the supplementary material}), although it is certainly not the case that the curl of the map vanishes. In figures \ref{fig:DivCurldelta0p25}(\textit{a}) and \ref{fig:DivCurldelta0p25}(\textit{c}), $\bm{\nabla}_{\mathbf{x}_0}\cdot (\mathbf{X}-\mathbf{x}_0)<0$ represents surface areas with net compression, and $\bm{\nabla}_{\mathbf{x}_0}\cdot (\mathbf{X}-\mathbf{x}_0)>0$ represents areas with net expansion by the map. Areas within the deposits expand, as do area elements connecting the largest deposit with the smaller deposits, and the corner regions compress. Saddle-like area elements directly between each of the initial deposit locations, and between each deposit and the nearest boundary, are compressed and expanded in orthogonal directions. In figures \ref{fig:DivCurldelta0p25}(\textit{b}) and \ref{fig:DivCurldelta0p25}(\textit{d}), positive values of $(\bm{\nabla}_{\mathbf{x}_0}\times(\mathbf{X}-\mathbf{x}_0))_{\perp}$ refer to anti-clockwise net local rotation (twist) by the map, and negative values for clockwise twist. It is notable that weak twisting motions arise where interfaces spread towards a nearby boundary, or near another drop interface, with regions of oppositely oriented twisting typically appearing in pairs. The most intense twisting appears to be confined to regions immediately outside the boundaries of the exogenous surfactant drops.

\begin{figure}
	\centering
 \begin{subfigure}{0.49\textwidth}\centering
		\includegraphics[width=1.1\textwidth]{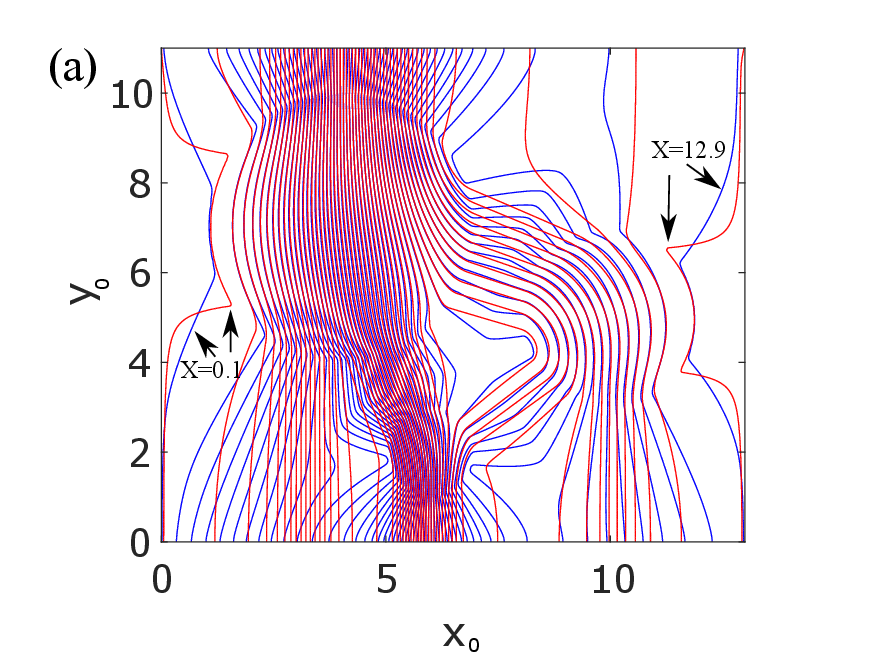}
	\end{subfigure}
 \begin{subfigure}{0.49\textwidth}
		\includegraphics[width=1.1\textwidth]{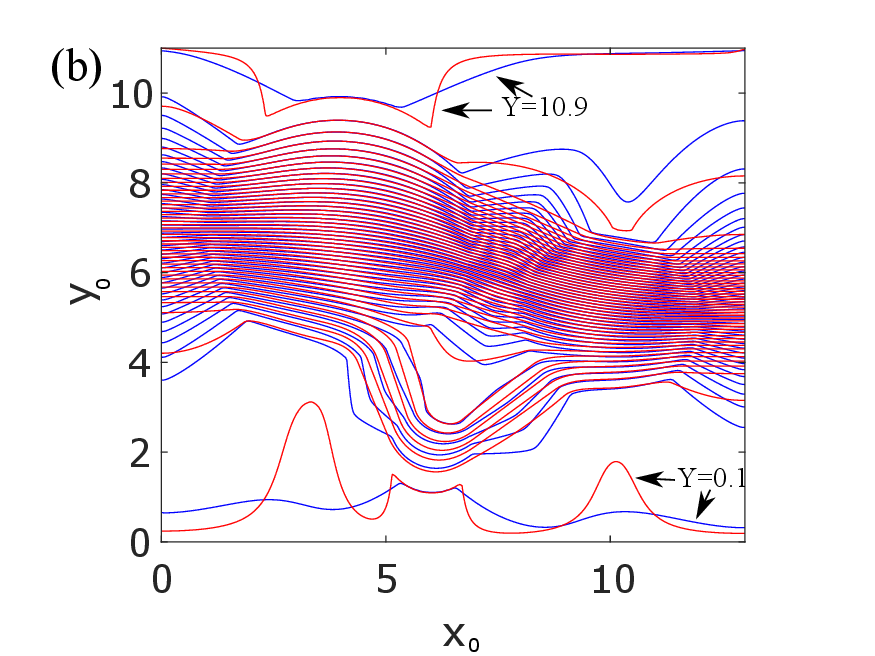}
	\end{subfigure}
 \begin{subfigure}{0.48\textwidth}
    \centering
\includegraphics[keepaspectratio=true,width=1.1\textwidth]{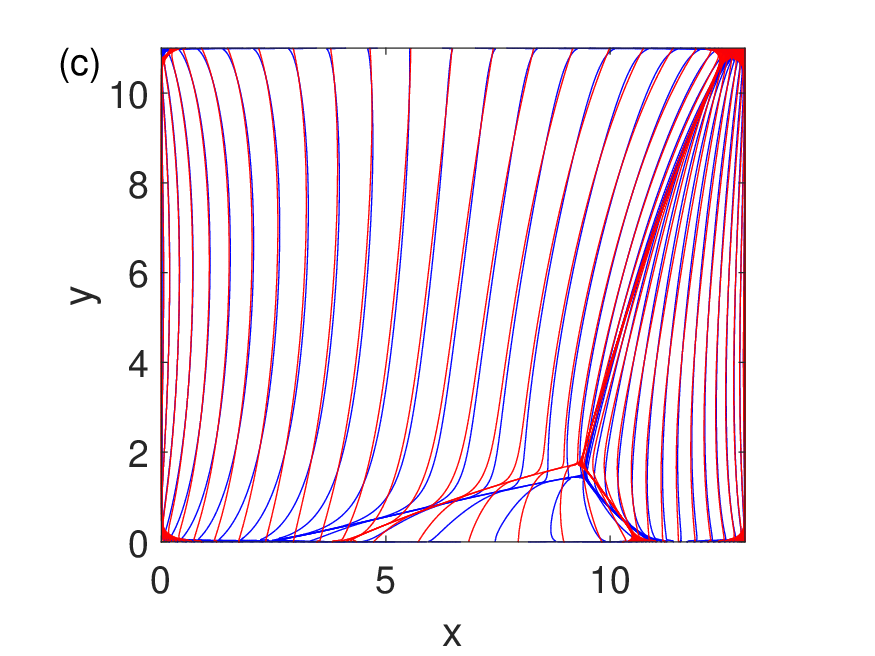}
    \end{subfigure}
     \begin{subfigure}{0.48\textwidth}
    \centering
\includegraphics[keepaspectratio=true,width=1.1\textwidth]{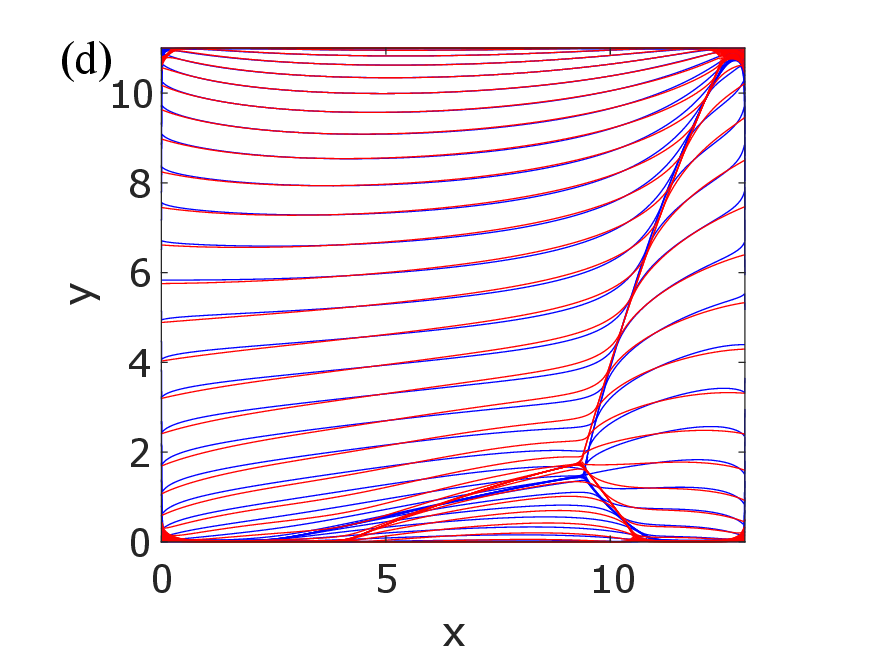}
    \end{subfigure}
    \begin{subfigure}{0.48\textwidth}
    \centering
\includegraphics[keepaspectratio=true,width=1.1\textwidth]{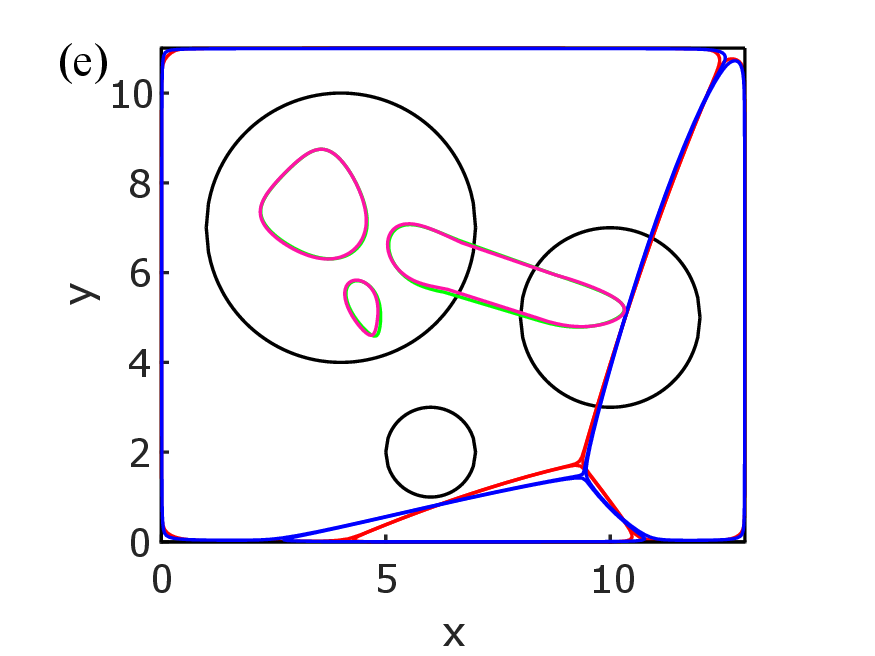}
    \end{subfigure}
    \begin{subfigure}{0.48\textwidth}
    \centering
\includegraphics[keepaspectratio=true,width=1.1\textwidth]{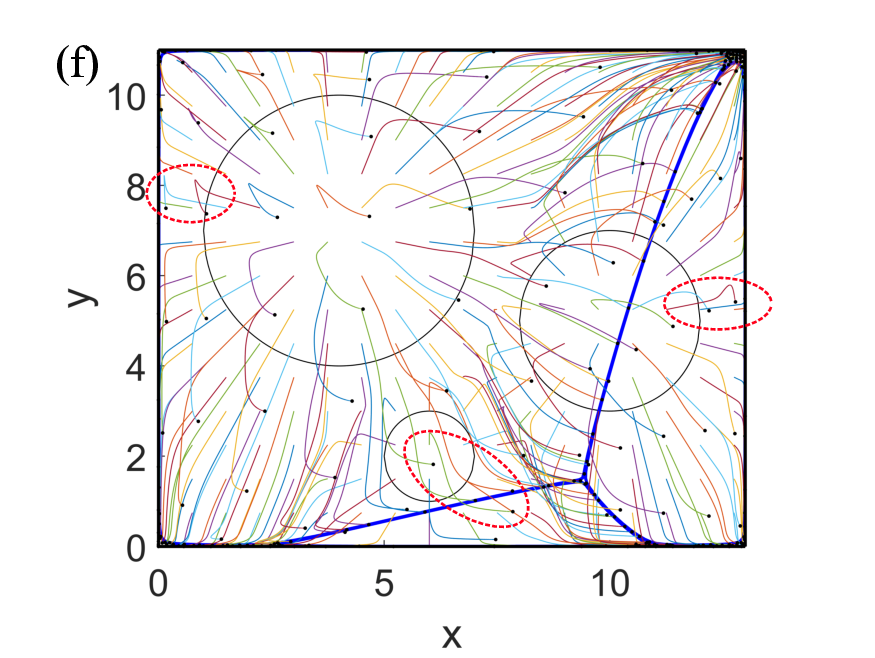}
    \end{subfigure}
	\caption{Plots showing a comparison between MA[0.002] and Eul[0.002]. (a) Contour plots of $X_{MA}$ taken from MA[0.002] (red) overlaid with the same valued contours from $X_{EU}$ (blue) taken from Eul[0.002]. (b) The same for $Y_{MA}$ (red) and $Y_{EU}$ (blue). (c) The same scheme for the inverse maps $X_{MA}^{-1}$ (red) and $X_{EU}^{-1}$ (blue); (d) shows $Y_{MA}^{-1}$ (red) and $Y_{EU}^{-1}$ (blue). 
 (e) An overlay of the final deposit boundaries from MA[0.002] (red) and Eul[0.002] (blue). (f) Particle trajectories, each given by a thin coloured line terminating at a black dot, from Eul[0.002] (see movie 3 of the supplementary material). The three red dashed ellipses each contain a complete particle trajectory which involves two sharp changes of direction.}
	\label{fig:OverlayMapsdelta0p002}
\end{figure}

\begin{figure}
	\centering
 \begin{subfigure}{0.49\textwidth}\centering
		\includegraphics[width=\textwidth]{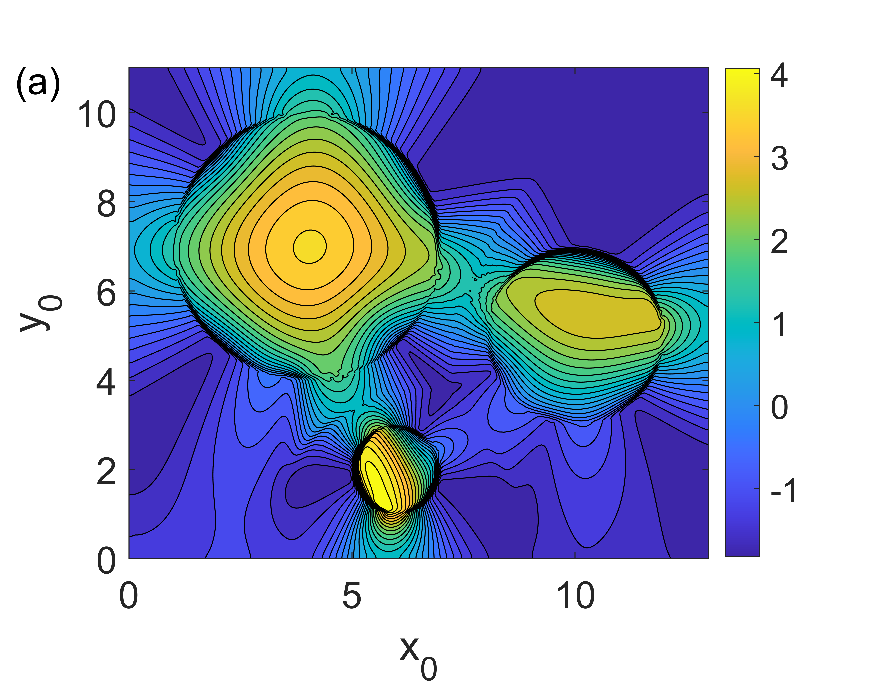}
	\end{subfigure}
 \begin{subfigure}{0.49\textwidth}
		\includegraphics[width=\textwidth]{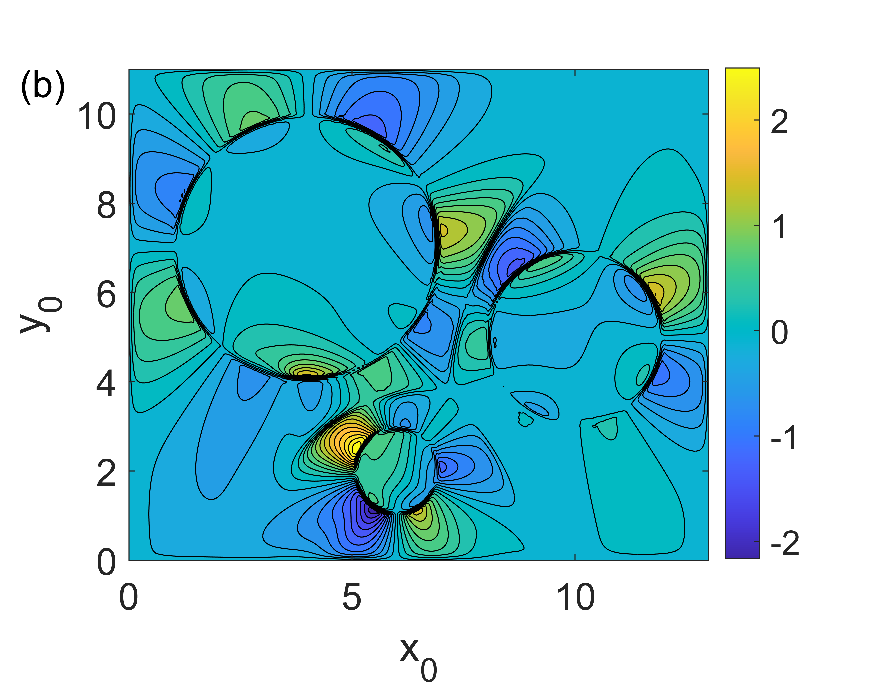}
	\end{subfigure}
 \begin{subfigure}{0.49\textwidth}
    \centering
    \includegraphics[keepaspectratio=true,width=\textwidth]{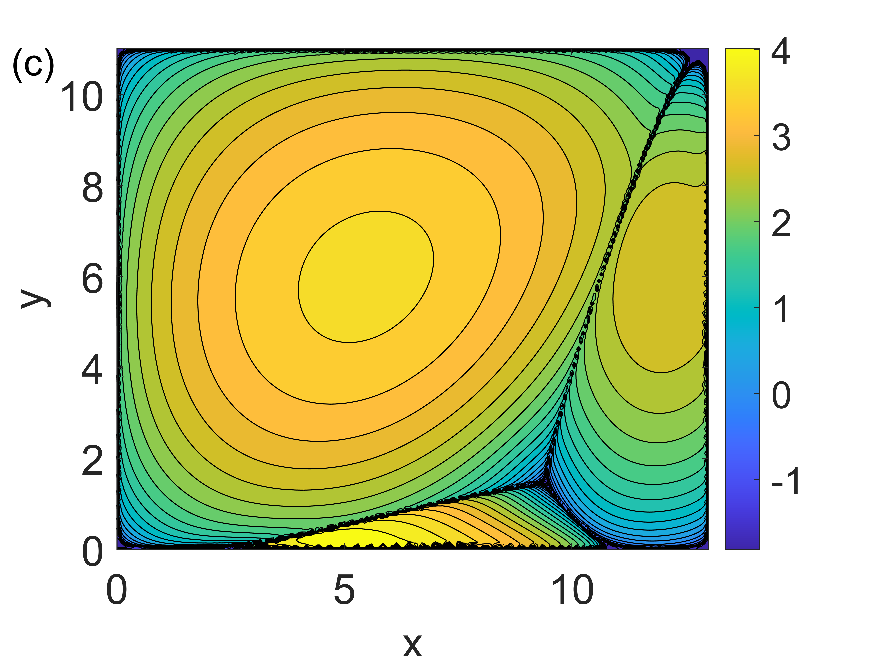}
    \end{subfigure}
     \begin{subfigure}{0.49\textwidth}
    \centering
    \includegraphics[keepaspectratio=true,width=\textwidth]{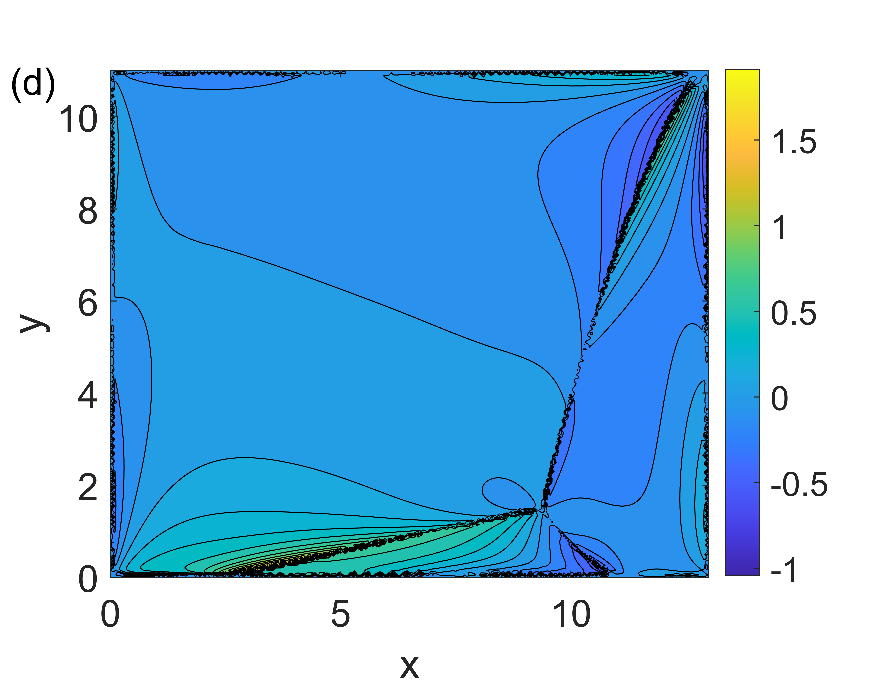}
    \end{subfigure}
	\caption{Contour plots showing the divergence and curl of the vector field from initial to final particle location taken from Eul[0.002]. (a) $\bm{\nabla}_{\mathbf{x}_0}\cdot(\mathbf{X}-\mathbf{x}_0)= \nabla_{\mathbf{x}_0}^2\phi-2$ in Lagrangian coordinates. (b) $(\bm{\nabla}_{\mathbf{x}_0}\times(\mathbf{X}-\mathbf{x}_0))_{\perp} = -\nabla_{\mathbf{x}_0}^2\psi$ in Lagrangian coordinates. (c) $\bm{\nabla}_{\mathbf{x}_0}\cdot(\mathbf{X}-\mathbf{x}_0) = \nabla_{\mathbf{x}_0}^2\phi-2$ in Eulerian coordinates. (d) $(\bm{\nabla}_{\mathbf{x}_0}\times(\mathbf{X}-\mathbf{x}_0))_{\perp} = -\nabla_{\mathbf{x}_0}^2\psi$ in Eulerian coordinates.}
	\label{fig:DivCurldelta0p002}
\end{figure}

\subsubsection{Weak endogenous surfactant}

Many fluids in the environment have low levels of contaminant surfactant, and in some cases (especially in controlled laboratory conditions) vanishingly small endogenous surfactant concentrations, and so we would like to understand the Lagrangian motion of surface particles in the limit of very small $\delta$. 
The Lagrangian dynamic method (\S\ref{sec:LagNum}) is not capable of handling small values of $\delta$, because the deformed Lagrangian domain becomes extremely stretched. The Eulerian particle-tracking method (\S\ref{sec:EulNum}), and the Monge--Amp\`ere approximation (\S\ref{sec:MethodsSteadyNum}), however, can both find well-behaved results with extremely small $\delta$. 

We present the solution to the Monge--Amp\`ere approximation MA[0.002] overlaid with the solution for the Eulerian particle tracking method Eul[0.002] in figure \ref{fig:OverlayMapsdelta0p002}. Figures \ref{fig:OverlayMapsdelta0p002}(\textit{a}) and \ref{fig:OverlayMapsdelta0p002}(\textit{b}) show the overlaid contour plots of the approximations for the steady-state solutions for $X$ and $Y$ in the Lagrangian coordinates, and  figures \ref{fig:OverlayMapsdelta0p002}(\textit{c}) and \ref{fig:OverlayMapsdelta0p002}(\textit{d}) show the inverse mapping in the Eulerian coordinates. Figure \ref{fig:OverlayMapsdelta0p002}(\textit{e}) shows an overlay of the prediction for the final boundary locations for the three deposits using both methods. 

Figure \ref{fig:OverlayMapsdelta0p002} shows that when $\delta$ is small, the Monge--Amp\`ere method does less well in approximating the solution, which is expected as particles spread further leading to large discrepancies between the Lagrangian and Eulerian curl of the maps; however, it is still credible as a first-order approximation. In particular, for particles that are initially located within the three deposits, the approximation is still accurate, with the largest discrepancy occurring for particles of endogenous surfactant. This is shown clearly when comparing figures \ref{fig:OverlayMapsdelta0p002}(\textit{c}) and \ref{fig:OverlayMapsdelta0p002}(\textit{d}), which show the inverse map in the Eulerian coordinates, to figures \ref{fig:OverlayMapsdelta0p002}(\textit{a}) and \ref{fig:OverlayMapsdelta0p002}(\textit{b}) which show the map in the Lagrangian coordinates. \rmc{(This is also shown  clearly in the plot of the absolute and relative error between the predictions for final particle locations in \S{S4}, figure S3(b,d), of the supplementary material.)} Contours coincide over an appreciably greater region of the final configuration in comparison to the initial configuration. \rmc{In Appendix \ref{sec:Differencedelta}, we present box-and-whisker plots of the absolute error between final particle locations predicted by the Monge--Amp\`ere and Eulerian particle tracking methods for several values of $\delta$, which show how the maximum discrepancy grows as $\delta\to 0$ (10\% of the domain length for $\delta=0.002$), but the median discrepancy remains an order of magnitude or more smaller than the maximum discrepancy for $\delta=0.002$.} 

The final locations of particles beginning on the region occupied by endogenous surfactant are compressed into effectively three lines, as shown by the blue curves in figure \ref{fig:OverlayMapsdelta0p002}(\textit{e}) for Eul[0.002]. The Monge--Amp\`ere approximation (MA[0.002]) for the location of the lines (in red) is relatively accurate in most places. Figure \ref{fig:OverlayMapsdelta0p002}(\textit{f}) and movie 3 of the supplementary material shows the particle trajectories from the Eulerian particle-tracking method. The variety of trajectories is remarkable, with many particles having two sharp changes in direction during the spreading (see for example the three trajectories highlighted by red dashed ellipses). This likely reflects the fact that particle trajectories can be influenced by different deposits at different times.

Figure \ref{fig:DivCurldelta0p002} shows the divergence and curl of the map computed from Eul[0.002]. The fact that the Monge--Amp\`ere approximation is less accurate for small $\delta$ is reflected in the fact that $|\nabla_{\mathbf{x}_0}^2\phi|$ and $|\nabla_{\mathbf{x}_0}^2\psi|$ are shown to be the same order of magnitude in certain places within the initial configuration (\rmc{the ratio $\nabla^2_{x_0}\psi/\nabla^2_{x_0}\phi$ is plotted in figure S4 in \S{S4} of the supplementary material}). However, figures \ref{fig:DivCurldelta0p002}(\textit{c}) and \ref{fig:DivCurldelta0p002}(\textit{d}), plotted in the final configuration, once again show that, for particles that start within the deposits, it is still the case that \eqref{eq:mapassumption} holds.  Once again $\bm{\nabla}_{\mathbf{x}_0}\cdot (\mathbf{X}-\mathbf{x}_0)<0$ represents area elements with net compression, and $\bm{\nabla}_{\mathbf{x}_0}\cdot (\mathbf{X}-\mathbf{x}_0)>0$ represents area elements with net expansion by the map. Areas within the deposits expand, as do area elements connecting the largest deposit with the smaller deposits, and we also now see expansion for areas connecting the deposits with the boundaries. In figure \ref{fig:DivCurldelta0p002}(\textit{b}), the patterns created by $(\bm{\nabla}_{\mathbf{x}_0}\times (\mathbf{X}-\mathbf{x}_0))_{\perp}$ are similar to the case where $\delta=0.25$, although the modulus of twist is greater in magnitude. 

\subsubsection{Self-similarity of corner regions}

\begin{figure}
\begin{subfigure}{\textwidth}
    \centering\includegraphics[width=\textwidth]{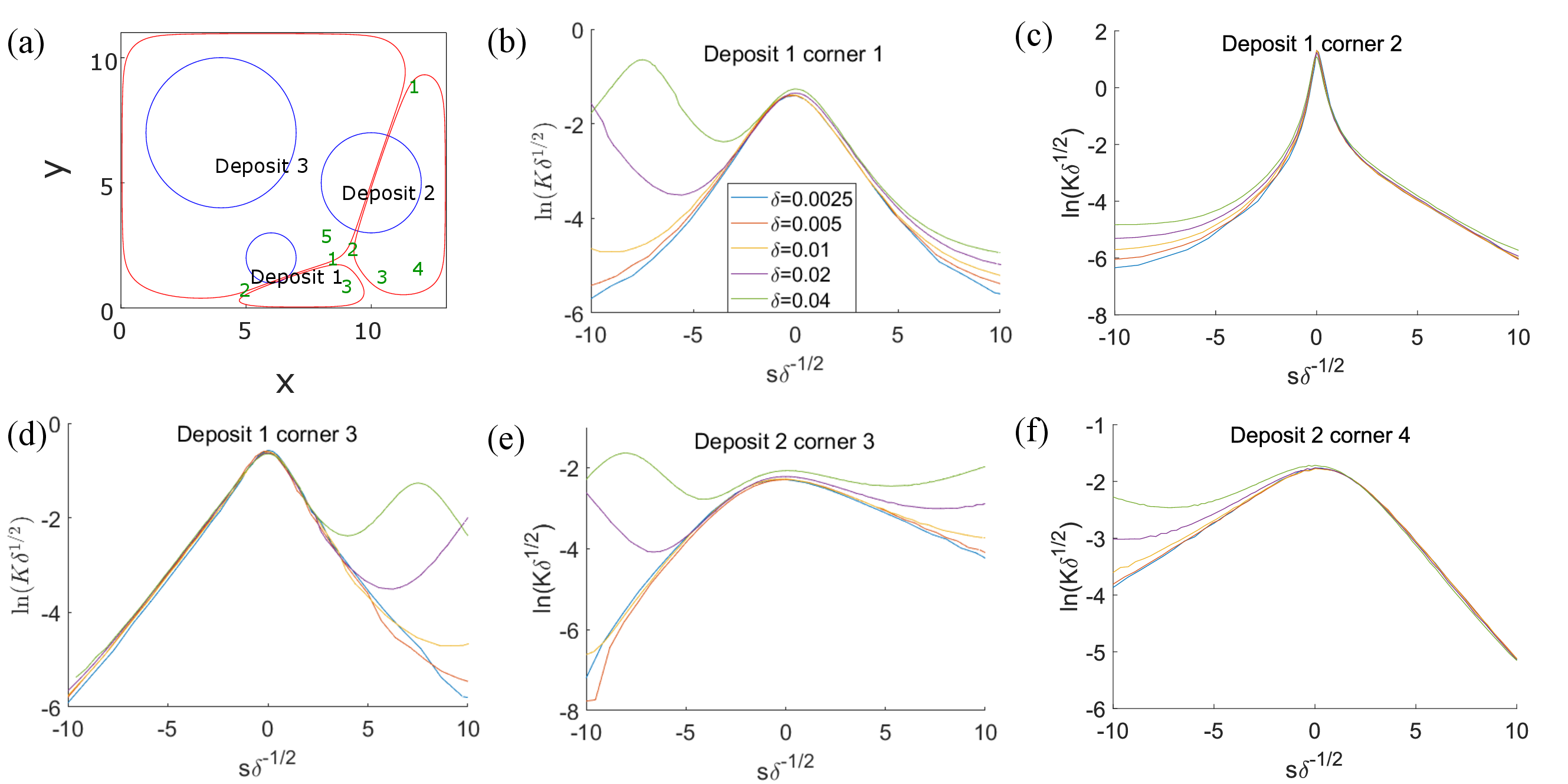}
    \end{subfigure}
    	\caption{ Scaled curvature plots for a selection of the corners of the boundaries of the three circular deposits at the steady state for small values of $\delta$. (\textit{a}) A graph of the solution from MA[0.04] that shows in green our numbering system for corners inside each deposit. (\textit{b})--(\textit{f}) Plots of the natural logarithm of the curvature scaled by $\delta^{1/2}$ against the arc length scaled by $\delta^{-1/2}$, where $s=0$ identifies the vertex of the corner in each case.}
\label{fig:ThreeBlobsCurvature}
\end{figure}

Computationally, the Monge--Amp\`ere solution is significantly cheaper to solve compared to either of the particle-tracking solutions. We take advantage of this to analyse the shape of the corners of deposits as $\delta \to 0$ using finely discretised Monge--Amp\`ere approximations. \jl{We note that the error found between the Eulerian and  Monge--Amp\`ere predictions are  small near the corner regions, even with $\delta=0.002$ (see figure S3 in the supplementary material). This suggests that predictions from the Monge--Amp\`ere solution remain accurate in the corner regions in the limit $\delta \to 0$.} Examples of these corner regions are numbered in green in figure \ref{fig:ThreeBlobsCurvature}(\textit{a}), which shows the results of MA[0.04]. To compute the deposit edges as sufficiently smooth curves to accurately analyse curvature of the edges was not possible from the particle-tracking solutions with available computational power, but is possible with the Monge--Amp\`ere method. During spreading, endogenous surfactant at initial concentration $\delta$, occupying an $O(1)$ area, is compressed into narrow threads (between pairs of drops) and into what resemble seven Plateau borders (at corners between drops, and at the domain boundary), at final concentration of order unity. Assuming the bulk of the endogenous surfactant is driven into these seven regions, and noting that each has an $O(1)$ aspect ratio, we can expect each drop corner to curve over a length scale of $O(\delta^{1/2})$. \oej{This scaling is motivated by the observation that an area with an O(1) length-scale and  with an initial concentration of endogenous surfactant $\delta$ (giving a total mass of $O(\delta)$) is ultimately compressed into an area with final concentration $\bar{\Gamma}\sim 1$ over an $O(\delta^{1/2})$ length-scale (preserving the total mass of $O(\delta)$).} 

Figure \ref{fig:ThreeBlobsCurvature} shows how the equilibrium shapes of the deposits calculated by Monge--Amp\`ere become more and more polygonal in the limit $\delta \to 0 $, with corner regions adopting a self-similar form.   For several values of small $\delta$, we calculate the curvature $K$ of the equilibrium boundary of each deposit as a function of arc length $s$. To show how each of a selection of corners sharpen as $\delta \to 0$, we set $s=0$ at the local curvature maximum and plot $\log{(\delta^{1/2}K)}$ against $s\delta^{1/2}$.     Figures \ref{fig:ThreeBlobsCurvature}(\textit{b}) to \ref{fig:ThreeBlobsCurvature}(\textit{f}) show collapse of the data, illustrating how, as $\delta \to 0$, the curvature of corners of the polygons (numbered in green in figure \ref{fig:ThreeBlobsCurvature}(\textit{a})) is proportional to $\delta^{-1/2}$, with this curvature occurring over arc-lengths proportional to $\delta^{1/2}$, and the edges of the deposits becoming effectively straight lines away from the corners. We see that in small regions away from $s=0$, the quantity $\ln{(\delta^{1/2}K)}$ becomes linear, indicating a functional dependence such as $\delta^{1/2}K \sim e^{-\lambda \delta^{-1/2}|s|}$ for some constant $\lambda$. In practice, the boundaries between the deposits have a small curvature in the Eulerian particle-tracking solution (figure \ref{fig:OverlayMapsdelta0p002}e), and so the deposits instead resemble slightly distorted polygons, with corner regions slightly distorted from the Monge--Amp\`ere prediction. \jl{As discussed above, the self-similarity observed for the internal corners and the appearance of a characteristic length scale of order $\delta^{1/2}$ is based on the fundamental physical principle of mass conservation. Since this principle is independent of the choice of method, we expect that the results observed using the Monge--Amp\`ere method will remain valid with the Eulerian and Lagrangian methods. This provides further evidence that the self-similarity behaviour observed in figure \ref{fig:ThreeBlobsCurvature} is physically grounded.}

\begin{figure}
    \centering\includegraphics[width=\textwidth]{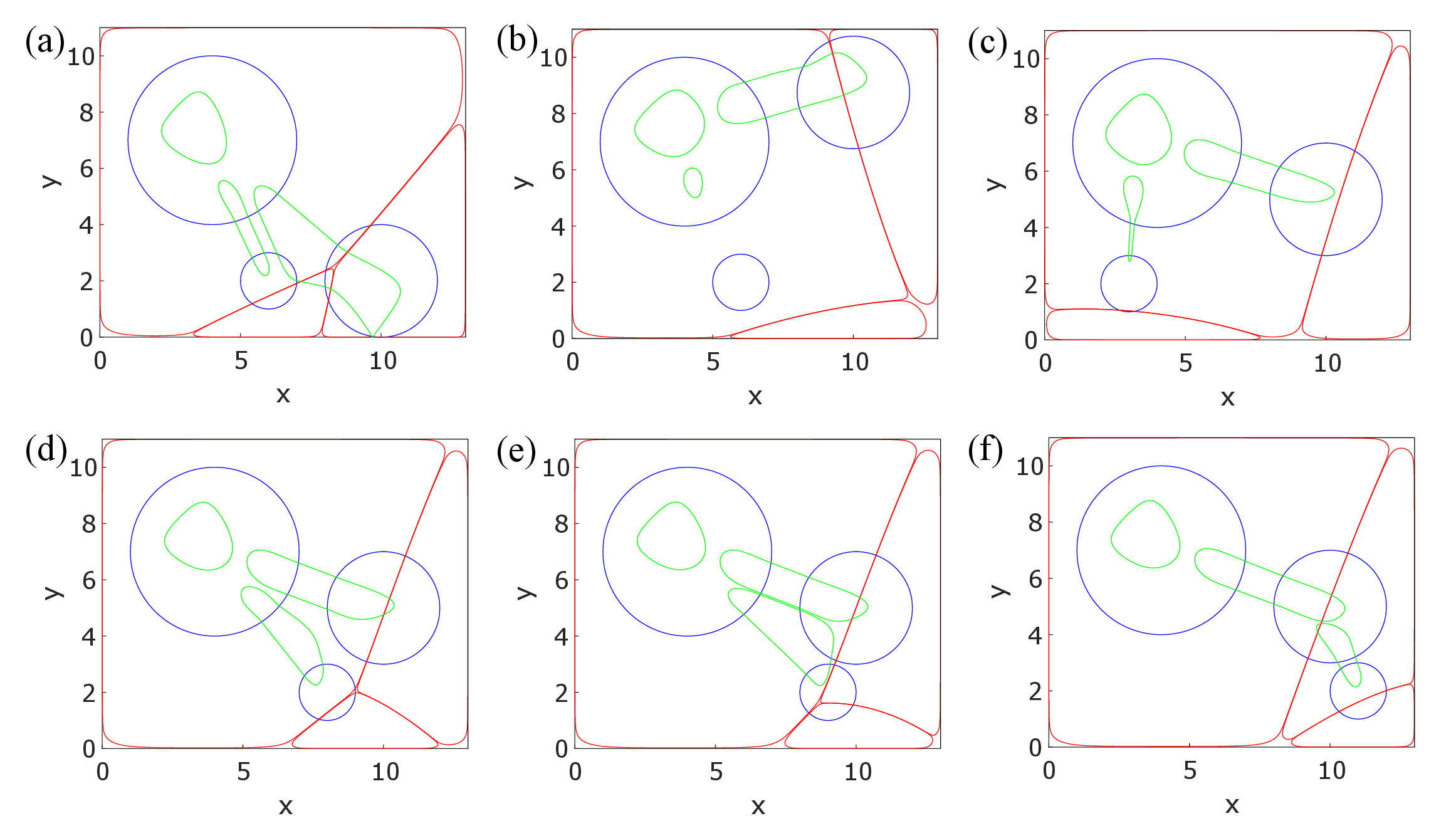}
    \caption{Figures showing the steady-state mapping and inverse mapping for $\delta=0.005$ with varying initial locations for some of the circular deposits 1 and 2 (deposit 3, as shown in figure~\ref{fig:ThreeBlobsCurvature}\textit{a}, remains fixed). Initial locations of the boundaries of the circular deposits are in blue, with the final locations of those boundaries in red. The green curves map to the blue circles under the same map. Panels (a) to (f) show the results for MA[0.005]Alt1 to MA[0.005]Alt6 from the key shown in table \ref{tab:SimulationSummary}.}
    \label{fig:MovingBlobs}
\end{figure}

Figure \ref{fig:MovingBlobs} shows the Monge--Amp\`ere approximation of the three-deposit problem with $\delta = 0.005$ (MA[0.005]Alt1-Alt6), using multiple different centres of the deposits (shown in the caption), revealing a variety of near-polygonal structures. Each deposit approaches either a triangle, a quadrilateral, or in the case of the largest deposit (figures \ref{fig:MovingBlobs}b, \ref{fig:MovingBlobs}d), a hexagon. 
The final configuration approaches a structure which can be characterised by up to four coordinates: the locations of corners shared by multiple deposits, which we call the characteristic coordinates of a given initial configuration. As in figure \ref{fig:ThreeBlobsCurvature}, we expect the Monge--Amp\`ere approximation to provide a good first-order estimate of equilibrium drop shape predicted by the full dynamic problem. However, at the present time we are unable to offer any simple strategy for determining the characteristic coordinates directly from the initial conditions.

\section{Discussion}
\label{sec:Discussion}

In this study we have evaluated the trajectories of \jl{passive} surface particles confined in a rectangular domain under the action of surfactant spreading on a thin film in a large-gravity \jl{and large-Péclet-number} limit using two separate methods (Eulerian particle tracking, outlined in \S\ref{sec:Eul} and Lagrangian particle tracking outlined in \S\ref{sec:MethodsLag}), the results of which corroborate each other. We have also identified a direct method (using the Monge--Amp\`ere equation outlined in \S\ref{sec:MethodsSteadyState}) for approximating the equilibrium configuration as $t\to \infty$. The solutions to this problem show how confinement and drop--drop interactions can lead to drift of surface particles (in addition to spreading) and how drop--boundary interactions lead to transient flow reversals. It is striking the way that some particles move a significant distance from their initial location, before moving back close to where they started, and that in figure \ref{fig:OverlayMapsdelta0p002}(\textit{f}) some particles sharply change directions twice, which is reminiscent of the multiple regimes identified for multiple surfactant sources noted by \cite{iasella2023interaction}. Predicting the equilibrium location of the interface between exogenous and endogenous surfactants is straightforward for spreading in one spatial dimension (relying on mass conservation arguments). \jl{However, in two-dimensions (as here) it becomes a non-trivial task, even for  simple scenarios with one or two initial deposits, particularly when symmetry is broken. With three initial deposits, new topological features appear in the form of internal corners between deposits. Nevertheless, equilibrium shapes possess asymptotic structures, such as the self-similar geometry of the internal corners in the limit of small initial endogenous surfactant concentration (figures \ref{fig:ThreeBlobsCurvature}, \ref{fig:MovingBlobs}). Such observations might offer a route to explicit predictions.}

\jl{In order to compute the particle trajectories involved in the surfactant-driven spreading, we constructed a finite-difference numerical method on a regular grid. We used an interpolated concentration gradient to determine the local velocity of the particles. The surfactant concentration gradients were  calculated using a finite-difference solution from a second grid.} In addition, we reformulated the Eulerian surfactant transport equation \eqref{eq:nonlineardiff} and the Lagrangian particle-transport equation \eqref{eq:particlevelocity} into a single Lagrangian vector equation \eqref{eq:KeyEqn2ndForm}, which describes the dynamics of material particles. This was achieved by choosing a Lagrangian coordinate system for which the surfactant concentration is initially uniform in the deformed Lagrangian domain (i.e. having uniform mass per unit Lagrangian area), and choosing this domain such that the initial conditions are simple to compute from equation \eqref{eq:DynInitConds}. Although the calculation of surfactant concentration is bypassed by this approach, the concentration can still be recovered at any time from the Jacobian via \eqref{eq:alphgam1}. 

\jl{From a computational point of view, the Eulerian and Lagrangian numerical methods used to compute the particle trajectories have each some advantages and drawbacks. They are similar in computational expense; however, the Eulerian method introduces additional errors at every time step by first  approximating $\bm{\nabla}_{\mathbf{x}}\Gamma$ on the Eulerian grid, and then by interpolating this quantity on to the current position of surface particles on the Lagrangian grid.} The Eulerian method also requires storage of two solutions (Eulerian surfactant concentration and Lagrangian particle position) on two separate grids, whereas the Lagrangian method only needs storage of a single solution. Some of the drawbacks of the Lagrangian method are that: the deformed domain needs to be calculated beforehand (although this only needs to be done once); the Lagrangian domain has an irregular shape which makes numerical resolution of \eqref{eq:KeyEqn2ndForm} difficult; and the domain can become  so deformed for small $\delta$ that numerical resolution of \eqref{eq:KeyEqn2ndForm} is not practical. For moderate $\delta$ however, the Lagrangian formulation is a mathematically elegant approach which finds the solution without needing to introduce interpolation errors, and the close agreement of the results between the two methods (figures \ref{fig:OverlayContourMapsdelta0p25} \jl{and S1}) provides corroboration of the particle-tracking method, lending support to our calculations of Eulerian solutions for small $\delta$.

\jl{Since equilibrium drop shapes, obtained in the limit of large time, can be computationally expensive to calculate with the Eulerian and Lagrangian methods, we have also explored how a Monge--Amp\`ere equation can be used to find approximations of the equilibrium configurations}. This was achieved by approximating the map between initial and final configuration as the gradient of a scalar potential, and \jl{neglecting} the rotational part of the map. The computational benefits of this approach are beyond dispute, as only a single, scalar, time-independent PDE needs to be solved. For example, the Monge--Amp\`ere solution shown in figure \ref{fig:OverlayContourMapsdelta0p25} can be computed for \jl{approximately $2.3\times 10^7$ gridpoints in a shorter time than a solution for approximately $3\times 10^5$ gridpoints using the Eulerian particle-tracking method from \S\ref{sec:LagNum} for the same parameters}. Beyond computational issues, the Monge--Amp\`ere method reveals a counter-intuitive feature of surfactant-spreading dynamics in two-dimensions: despite always being transported by an instantaneously irrotational flow \eqref{eq:velocityexp}, in the Eulerian sense, the mapping of particles from their initial to their current state can accumulate a weak rotational component \jl{with respect to Lagrangian coordinates}, as illustrated in figures \ref{fig:DivCurldelta0p25}(\textit{b},\textit{d}) and \ref{fig:DivCurldelta0p002}(\textit{b},\textit{d}). \jl{This small Lagrangian rotational component} can lead to weak distortions of final equilibrium configurations between the Monge--Amp\`ere predictions and the Eulerian and Lagrangian predictions (figure \ref{fig:OverlayContourMapsdelta0p25}). \rmc{In the supplementary material, we explored a variety of alternate exogenous configurations to analyze further the accuracy of the Monge--Amp\`ere predictions. These show how symmetry of the initial conditions reduces the median error between the Monge--Amp\`ere approximation and the Eulerian solution. In general, the error between the Monge--Amp\`ere approximation and the Eulerian solution reduces with larger initial endogenous surfactant concentration $\delta$ and with having fewer initial deposits. It is worth noting that when spreading is one-dimensional, then the Monge--Amp\`{e}re equation gives the exact solution. This is due to the fact that topologically, no rearrangement of the particles can be performed in one dimension without an energetic cost captured by the evolution equation. In contrast, in two dimensions, particles can re-arrange through a divergence-free stress field (such as blowing by a Suminagashi artist after reaching equilibrium), which has no energetic cost for the evolution equation (\ref{eq:nonlineardiff}).}

\jl{Furthermore,} the Monge--Amp\`ere formulation makes a connection between the theory of surfactant-driven transport and the theory of optimal transport. Monge--Amp\`ere equations arise for optimal transport problems where the transport satisfies the so-called quadratic Monge--Kantorovich optimal transport problem (qMK). If surfactant were transported optimally according to qMK, then the map $\mathbf{X}$ would satisfy
\begin{equation}\label{eq:qMK}
\min_{\mathbf{X}:\Gamma_0\to \bar{\Gamma}} \int_{\Omega}|\mathbf{X}-\mathbf{x}_0|^2\Gamma_0 \ \dee A_{\mathbf{x}_0},
\end{equation}
where $\min_{\mathbf{X}:\Gamma_0\to \bar{\Gamma}}$ means that we choose the minimum over all possible maps $\mathbf{X}$ which transport the initial surfactant concentration profile $\Gamma_0$ to the final uniform profile $\bar{\Gamma}$ (equivalently from all possible maps $\mathbf{X}$ which satisfy \eqref{eq:SSproblemwithoutphi}). It is known that such maps $\mathbf{X}$, which satisfy \eqref{eq:qMK}, are the gradients of (convex) scalar potentials which satisfy the Monge--Amp\`ere equation \citep{rockafellar1970convex, caffarelli2010free}. Unfortunately, the surfactant problem (\ref{eq:nonlineardiff}) does not satisfy \eqref{eq:qMK} precisely, as we can see by the discrepancies in figure \ref{fig:OverlayMapsdelta0p002}, for example, which shows a different prediction for the final configuration compared to the Eulerian particle-tracking method. However, \cite{otto2001geometry} showed that solutions to equations of porous medium type (a class of equations to which \eqref{eq:nonlineardiff} belongs) are gradient flows on the function space  $\mathcal{M}$ of possible solutions; this function space can be shown to satisfy the definition of a Riemannian manifold when distances on the manifold are measured by the Wasserstein-2 distance, defined variationally as
\begin{equation}\label{eq:W2} W_2(\Gamma_0,\bar{\Gamma}) = \sqrt{
\min_{\mathbf{X}:\Gamma_0\to \bar{\Gamma}} \int_{\Omega}|\mathbf{X}-\mathbf{x}_0|^2\Gamma_0 \ \dee A_{\mathbf{x}_0} }.
\end{equation}
The Wasserstein-2 distance \eqref{eq:W2} is simply the square root of qMK \eqref{eq:qMK}. \jl{Hence}, optimal maps which satisfy \eqref{eq:qMK} must follow the shortest path between $\Gamma_0$ and $\bar{\Gamma}$ on $\mathcal{M}$ (a geodesic). The question of how closely the Monge--Amp\`ere method approximates the correct steady-state solution for surfactant spreading is therefore equivalent to the question of how far the gradient flow deviates from the geodesic on $\mathcal{M}$. This leads to the possibility that a quantification of how well the Monge--Amp\`ere method approximates the correct solution could be answered with variational analysis.  We do not pursue this further here, \rmc{except to recall the degeneracy in the evolution equation revealed in Appendix A, namely that (\ref{eq:nonlineardiff}) does not account for the dissipation associated with any flow that preserves surface concentration: this raises the possibility that such flows may account for differences between equilibria predicted by the Monge--Amp\`ere and Eulerian descriptions.} Similarly, we leave as open the question of whether the solutions as $\delta\to 0$ can be used to infer the behaviour of the final state with an initially clean interface ($\delta=0$). 

\jl{In the limit of vanishing endogenous surfactant concentration $\delta\to 0$, using the Monge--Amp\`ere approximation, we find that exogenous deposits approach equilibrium shapes with polygonal boundaries (figure \ref{fig:MovingBlobs}). Moreover, the internal corners between deposits present self-similar geometries, with a typical spatial extent scaling as $\delta^{1/2}$  (figure \ref{fig:ThreeBlobsCurvature}). This self-similar behaviour can be explained by mass conservation.} 
\jl{In the limit of $\delta \to 0$, we also observe that the topology at equilibrium is reduced to a small finite number of characteristic coordinates which are the intersections between the polygonal final shapes of the deposits. Predicting these coordinates \textit{a priori} remains an open problem. }

\begin{figure}
    \centering
    \begin{subfigure}{\textwidth}
\includegraphics[width=\textwidth]{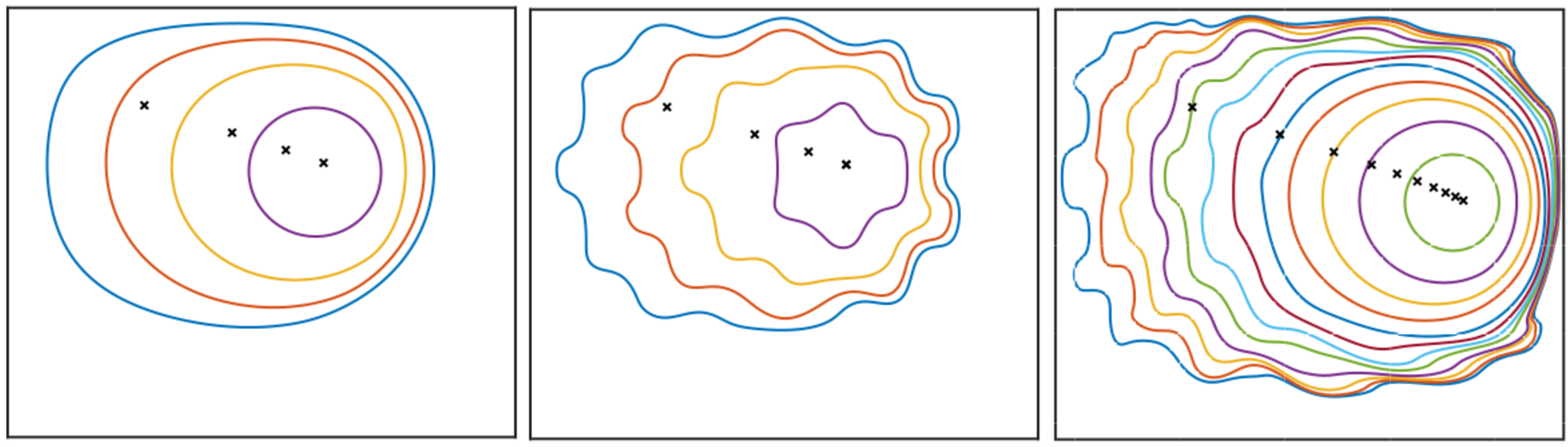}
\end{subfigure}
\begin{subfigure}{\textwidth}
\includegraphics[width=\textwidth]{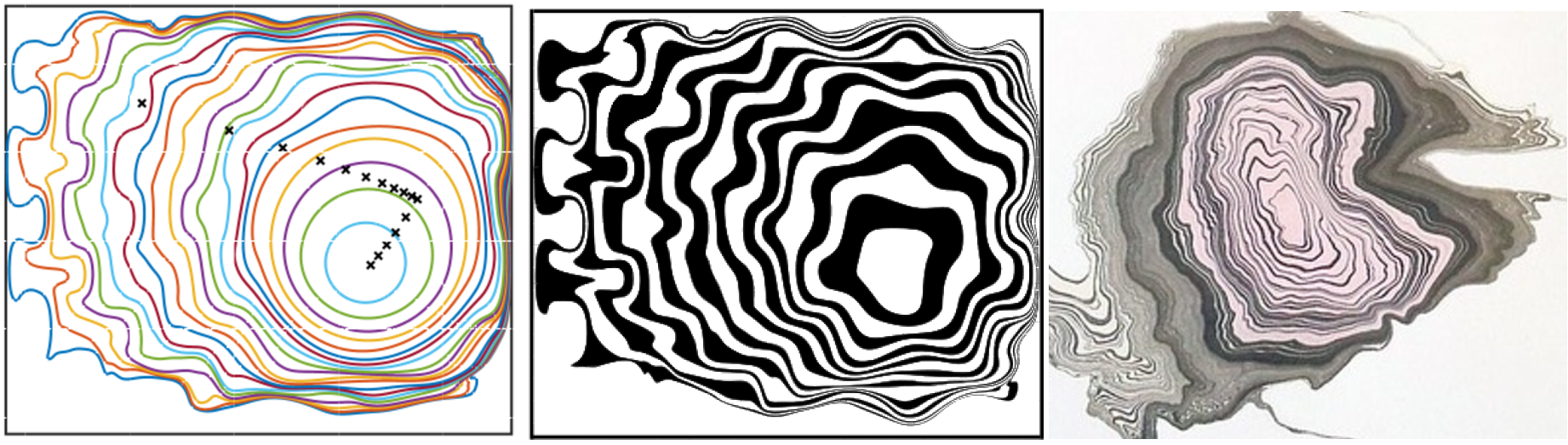}
\end{subfigure}
    \caption{\jl{Successive solutions of the Monge--Amp\`ere equation are shown in the first five panels, as detailed in Appendix \ref{sec:appSumiMethod}, showing a final pattern which is reminiscent of a Suminagashi pattern (final panel) (see also movie 4 of the supplementary material)}. The image was created by 20 solutions of the Monge--Amp\`ere equation with a stirring or blowing step after the release of every four deposits. The top images show the creation of the pattern after $4$ depositions, after the first stirring step, and after $12$ depositions, respectively. The bottom three images show the solution just before the final stirring step after 20 depositions, the final result where a monochrome colour scheme is added, and a Suminagashi pattern made by Bea \cite{beamehan} for qualitative comparison. }
    \label{fig:SuminagashiAlt}
\end{figure} 

\jl{In figure \ref{fig:SuminagashiAlt} and movie 4 of the supplementary material, we explore how the Monge--Amp\`ere method could be used to create a pattern resembling Suminagashi art. We have used divergence-free maps to mimic the blowing process used by Suminagashi artists (the details for the creation of this picture are in Appendix \ref{sec:appSumiMethod}).} Such a computation, which involves 20 consecutive calculations of the Monge--Amp\`ere equation, would be extremely expensive to run using a full dynamic solution. The choice of a divergence-free map found by time-stepping \eqref{eq:blowingtimestep} for the blowing step used by artists is credible, as gentle blowing on a surfactant-laden surface would likely deform the surface in such a way as to not create concentration gradients. \jl{Indeed, at equilibrium, the surfactants prevents the formation of concentration gradients \citep{manikantan2020surfactant}, because the Marangoni force opposes them (Appendix \ref{sec:appnonldiff})}.  \oej{Moreover, we exploited the fact that gravitational forces in the Suminagashi process are sufficiently strong to suppress deformations of the gas--liquid interface, and that surface diffusion is sufficiently weak compared to advection to avoid smearing of the edge deposits.  In other contexts, Marangoni flows can induce surface deformations.  Despite the fact that the coupled evolution equations between the bulk and the surface for such problems can be written in a gradient-flow formulation \citep{thiele2018}, the link between such flows and optimal transport is not clear; the bulk flow that transports fluid is distinct from the the surface flow that transports surfactant, such that distinct maps may be needed to characterize such problems.} 
\jl{The numerical results from the Monge--Amp\`ere method presented in figure \ref{fig:SuminagashiAlt} are only suggestive in their reproduction of Suminagashi art, nevertheless we hope they will encourage experimentalists to explore this link further.}

Many applications involving surfactants are concerned with the transport of passive solutes by the Marangoni effect in confined geometries, ranging from pharmaceutical delivery to the human lung to the creation of artistic patterns using Suminagashi techniques. We have \jl{presented} methods for finding the dynamic behaviour of material surface particles, and for finding an approximation of the equilibrium location of material particles without having to solve for transient dynamics. The equilibrium approximation was achieved by showing a connection between surfactant dynamics and the theory of optimal transport, a research area at the forefront of modern mathematics, through a Monge--Amp\`ere equation associated with the surfactant-driven transport. We hope this connection, and the methods presented here, will spark the imaginations of researchers interested in both fundamental understanding and practical work related to  \jl{surfactant-induced Marangoni flows carrying solutes}.

\begin{acknowledgments}

JRL and OEJ acknowledge financial support from EPSRC grant EP/T030739/1.  For the purpose of Open Access, the authors have applied a Creative Commons Attribution (CC BY) licence to any Author Accepted Manuscript version arising. \rmc{The authors thank all anonymous reviewers for constructive comments, and RM wishes to thank Thomas Bickel and Matthias Heil for constructive discussions.}

\end{acknowledgments}

\appendix

\section{Derivation of the nonlinear diffusion equation \eqref{eq:nonlineardiff}}
\label{sec:appnonldiff}

We consider a liquid layer of density $\rho^*$, viscosity $\mu^*$ and uniform thickness $h^*$, sitting on the horizontal plane $z^*=0$, subject to a restoring force (provided, for example, by a strong vertical gravitational field $g^*$) that suppresses out-of-plane deflections of the gas/liquid interface at $z^*=h^*$ (where stars denote dimensional quantities). We therefore disregard the normal stress condition in the equations below, but provide a condition on the relevant Bond number that permits this approximation.  An insoluble surfactant with concentration $\Gamma^*(\mathbf{x}^*_\parallel,t^*)$ occupies the interface, where $\mathbf{x}^*_\parallel$ denotes horizontal or in-plane coordinates such that $\mathbf{x}^*_\parallel\equiv(x^*,y^*)$.  The surfactant lowers surface tension via a linear equation of state, so that in-plane surface tension gradients are $-\mathcal{A}^*\bm{\nabla}^*_{\parallel} \Gamma^*$, with $\mathcal{A}^*>0$ \rmc{being} the surface activity of the surfactant \citep{manikantan2020surfactant}, $\bm{\nabla}^*_\parallel\equiv \hat{\mathbf{x}} \partial^*_x +\hat{\mathbf{y}} \partial^*_y$, and  where $(\hat{\mathbf{x}},\hat{\mathbf{y}},\hat{\mathbf{z}})$ are unit vectors in the three Cartesian coordinate directions.  Adopting lubrication theory, we assumed that the horizontal velocity field $\mathbf{u}^*_\parallel(\mathbf{x}^*_\parallel, z^*,t^*)$ in the liquid layer satisfies Stokes equation $\mu^* \mathbf{u}^*_{\parallel,z^*z^*}=\bm{\nabla}^*_\parallel p^*$ in $0\leq z^*\leq h^*$, where $p^*(\mathbf{x}^*_\parallel,t^*)$ is the leading-order pressure field.  The horizontal volume flux, $\mathbf{q}^*(\mathbf{x}^*_\parallel,t^*)=\int_0^{h^*}\mathbf{u}^*_\parallel\,\mathrm{d}z^*$ satisfies $\bm{\nabla}^*_\parallel\cdot \mathbf{q}^*=0$, in order to maintain uniformity of the layer thickness.  
Integrating the momentum equation, applying a no-slip boundary condition
$\mathbf{u}^*_\parallel=\mathbf{0}$ on $z^*=0$, and the tangential stress condition
\begin{equation}\label{eq:viscousMarangonistressbalance}
    \mu^* \mathbf{u}^*_{\parallel,z^*}=-\mathcal{A}^*\bm{\nabla}^*_\parallel \Gamma^* +\boldsymbol{\tau}^*\quad\mathrm{at}\quad z^*=h^*,
\end{equation}
where $\boldsymbol{\tau}^*(\mathbf{x}^*_\parallel,t^*)$ is an imposed shear stress from the gas phase, 
we obtain $\mu^*\mathbf{u}^*_\parallel =-\bm{\nabla}^*_\parallel p^* z^*(h^*-z^*/2)+(\boldsymbol{\tau}^*-\mathcal{A}^*\bm{\nabla}^*_\parallel \Gamma^*)z^*$, yielding the following expressions for the flux and surface velocity:
\begin{subequations}
\begin{align}
    \mu^*\mathbf{q}^* &=-\tfrac{1}{3} h^{^*3} \bm{\nabla}^*_\parallel p^* +\tfrac{1}{2}h^{^*2} (\boldsymbol{\tau}^*-\mathcal{A}^*\bm{\nabla}^*_\parallel \Gamma^*);\\
     \mu^* \mathbf{u}^*_s&=-\tfrac{1}{2}h^{^*2} \bm{\nabla}^*_\parallel p^* +h^*(\boldsymbol{\tau}^*-\mathcal{A}^*\bm{\nabla}^*_\parallel\Gamma^*).
\end{align}
\label{eq:qu}
\end{subequations}
Material particles at the interface are transported via $\mathrm{d}\mathbf{x}^*_\parallel/\mathrm{d}t^*=\mathbf{u}^*_s(\mathbf{x}^*_\parallel, t^*)$.

Using Helmholtz decomposition, we may write $\boldsymbol{\tau}^*=\bm{\nabla}^*_\parallel\varphi^*+\bm{\nabla}^*_\parallel\times (\kappa^*\hat{\mathbf{z}})$ for some scalar potentials $\varphi^*(\mathbf{x}^*_\parallel, t^*)$ and $\kappa^*(\mathbf{x}^*_\parallel,t^*)$, where $\bm{\nabla}^*_\parallel \cdot \boldsymbol{\tau}^*=\bm{\nabla}^{*2}_\parallel \varphi^*$ and $\bm{\nabla}^*_\parallel\times\boldsymbol{\tau}^*=\bm{\nabla}^{*2}_\parallel \kappa^*$.  Then the mass conservation constraint $\bm{\nabla}^*_\parallel\cdot\mathbf{q}^*=0$ implies
\begin{equation}
    0=\bm{\nabla}^{*2}_\parallel\left[ -\tfrac{1}{3}h^{*3}  p^{*} +\tfrac{1}{2}h^{*2} ( \varphi^{*}- \mathcal{A}^{*} \Gamma^{*})\right].
    \label{eq:d2}
\end{equation}
We impose no-flux (Neumann) conditions on the pressure, surfactant concentration and stress potential $\varphi^*$ at the periphery of the domain, so that on the boundary $\mathbf{n}_b\cdot \bm{\nabla}^{*}_\parallel p^{*}=\mathbf{n}_b\cdot \bm{\nabla}^{*}_\parallel \varphi^{*} =\mathbf{n}_b\cdot \bm{\nabla}^{*}_\parallel \Gamma^{*}=0$ where $\mathbf{n}_b$ is the unit outward normal.  
Integrating (\ref{eq:d2}), the pressure gradient satisfies
\begin{equation}
    h^{*}\bm{\nabla}^{*}_\parallel p^{*}=\tfrac{3}{2}\bm{\nabla}^{*}_\parallel\left(  \varphi^{*} - \mathcal{A}^{*} \Gamma^{*}\right ).
    \label{eq:pg}
\end{equation}
Thus, the surface velocity field (\ref{eq:qu}b) becomes
\begin{equation}
    \mu^{*} \mathbf{u}^{*}_s=
    \tfrac{1}{4}h^{*}(\bm{\nabla}^{*}_\parallel \varphi^{*} - \mathcal{A}^{*}\bm{\nabla}^{*}_\parallel \Gamma^{*}) + h^{*} \bm{\nabla}^{*}_\parallel \times (\kappa^{*} \hat{\mathbf{z}}),
    \label{eq:us}
\end{equation}
which can be inserted into the surfactant transport equation \begin{equation}
    \Gamma^{*}_t+\bm{\nabla}^{*}_\parallel \cdot (\mathbf{u}^{*}_s \Gamma^{*})=0
\label{eq:strans}
\end{equation} 
(neglecting surface diffusion).  We highlight two special cases of the resulting evolution equation for the surfactant concentration. 

First, in the absence of an imposed shear stress ($\boldsymbol{\tau}^*=0$), (\ref{eq:us}) and (\ref{eq:strans}) yield a generalisation to two dimensions of the nonlinear diffusion equation derived in \cite{jensen1998stress1},
\begin{equation}\label{eq:dimsurfactanttransport}
    \Gamma^{*}_t=\frac{\mathcal{A}^{*}h^{*}}{4\mu^{*}}\bm{\nabla}^{*}_\parallel \cdot (\Gamma^{*}\bm{\nabla}^{*}_\parallel \Gamma^{*}).
\end{equation}
As assumed previously, the condition for the interface to remain flat is that horizontal pressure gradients generated by interfacial deflections (through gravity forces for instance) dominate those arising from Marangoni effects in (\ref{eq:pg}), namely that the Bond number is large
\begin{equation}
1 \ll \frac{\rho^{*} g^{*} h^{*2}}{\mathcal{A}^{*} \Delta \Gamma^{*}},   
\label{eq:bond}
\end{equation}
where $\Delta \Gamma^{*}$ represent characteristic surfactant concentration differences driving the flow. \oej{To neglect the effects of surface diffusivity $D^*$ from (\ref{eq:dimsurfactanttransport}), the surface P\'eclet number should be large, or
\begin{equation}
\frac{D^* \mu^*}{\mathcal{A}^*\Delta \Gamma^* h^*}\ll 1.
\label{eq:surdif}
\end{equation}
}

Second, retaining the imposed shear stress $\boldsymbol{\tau}^*$ but decomposing the surfactant concentration such that $\Gamma^{*}=\overline{\Gamma}^{*}+\hat{\Gamma}^{*}(\mathbf{x}^{*}_\parallel,t^{*})$ and linearising about a uniform state $\overline{\Gamma}^{*}>0$ (assuming $\vert \hat{\Gamma}^{*}\vert \ll \overline{\Gamma}^{*}$), (\ref{eq:us}) and (\ref{eq:strans}) become
\begin{equation}
    \hat{\Gamma}^{*}_t- \frac{\mathcal{A}^{*}h^{*}\overline{\Gamma}^{*}}{4\mu^{*}}\bm{\nabla}^{*2}_\parallel \hat{\Gamma}^{*} = - \frac{h^{*}\overline{\Gamma}^{*}}{4\mu^{*}}\bm{\nabla}^{*}_\parallel \cdot \boldsymbol{\tau}^{*}.
    \label{eq:lin}
\end{equation}
The rotational component of the shear stress (associated with $\kappa^{*}$) moves material particles (via (\ref{eq:qu}b)) but does not change surface concentration in (\ref{eq:lin}) (in this linear approximation).  Thus, based on \eqref{eq:us}, there is unrestricted transport of material elements under 
\begin{equation}\label{eq:rotationalvelocity}
    \frac{\mathrm{d}\mathbf{x}^{*}_\parallel}{\mathrm{d}t^{*}}=\frac{h^{*}}{\mu^{*}} \left[\bm{\nabla}^{*}_\parallel \times (\kappa^{*} \hat{\mathbf{z}})\right]\Big\vert_{\mathbf{x}^{*}_\parallel},
\end{equation}
a feature that we will exploit to create Suminagashi patterns in Appendix \ref{sec:appSumiMethod}. In contrast, shear stress with non-zero divergence acts as a forcing to the transport equation governing surfactant concentration in (\ref{eq:lin}), which responds diffusively in the linear approximation.   

The governing non-dimensional surfactant transport equation \eqref{eq:nonlineardiff} of the problems studied in this paper (see \S\ref{sec:ModelandMethods}) is based on the dimensional equation \eqref{eq:strans}. Considering the initial condition illustrated in figure \ref{fig:Schematic}, we non-dimensionalize \eqref{eq:strans} using a characteristic horizontal length scale $r_1^*$, representing the initial radius of a deposit of exogenous surfactant. We impose that the liquid height $h^*$ is small compared to this length scale, with the ratio given as the small parameter $\epsilon = h^*/r_1^* \ll 1$. We impose that the ratio of horizontal lengths $L_1^*/L_2^*$ are $O(1)$ with respect to $\epsilon$ (this is to ensure that the ratio of length scales do not affect the asymptotics). Surface tension $\gamma^*=\gamma_0^*-\mathcal{A}^* \Gamma^*$ is non-dimensionalised by $\gamma = (\gamma^*-\gamma_c^*)/S^*$, where $S^* = \gamma_0^*-\gamma_c^*= \mathcal{A}^* \Gamma_c^*$, where $\gamma_c^*$ is the surface tension when $\Gamma^*=\Gamma^*_c$, a characteristic  concentration used to non-dimensionalise all surfactant concentrations, and $\gamma_0^*$ is the nominal surface tension when $\Gamma^*=0$. The surface velocity scale is obtained from the viscous--Marangoni stress balance in the dynamic boundary condition at the liquid surface \eqref{eq:viscousMarangonistressbalance}, which gives characteristic surface velocity $\epsilon S^*/\mu^*$. The scale for the vertical velocity component $w^*$ is found from non-dimensionalising the continuity equation $\bm{\nabla}\cdot \mathbf{u}$ ($\bm{\nabla}$ and $\mathbf{u}$ are the full three-dimensional nabla operator and velocity field) by the horizontal velocity and length scales, yielding $\epsilon^2 S^*/\mu^*$. The time scale is found by combining a horizontal velocity scale with the length scale $r_1^*$, yielding $\mu^*r_1^*/(\epsilon S^*)$. The pressure scale we use is $S^*/(\epsilon r_1^*)$. In summary, we relate unstarred dimensionless variables to starred dimensional variables by
\begin{equation*}
    x = \frac{x^*}{r_1^*}, \qquad
    y = \frac{y^*}{r_1^*}, \qquad
    z = \frac{z^*}{\epsilon r_1^*}, 
    \qquad 
L_i = \frac{L_i^*}{r_1^*} \quad \text{for } i=1,2,
\end{equation*}
\begin{equation*}
  u = \frac{u^* \mu^* }{\epsilon S^*},
  \qquad v = \frac{v^* \mu^* }{\epsilon S^*},
  \qquad w = \frac{w^* \mu^* }{\epsilon^2 S^*},
  \qquad 
  \gamma = \frac{\gamma^*-\gamma_c^*}{S^*},
\end{equation*}
\begin{equation} \label{eq:scales}
t = \frac{\epsilon t^* S^*}{\mu^*r_1^*}\qquad
  \Gamma = \frac{\Gamma^*}{\Gamma_c^*},
  \qquad 
  \delta = \frac{\delta^*}{\Gamma_c^*}, \qquad
  \bar{\Gamma} = \frac{\bar{\Gamma}^*}{\Gamma_c^*},
  \qquad
  p = \frac{p^* \epsilon r_1^*}{S^*}.
\end{equation}
Using the scales above, we can non-dimensionalize the governing surfactant transport equation \eqref{eq:dimsurfactanttransport} to obtain equation \eqref{eq:nonlineardiff}.

\section{Late time correction for the mapping $\mathbf{X}$}\label{sec:AppX_cr}

We run the Eulerian and Lagrangian methods for simulating particle trajectories from $t=0$ to some final time $t_f$, where $t_f$ needs to be chosen such that it gives an accurate approximation of the steady state within a certain tolerance. To find such $t_f$, we analyse the error between the mapping calculated numerically at $t_f$, and an estimate of the mapping computed theoretically using a linearised version of the governing equation as $t \to \infty$. At late time the nonlinear diffusion equation \eqref{eq:nonlineardiff} can be approximated by the linear diffusion equation
\begin{equation}\label{eq:diffeqn}
	\Gamma_t = \frac{\bar{\Gamma}}{4} \nabla_{\mathbf{x}}^2\Gamma,
\end{equation} 
with the no-flux boundary condition \eqref{eq:EulBoundConds} and  where $\bar{\Gamma}$ is defined in \eqref{eq:GammaBar}.
Using separation of variables, the solution of \eqref{eq:diffeqn} can be found analytically as the double series
\begin{equation}
	\Gamma(x,y,t) =  \sum_{m=0}^{\infty} \sum_{n=0}^{\infty} \sigma_{mn} \cos{\left(\frac{m \pi x}{L_1}\right)} \cos{\left(\frac{n \pi y}{L_2}\right)} e^{-\omega_{mn}(t-t_f)},
\end{equation}
for a series of constants $\sigma_{mn}$ and where
\begin{equation}\label{eq:omega_mn}
 \omega_{mn} = \frac{\bar{\Gamma}}{4} \left(\frac{m^2\pi^2}{L_1^2} + \frac{n^2\pi^2}{L_2^2}\right).
\end{equation}
Thus, the surface velocity field is given by
\begin{equation}
	\mathbf{u}_s = - \frac{1}{4}\bm{\nabla}_{\mathbf{x}}\Gamma = \begin{pmatrix}
		\sum_{m=0}^{\infty} \sum_{n=0}^{\infty} \frac{m \pi \sigma_{mn}}{4L_1} \sin{(m\pi x/L_1)}\cos{(n \pi y/L_2)}e^{-\omega_{mn}(t-t_f)}
		\\
		 \sum_{m=0}^{\infty}\sum_{n=0}^{\infty}\frac{n \pi \sigma_{mn}}{4L_2} \cos{(m\pi x/L_1)}\sin{(n \pi y/L_2)}e^{-\omega_{mn}(t-t_f)}
	\end{pmatrix}.
\end{equation}
We integrate this expression from $t=t_f$ to $t\to \infty$ giving a linear correction for the map of particle trajectories
\begin{equation}\label{eq:X_cr}
	\mathbf{X}_{cr} = \begin{pmatrix}
		\sum_{m=1}^{\infty}\sum_{n=0}^{\infty}\frac{m \pi \sigma_{mn}}{4L_1 \omega_{mn}} \sin{(m\pi x/L_1)}\cos{(n \pi y/L_2)}
		\\
		\sum_{m=0}^{\infty} \sum_{n=1}^{\infty} \frac{n \pi \sigma_{mn}}{4L_2 \omega_{mn}} \cos{(m\pi x/L_1)}\sin{(n \pi y/L_2)}
	\end{pmatrix}.
\end{equation}
Hence, the equilibrium mapping between $t=0$ and $t\to \infty$ can be approximated by $\mathbf{X} \approx \mathbf{X}_{t_f}+\mathbf{X}_{cr}$, where $\mathbf{X}_{t_f}$ is the mapping solution from $t=0$ to a large time $t_f$, and where the coefficients $\sigma_{mn}$ are given by
\begin{equation}\label{eq:sigma_mn}
	\sigma_{mn} = 
	\begin{cases}
		\frac{1}{L_1L_2}\int_{x=0}^{L_1}\int_{y=0}^{L_2}  \Gamma(x,y,t_f) \dy \dx = \bar{\Gamma}  \qquad \text{for } m,n = 0,
		\\
		\frac{2}{L_1L_2}\int_{x=0}^{L_1}\int_{y=0}^{L_2} \cos{\left(\frac{n \pi y}{L_2}\right)} \Gamma(x,y,t_f) \dy \dx  \qquad \text{for }  n>0, m=0, 
		\\
		\frac{2}{L_1L_2}\int_{x=0}^{L_1}\int_{y=0}^{L_2} \cos{\left(\frac{m \pi x}{L_1}\right)}  \Gamma(x,y,t_f) \dy \dx  \qquad \text{for } m>0, n=0,
		\\
		\frac{4}{L_1L_2}\int_{x=0}^{L_1}\int_{y=0}^{L_2} \cos{\left(\frac{m \pi x}{L_1}\right)} \cos{\left(\frac{n \pi y}{L_2}\right)} \Gamma(x,y,t_f) \dy \dx  \qquad \text{for } m,n>0.
	\end{cases}
\end{equation}
It is relevant to our arguments for using the approximation \eqref{eq:mapassumption} to point out that the late time mapping approximation \eqref{eq:X_cr} is exactly the gradient of the scalar potential
\begin{equation}\label{eq:phi_cr}
    \phi_{cr} = \left(\sum_{m=0}^{\infty}\sum_{n=0}^{\infty}\frac{ \sigma_{mn}}{4 \omega_{mn}} \cos{(m\pi x/L_1)}\cos{(n \pi y/L_2)} 
    \right) - \frac{ \sigma_{00}}{4 \omega_{00}}. 
\end{equation}

At a large time $t=t_f$, we expect the three leading order terms for $\Gamma(x,y,t_f)$ to be
\begin{equation}\label{eq:concapprox}
	\Gamma(x,y,t_f) \approx \bar{\Gamma} + \sigma_{01}\cos{\left(\frac{\pi y}{L_2}\right)} + \sigma_{10}\cos{\left(\frac{\pi x}{L_1}\right)},
\end{equation}
as higher order modes are subject to exponential decay over a smaller time-scale through \eqref{eq:omega_mn}.  (Taking $t_f=1047.8$ with $\delta=0.25$, for example, we found that the next largest coefficient $\sigma_{mn}$ was a factor of $O(10^{-3})$ smaller than either $\sigma_{01}$ or $\sigma_{10}$, justifying \eqref{eq:concapprox}.)  If we call $\Gamma_a$ the concentration at $(0,L_2)$ and $\Gamma_b$ the concentration at $(L_1,L_2)$, then 
\begin{equation}
	\Gamma_a \approx \bar{\Gamma} + \sigma_{10} - \sigma_{01}, \qquad \Gamma_b \approx \bar{\Gamma} - \sigma_{10} - \sigma_{01}  ,
\end{equation}
which means
\begin{equation}
	\sigma_{10} \approx \frac{1}{2}(\Gamma_a-\Gamma_b), \qquad \sigma_{01} \approx \frac{1}{2}(2\bar{\Gamma}-\Gamma_a-\Gamma_b) .
\end{equation}
The leading-order terms for the correction to the mapping $\mathbf{X}_{cr}$ which maps particles from $t_f$ to $t\to \infty$ can therefore be approximated as
\begin{equation}
	\mathbf{X}_{cr} \approx \begin{pmatrix}
		\frac{ L_1(\Gamma_a-\Gamma_b)}{2\pi \bar{\Gamma}} \sin{(\pi x/L_1)}
		\\
		\frac{ L_2(2\bar{\Gamma}-\Gamma_a-\Gamma_b)}{2\pi \bar{\Gamma}} \sin{( \pi y/L_2)}
	\end{pmatrix}.
\end{equation}
We run a given Eulerian particle-tracking simulation to a time $t_f$ where the leading order amplitude of the correction is below a pair of chosen tolerance levels $[X_{tol},Y_{tol}]$, or in other words until
\begin{equation}\label{eq:XYtol}
\text{abs}\left|	\frac{ L_1(\Gamma_a-\Gamma_b)}{2\pi \bar{\Gamma}} \right| < X_{tol}, \qquad
	\text{abs}\left|	\frac{ L_2(2\bar{\Gamma}-\Gamma_a-\Gamma_b)}{2\pi \bar{\Gamma}} \right| <Y_{tol}.
\end{equation}

\section{Behaviour of solutions to the nonlinear diffusion equation \eqref{eq:nonlineardiff} near sharp corners}\label{sec:Appsharpcorner}

We wish to understand the behaviour of the two-dimensional nonlinear diffusion equation \eqref{eq:nonlineardiff} subject to boundary conditions \eqref{eq:EulBoundConds}, near a sharp corner of a wedge-shaped domain. We introduce a polar coordinate system $(r,\theta)$ with the origin at the corner, with one boundary located at $\theta=0$, and the other at $\theta=\Phi$. The surfactant concentration $\Gamma$ must neither diverge, nor go to zero at the  corner, and so we look for expansions of the form
\begin{equation}
	\Gamma_{(m)} = A_{m,0}(\theta,t) + A_{m,1}(\theta,t)r^{a_{m,1}}+
	A_{m,2}(\theta,t)r^{a_{m,2}}+\dots,
\end{equation}
and where $0<a_{m,1}<a_{m,2}<\dots$, and $m$ is an index as we anticipate multiple expansions\rmc{, each indexed by a different value of $m$. The result will be a sum of asymptotic series, a technique used to derive a corner expansion for surfactant-induced flow in \cite{mcnair2022surfactant}.} Substituting an expansion \rmc{for a given $m$} into \eqref{eq:nonlineardiff} we obtain
\begin{multline}\label{eq:expansion}
4\left(	\frac{\partial A_{m,0}}{\partial t} + r^{a_{m,1}}\frac{\partial A_{{m,1}}}{\partial t} + r^{a_{m,2}}\frac{\partial A_{{m,2}}}{\partial t}+ \dots \right) 
\\
= \frac{1}{2r^2}\frac{\partial^2 A_{m,0}^2}{\partial \theta^2} + a_{m,1}^2A_{m,0}A_{{m,1}}r^{a_{m,1}-2} + r^{a_{m,1}-2}\frac{\partial^2}{\partial \theta^2} (A_{m,0} A_{{m,1}}) +
	\\
	\frac{1}{2}r^{2a_{m,1}-2}\frac{\partial^2 A_{m,1}^2 }{\partial \theta^2} + A_{m,1}^2 (2a_{m,1})^2 r^{2a_{m,1}-2}
	+ r^{a_{m,2}-2}\frac{\partial^2}{\partial \theta^2}(A_{m,0}A_{m,2})
 \\
 + a_{m,2}^2A_{m,0}A_{m,2}r^{a_{m,2}-2}+ \dots .
\end{multline}
As $a_{m,1}>0$, the leading-order equation is $\partial^2 A_{m,0}^2/\partial \theta^2 = 0 $,
which, when solved subject to boundary conditions of $\partial A_{m,0}/\partial \theta=0$ (from \eqref{eq:EulBoundConds}) gives $A_{m,0}=A_{m,0}(t)$. The balance at the next order is
\begin{equation}\label{eq:secondorder}
4	\frac{\partial A_{m,0}}{\partial t}  = a_{m,1}^2A_{m,0}A_{m,1}r^{a_{m,1}-2} + r^{a_{m,1}-2}\frac{\partial^2}{\partial \theta^2} (A_{m,0} A_{m,1}).
\end{equation}
One possible exponent, \rmc{which we index with $m=0$,} in the expansion is $a_{0,1}=2$, yielding an expansion driven by $\partial A_{0,0}/\partial t$
\begin{equation}\label{eq:Gamma_0}
	\Gamma_{(0)} = A_{0,0}(t) +\frac{\partial A_{0,0}/\partial t}{A_{0,0}}r^2+ O(r^4),
\end{equation}
representing a purely radial flow that drives surfactant into or out of the corner, leading to changes in the corner concentration $A_{0,0}(t)$.  The fact that the series goes in powers of $r^{2n}$ can be obtained by examining \eqref{eq:expansion} at the next order.

\rmc{We now look for other possible expansions indexed by $m=1,2,3\dots$, and} to avoid duplication of the primary flow contribution (\ref{eq:Gamma_0}), we set $A_{1,0}=A_{2,0}=\dots=0$.  (When $\Phi=\pi/2$, again assuming $a_{0,1}=2$, (\ref{eq:secondorder}) also possesses a homogeneous solution $A_{0,1}=f_{0,1}(t) \cos(2\theta)$, representing a stagnation-point flow in the corner, with strength $f_{0,1}(t)$ determined by conditions far from the corner.) More generally, \eqref{eq:secondorder}  possesses homogeneous solutions satisfying, for $m=1,2, \dots$, 
\begin{equation}
	 a_{m,1}^2A_{m,1} + \frac{\partial^2}{\partial \theta^2} ( A_{m,1}) = 0 .
\end{equation}
This means that
$A_{{m,1}} = f_{{m,1}}(t)\cos{(a_{m,1}\theta)}$,
after applying $\partial A_{m,1}/\partial \theta=0$ at $\theta=0$. Applying $\partial A_{m,1}/\partial \theta=0$ at $\theta=\Phi$ means that
\begin{equation}
	a_{m,1} = \frac{m\pi}{\Phi},
 \label{eq:cornerexp}
\end{equation} for any integer $m$, and this provides an infinite set of possible exponents. (When $\Phi=\pi/2$, (\ref{eq:cornerexp}) with $m=1$ recovers the stagnation point flow cited above with amplitude $f_{0,1}$.)  By examining the next order balance in \eqref{eq:expansion}, it  must also be the case that $a_{m,2} = m\pi/\Phi+2$, and so on. Therefore, the full solution is given by the sum of asymptotic series
\begin{equation}
	\Gamma = \Gamma_{(0)} +\Gamma_{(1)}+\Gamma_{(2)}+\Gamma_{(3)}+\dots,
\end{equation}
where $\Gamma_{(0)}$ is given by \eqref{eq:Gamma_0}, and
\begin{equation}
	\Gamma_{(1)} = f_{1,1}(t)\cos{(\pi \theta/\Phi)}r^{\pi/\Phi}+ A_{1,2}(\theta,t)r^{\pi/\Phi+2}+ A_{1,3}(\theta,t)r^{\pi/\Phi+4}+ \dots
\end{equation}
\begin{equation}
	\Gamma_{(2)} = f_{2,1}(t)\cos{(2\pi \theta/\Phi)}r^{2\pi/\Phi}+ A_{2,2}(\theta,t)r^{2\pi/\Phi+2}+ A_{2,3}(\theta,t)r^{2\pi/\Phi+4}+ \dots
\end{equation}
\begin{equation}
	\Gamma_{(3)} = f_{3,1}(t)\cos{(3\pi \theta/\Phi)}r^{3\pi/\Phi}+ A_{3,2}(\theta,t)r^{3\pi/\Phi+2}+ A_{3,3}(\theta,t)r^{3\pi/\Phi+4}+ \dots ,
\end{equation}
and so on. The series can be summarised as
\begin{equation}
	\Gamma(r,\theta,t) = \sum_{n=0}^{\infty}\sum_{m=0}^{\infty}A_{m,n}(\theta,t)r^{m\pi/\Phi+2n}.
\end{equation}
The velocity field is $\mathbf{u}_s = -\bm{\nabla}_{\mathbf{x}}\Gamma/4$, and so as $r\to 0$ it is a combination of a radial and a stagnation-point flow,
\begin{equation}
    \mathbf{u}_s\approx \hat{\mathbf{r}}\left(2\frac{\partial A_{0,0}/\partial t}{A_{0,0}}r+\frac{\pi}{4\Phi} f_{1,1}(t)\cos\left(\frac{\pi\theta}{\Phi}\right) r^{(\pi/\Phi)-1}\right)-\hat{\boldsymbol{\theta}} f_{1,1}(t) \frac{\pi}{4\Phi}\sin\left(\frac{\pi\theta}{\Phi}\right) r^{(\pi/\Phi)-1}.
\end{equation}
For internal angles in a convex domain (when $\Phi<\pi$), the velocity is proportional to $r$ to a positive power, so that it goes to zero at the corner.  In particular, when $\Phi=\pi/2$, a particle on the boundary $\theta=0$ or $\theta=\pi/2$ has an inward radial velocity bounded above by $Fr$ for some finite $F>0$.  A particle starting at $r=r_0$ at $t=0$ will then lie in $r>r_0\exp[-F t ]$ for $t>0$, never reaching the corner in finite time, supporting the use of (\ref{eq:BoundCondsinxy}).  For wedge angles $\Phi<\pi/2$, the radial flow is dominant as $r \rightarrow 0$ and (\ref{eq:BoundCondsinxy}) again applies.  For wedge angles satisfying $\Phi\in(\pi/2,\pi)$, the stagnation-point flow dominates as $r\rightarrow 0$.  In this case, although the velocity field along a boundary vanishes as $r\rightarrow 0$, a sustained inward flow $\mathrm{d}r/\mathrm{d}t=-Fr^{(\pi/\Phi)-1}$ for constant $F>0$ has the potential to drive particles to the corner in finite time, with $r=[F(t_0-t)(2-(\pi/\Phi))]^{1/(2-(\pi/\Phi))}$ as $t\rightarrow t_0$ for some $t_0$, calling into question the validity of (\ref{eq:BoundCondsinxy}) in this case.  For even larger wedge angles ($\Phi>\pi$), regularity of the velocity field demands that $f_{1,1}=0$. This highlights the influence of any smoothing over a lengthscale \jl{of order} $\delta\ll 1$ that might be applied to computations of flows around such corners in non-convex domains, where velocities of magnitude $1/\delta^{1-(\pi/\Phi)}$ can be anticipated.



\section{The numerical scheme to solve the Monge--Amp\`ere equation}
\label{sec:AppNumScheme}

\subsection{Outline of the scheme}\label{sec:AppNumSchemeOutline}

\rmc{We approximate \eqref{eq:MongeAmperebeta} and \eqref{eq:MABoundConds} on an $(M+2)\times (N+2)$ rectangular grid. The points on the four boundaries we consider as known, as they can be expressed as functions of the interior points by the boundary conditions using one-sided second-order differences. This leaves us with an $M\times N$ grid of unknowns. We create a $MN\times 1$ vector of the values of the solution $\phi$ at these grid points by stacking the rows of the grid of unknowns into a column vector $\boldsymbol{\phi}$ starting from the top and proceeding downwards. So, for example, $\phi_{i+N}$ refers to the grid-point directly below $\phi_i$, and $\phi_{i+1}$ refers to the grid-point directly to the right of $\phi_i$ unless $i=pN$ for some integer $p$.}

\rmc{We approximate \eqref{eq:MongeAmperebeta} using second-order differences at the $i$th grid point as  
\begin{multline} \label{eq:fdef}
f_i(\boldsymbol{\phi})=
	\frac{(\phi_{i+1}-2\phi_{i}+\phi_{i-1})(\phi_{i+N}-2\phi_{i}+\phi_{i-N})}{\Delta x^2 \Delta y^2} 
 \\
 -
	\left(\frac{-\phi_{i+N+1}-\phi_{i-N-1}+\phi_{i+N-1}+\phi_{i-N+1}}{4\Delta x \Delta y}\right)^2
 -\mathcal{G}_{i} = 0,
\end{multline} which we define to be the $i$th component of the vector $\mathbf{f}(\boldsymbol{\phi})$, and $\Delta x$ and $\Delta y$ refer to the grid spacing in the $x_0$ and $y_0$ coordinate directions respectively, and we define
\begin{equation}
    \mathcal{G} = 1 - \beta_j \left(1 -  \frac{\Gamma_0(x_0,y_0)}{\bar{\Gamma}} \right).
\end{equation} }

The $k$th iteration of the Newton scheme for the vector $\boldsymbol{\phi}$ of the values of the function $\phi$ approximated at the grid points is
\begin{equation}\label{eq:NewtRaph}
\boldsymbol{\phi}^{k} = \boldsymbol{\phi}^{k-1} - \left(\boldsymbol{\nabla}_{\boldsymbol{\phi}}\mathbf{f}(\boldsymbol{\phi^{k-1}})\right)^{-1}\mathbf{f}(\boldsymbol{\phi^{k-1}}) ,
\end{equation}
where the Jacobian $\boldsymbol{\nabla}_{\boldsymbol{\phi}}\mathbf{f}(\boldsymbol{\phi})$ is the sparse matrix of derivatives of the components $f_i$ with respect to solution at each grid point $\phi_j$, so that the element $\partial f_i /\partial \phi_j$ is zero, except for nine diagonals given by
\begin{subequations}\label{eq:Jacdef}
\begin{align}
	\frac{\partial f_i}{\partial \phi_{i}} &= \frac{2\left(4\phi_{i}-\phi_{i+1}-\phi_{i-1}-\phi_{i+N}-\phi_{i-N}\right) }{\Delta x^2 \Delta y^2} ,
\\
	\frac{\partial f_i}{\partial \phi_{i+1}} &= 
	\frac{\phi_{i+N}-2\phi_{i}+\phi_{i-N}}{\Delta y^2 \Delta x^2},
	\\
	\frac{\partial f_i}{\partial \phi_{i-1}} &= 
	\frac{\phi_{i+N}-2\phi_{i}+\phi_{i-N}}{\Delta y^2 \Delta x^2},
	\\
	\frac{\partial f_i}{\partial \phi_{i+N}} &= 
	\frac{\phi_{i+1}-2\phi_{i}+\phi_{i-1}}{\Delta y^2 \Delta x^2},
	\\
	\frac{\partial f_i}{\partial \phi_{i-N}} &= 
	\frac{\phi_{i+1}-2\phi_{i}+\phi_{i-1}}{\Delta y^2 \Delta x^2},
	\\
	\frac{\partial f_i}{\partial \phi_{i+N+1}} &= 
	\frac{-\phi_{i+N+1}-\phi_{i-N-1}+\phi_{i+N-1}+\phi_{i-N+1}}{8\Delta x^2 \Delta y^2},
	\\
	\frac{\partial f_i}{\partial \phi_{i-N-1}} &= 
	\frac{-\phi_{i+N+1}-\phi_{i-N-1}+\phi_{i+N-1}+\phi_{i-N+1}}{8\Delta x^2 \Delta y^2},
	\\
	\frac{\partial f_i}{\partial \phi_{i+N-1}} &= -
	\frac{-\phi_{i+N+1}-\phi_{i-N-1}+\phi_{i+N-1}+\phi_{i-N+1}}{8\Delta x^2 \Delta y^2},
	\\
	\frac{\partial f_i}{\partial \phi_{i-N+1}} &= -
	\frac{-\phi_{i+N+1}-\phi_{i-N-1}+\phi_{i+N-1}+\phi_{i-N+1}}{8\Delta x^2 \Delta y^2}.
\end{align}
\end{subequations}
The definitions \eqref{eq:fdef} and \eqref{eq:Jacdef} are valid for every interior point of the $M\times N$ grid of unknowns, however for the values $i=1$ to $N$, $i=(M-1)N+1$ to $MN$, $(p-1)N+1$ and $pn$ for every integer $1<p<M$, the elements of $f$ and the Jacobian are slightly different to this as they incorporate the boundary conditions, but we omit them here for conciseness. We iterate \eqref{eq:NewtRaph} until the sum of absolute differences of the components of $\boldsymbol{\phi}$ from one iteration to the next falls below a small tolerance value \rmc{which we choose to be $10^{-6}$}.

We use this numerical method to perform the calculation using the method presented in \S\ref{sec:MethodsSteadyNum} for a relatively coarse grid until $\beta_J=1$, and then increase the refinement by interpolating the solution onto a more refined grid, and re-solving \eqref{eq:MongeAmperebeta} and \eqref{eq:MABoundConds} with $\beta_j=1$ (a multi-grid method). We repeat this second process until we have a sufficiently refined solution.

\subsection{Convergence of the numerical scheme}
\label{sec:AppConv}

Figure \ref{fig:conv}  demonstrates the convergence of the numerical scheme presented in detail in \jl{Appendix \ref{sec:AppNumSchemeOutline}} for the solution to the problem set out in \S\ref{sec:MethodsSteadyNum} with $\delta=0.25$, \rmc{$\delta=0.05$ and $\delta=0.002$}. For the purpose of illustration we have picked a point $(x_0,y_0)=(5,5)$ and displayed the error between an approximation for the correct solution, and solutions for $\phi(5,5)$, $X=\phi_{x_0}(5,5)$ and $Y=\phi_{y_0}(5,5)$ at various values for a uniformly spaced grid $h=\Delta x=\Delta y$. The approximation for the correct solution is found by selecting a $h\equiv h_{end}$ smaller than all of the other values of $h$ used, and using this as the grid spacing to solve for $\phi(5,5)\equiv \phi_{end}(5,5)$, $X(5,5)=\phi_{x_0}(5,5)\equiv X_{end}(5,5)$ and $Y(5,5)=\phi_{y_0}(5,5)\equiv y_{end}(5,5)$, \rmc{for each value of $\delta$}.

The solution for all three values in figure \ref{fig:conv} converge approximately with a slope of $2$ on the log scale \rmc{indicated by a dotted black line}, which is expected for our second-order finite-difference scheme. The graphs are only approximately showing this convergence, as there are numerous sources of \rmc{noise}, for example, the approximation we have used for the correct solution. Also, the code stops iterating once the sum of the absolute difference of each gridpoint in a solution between one iteration and the next falls below a certain small tolerance value. The exact locations of the edges of the deposits do not coincide precisely with gridpoints, and this is a further source of error as the initial conditions (and therefore the right hand side of \eqref{eq:MongeAmpere}) for differently discretised grids are effectively slightly different. The solution for $X$ and $Y$ also involve calculating a further derivative using a second order finite-difference approximation, introducing further error. However, the dominant error is clearly the square of the discretisation parameter $h$ used in the approximation of \eqref{eq:MongeAmpere}. \rmc{Solutions with $\delta=0.002$ are significantly more expensive computationally, which is why the curves terminate at larger values of $h$ than the other curves. The fact that we have used a larger value of $h$ to calculate $\phi_{end}(5,5)$ and its derivatives for $\delta=0.002$ might also partly explain why some of these curves are noisier than for larger values of $\delta$.}

\begin{figure}
 \centering
\begin{subfigure}{0.8\textwidth}
    \centering
    \includegraphics[width=\textwidth]{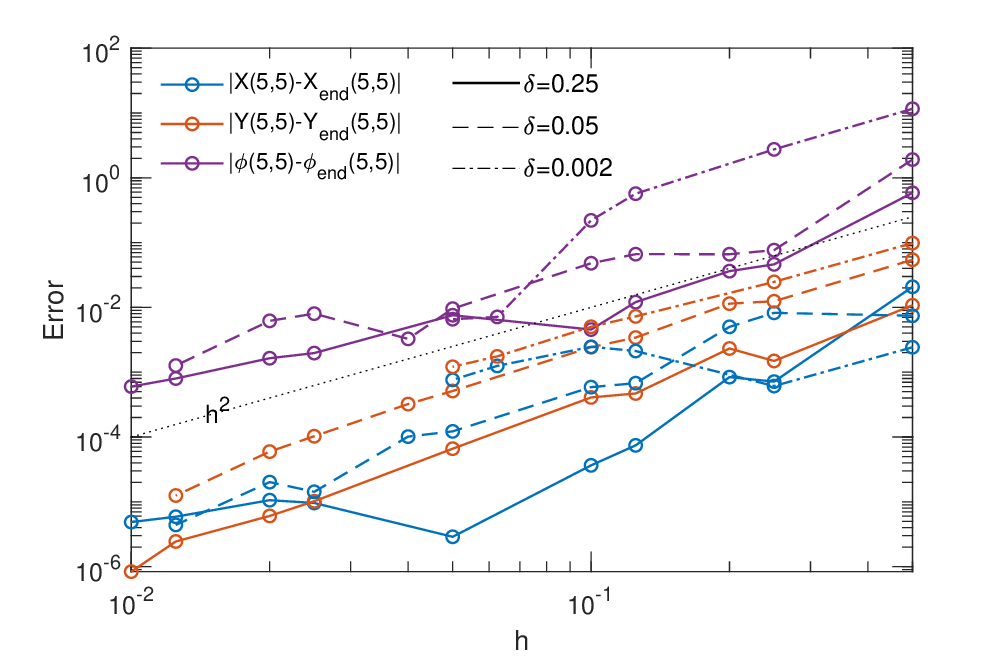}
\end{subfigure}
    \caption{Graph showing the convergence of the finite-difference approximation to the solution of the Monge--Amp\`ere equation \eqref{eq:MongeAmpere}. To illustrate convergence, we choose a point on the Lagrangian domain $(x_0,y_0)= (5,5)$ and calculate the solution for the three-deposits problem outlined in \S\ref{sec:MethodsSteadyNum} with $\delta=0.25$, \rmc{$\delta=0.05$ and $\delta=0.002$ (each $\delta$ is assigned a different linestyle, see legend)} for multiple values of a uniform grid spacing $h$ (data points are shown as circles). A set of values for each $\delta$, which we call $\phi_{end}(5,5)$, $X_{end}(5,5)$ and $Y_{end}(5,5)$, are calculated for a discretisation parameter $ h_{end}$ which is smaller than the rest of the values of $h$ used. We use this solution as our approximation to the correct solution. The curves show the difference between $X(5,5)$ (blue), $Y(5,5)$ (orange), and $\phi(5,5)$ (purple) and the approximated correct solution, \jl{showing convergence at a rate of order $h^2$ (dotted black line)}.}
    \label{fig:conv}
\end{figure}

\section{Simulating Suminagashi patterns}
\label{sec:appSumiMethod}

We create a Suminagashi pattern using successive solutions of the Monge--Amp\`ere equation. As described in the introduction (\S\ref{sec:Intro}), the creation of Suminagashi art patterns consists of repeated deposition of an ink surfactant mixture onto a liquid surface, followed by the artist blowing on the surface between ink depositions. We simulate this procedure using the Monge--Amp\`ere equation \eqref{eq:MongeAmpere} to simulate the final locations of ink depositions in the spreading steps, and a divergence-free map to create the effects of the blowing steps. We show in \eqref{eq:rotationalvelocity} how the rotational component of a surface shear stress can redistribute surfactant without changing its surface concentration.

\rmc{The physical assumptions we consider in this study for the Monge--Amp\`ere method are appropriate for Suminagashi.} \oej{Exploiting conditions (\ref{eq:bond}) and (\ref{eq:surdif}): for a typical 1~cm layer of water with a surface tension reduction of 1 to 10~g/s$^2$, we find that the Bond number is large and of the order of $10^{2}$ to $10^{3}$; surface diffusivity for most surfactants are of the order of $D^*= 10^{-10}$ to $10^{-9}~\mathrm{m}^2/s$ 
\citep{CHANG19951}, which leads to a very large surface P\'eclet number of the order of $10^{7}$ to $10^{9}$; furthermore, Suminagashi patterns can be transferred to paper within a minute, during which any boundary thickens under molecular (surface) diffusivity by  less than ${0.3}~\mathrm{mm}$.}

\subsection{Spreading steps}

Each spreading step starts from initial conditions \rmc{\eqref{eq:InitCondsGenNonD} with $\mathcal{F}(x_0,y_0) = \mathcal{C}_q(x_1,y_1,0.5,1-\delta)$}, and by imposing $L_1=13$, $L_2=11$, so that this is now a single deposit of exogenous surfactant spreading into endogenous surfactant. For the first spreading step, we set $\delta=0.01$, and solve \eqref{eq:MongeAmpere} \eqref{eq:MABoundConds}. We store the final location of the boundary of this initial deposit, which we call $C_{1,(1)}$, where the first index labels each deposit edge, and the second index refers to how many total deposits have been released. Next, we set up a new problem by replacing $\delta$ by $\bar{\Gamma}$ from the previous problem, and we shift $(x_3,y_3)$ to a displaced position (using an algorithm described below as \eqref{eq:algo1}), and solve again (\ref{eq:MongeAmpere},\ref{eq:MABoundConds}), finding the edge of the new deposit, and the new locations of the edges of previously released deposits, repeating the above process $J$ times. This process is summarised below.

\subsection*{Summary of the algorithm for the spreading steps}
For $j=1$ to $J$: solve (\ref{eq:MongeAmpere}, \ref{eq:MABoundConds}) subject to initial conditions \eqref{eq:InitCondsGenNonD} with $\mathcal{F}(x_0,y_0) = \mathcal{C}_q(x_1,y_1,0.5,1-\delta)$, using the solution method outlined in \S\ref{sec:MethodsSteadyNum} with $L_1=13$, $L_2=11$, $\Gamma_2=\Gamma_3=\delta$, $r_1=0.5$, $r_2=0$, $r_3=0$, with $(x_1,y_1)=(x_{1,j},y_{1,j})$ and $\delta=\delta_j$ where this last quantity is found using
\begin{equation}
    \delta_{j} = \bar{\Gamma}_{j-1}, \qquad \bar{\Gamma}_0 = 0.05,
\end{equation} and where at each step $\bar{\Gamma}_j$ is computed using \eqref{eq:GammaBar}. We find all curves at step $j$ by
\begin{equation}
    C_{i,(j+1)} = \mathbf{X}_{j}(C_{i,(j)} ) \qquad \text{for} \quad i=1,2\dots j,
\end{equation} where $\mathbf{X}_j$ is the solution of (\ref{eq:MongeAmpere},\ref{eq:MABoundConds}) at step $j$, and one new evolving curve is introduced at each step $j$ by
\begin{equation}
C_{j,(j)} = \{ (x,y) | (x-x_{1,j})^2+(y-y_{1,j})^2=0.5\}
\end{equation} The centres of the new deposits are chosen such that
\begin{equation}\label{eq:algo1}
    (x_{1,j+1},y_{1,j+1}) = (x_{c,j}-0.75,y_{c,j}-0.2), \qquad \text{where } (x_{c,j},y_{c,j}) = \max_x{(C_{j,(j)}}),
\end{equation} 
until $\max_{x}(C_{j,(j)})>12.9$ and then
\begin{equation}\label{eq:algo2}
    (x_{1,j+1},y_{1,j+1}) = (x_{c,j}-0.05,y_{c,j}+0.75), \qquad \text{where } (x_{c,j},y_{c,j}) = \min_y{(C_{j,(j)}}),
\end{equation}  
with
\begin{equation}
    (x_{1,1},y_{1,1}) = (3.5,8.5).
\end{equation} 
We vary the location of the centres of the new deposits, as we have observed Suminagashi artists to do this. The final picture is created by plotting $C_{i,(J+1)}$ for all $i=1$ to $J$.

\subsection{Blowing steps}

Between some of the spreading steps, we  introduce blowing steps, which are computed by finding a divergence-free map $\mathbf{X}_b$ by time stepping
\begin{equation}\label{eq:blowingtimestep}
 \frac{\dee}{\dt } \mathbf{X}_b = \sum_{n_s=1}^{\infty}A_{n_s}\begin{pmatrix}
     \frac{n_{s}\pi}{11}\sin{\left(\frac{n_{s}\pi x}{13}\right)}\cos{\left(\frac{n_{s}\pi y}{11} \right)}
     \\
-     \frac{n_{s}\pi}{13}\cos{\left(\frac{n_{s}\pi x}{13}\right)}\sin{\left(\frac{n_{s}\pi y}{11} \right)}
 \end{pmatrix},
\end{equation}
for some integer $n_{s}$, between $t=0$ and some $t_{s}$, where $A_{n_s}$ are the amplitudes of the modes, and where $\mathbf{X}_b=[x_0,y_0]$ at $t=0$. The resulting map $\mathbf{X}_b$ is divergence-free (such that concentrations do not change) and satisfies the required boundary conditions. The functions on the right hand side of \eqref{eq:blowingtimestep}  form a basis, such that any divergence-free map can be obtained by choosing a set of amplitudes $A_{n_s}$ and some choice of $t_s$. The new location of deposit edge $i$, after spreading step $j$ is given by $\mathbf{X}_b(C_{i,(j)})$.

\subsection{Sequential depositions: the Suminagashi patterns}
\label{sec:SuminagashiRes}

To illustrate the utility of the Monge--Amp\`ere approximation, a solution of the Suminagashi algorithm is presented in figure \ref{fig:SuminagashiAlt}. Here, for \eqref{eq:blowingtimestep},
we choose $A_{ns}=0$ for every $n_s$ except $N_s=4$, and we choose $A_4=1$. We run 20 spreading steps with a blowing step imposed after every four spreading steps with $t_s=0.15$. In figure \ref{fig:SuminagashiAlt} the creation of the final figure in the last panel is shown after $4$ spreading steps, after $4$ spreading steps and one blowing step, after $12$ spreading steps, after $12$ spreading steps and $3$ blowing steps, and lastly after $20$ spreading steps before the final blowing step. \jl{The final result is given in a monochrome colour scheme in the second to last panel in figure~\ref{fig:SuminagashiAlt}}.

\rmc{\section{Difference in the prediction between Eulerian particle tracking and Monge--Amp\`ere for different values of $\delta$}
\label{sec:Differencedelta}}

\begin{figure}
 \centering
\begin{subfigure}{0.49\textwidth}
    \centering
    \includegraphics[width=\textwidth]{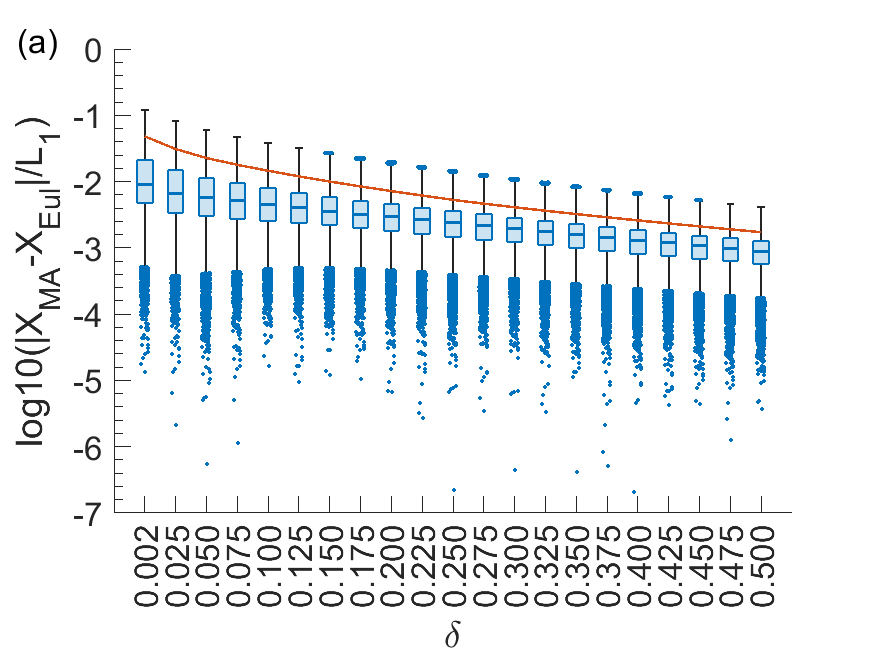}
\end{subfigure}
\begin{subfigure}{0.49\textwidth}
    \centering
    \includegraphics[width=\textwidth]{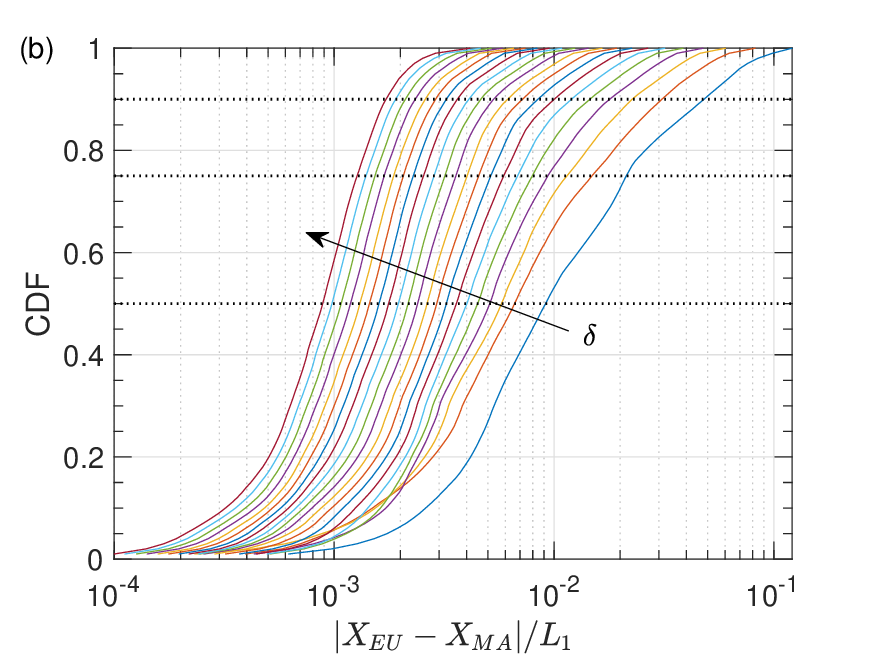}
\end{subfigure}
    \caption{\rmc{Graphs showing the normalised absolute error (\ref{eq:EucDist}) for several different values of $\delta$. The initial conditions are (\ref{eq:InitCondsGenNonD}, \ref{eq:InitCondsGenNonD2}), with $\mathcal{C}_q$ replaced by $\mathcal{C}_c$ defined in \eqref{eq:circlenotationcos} except for the first solution, which is taken from Eul[0.002] and MA[0.002] and has a quadratic initial profile. (\textit{a}) Shows the logarithm of the Euclidean distance between the predictions of final location, for each initial particle location from a $221\times261$ grid, represented as a data point. The whiskers show 1.5 times the interquartile range above and below the quartiles, and particles outside this range are considered outliers and plotted as a cloud of points in blue, with the 90th percentile plotted as an orange line. 
    (\textit{b}) Shows the same data as in (\textit{a}) plotted in the form of a cumulative distribution function, with $\delta$ increasing in the direction of the arrow. Horizontal dashed lines indicate the median, the 75th percentile and the 90th percentile, respectively.}}
    \label{fig:deltas}
\end{figure}

\rmc{Figure \ref{fig:deltas} shows a series of box and whisker plots of the normalised absolute error (\ref{eq:EucDist}), measuring the Euclidean distance between predictions for final particle location between Monge--Amp\`ere and Eulerian particle-tracking, for multiple values of $\delta$. Except for the solution with $\delta=0.002$, which has quadratic initial conditions (\ref{eq:InitCondsGenNonD}, \ref{eq:InitCondsGenNonD2}), the rest of the solutions have cosine shaped exogenous deposit profiles such that
\begin{equation}\label{eq:circlenotationcos}
 \mathcal{C}_c(\mathbf{x}_0;\mathbf{x}_c,r,\Gamma_{0,c}-\delta)  = \begin{cases}
   \left(\frac{\Gamma_{0,c}-\delta}{2}\right)\left(1+\cos\left( \frac{\pi(|\mathbf{x}_0-\mathbf{x}_c|^2)}{r^2} \right) \right)
   &\quad |\mathbf{x}_0-\mathbf{x}_c|\leq r \\
0 &\quad |\mathbf{x}_0-\mathbf{x}_c| > r\end{cases}
\end{equation} 
replaces $\mathcal{C}_q(\mathbf{x}_0;\mathbf{x}_c,r,\Gamma_{0,c}-\delta)$ in (\ref{eq:InitCondsGenNonD}, \ref{eq:InitCondsGenNonD2}). The shape of the exogenous deposit profiles does not have a significant effect on results as shown in figure S7 of \S{S5} of the supplementary material, but we change the profile shape here for the sake of variety.}

\rmc{For moderate $\delta$, the error between different methods remains  small, but increases rapidly as $\delta$ tends to zero.} \oej{However, for $\delta$ as small as 0.075, the normalised absolute error is below 0.01 for 75\% of particles, and even for $\delta=0.002$ the error is below 0.05 for 90\% of particles.}

\bibliographystyle{jfm}

\bibliography{Refs}

\end{document}


\maketitle

\section{Full definition of $\xi$ for the solution of the three-deposits problem using the Lagrangian method}\label{sec:Appxidef}

In the problem presented in \S2.2.2, to obtain the deformed Lagrangian coordinate $\xi$ (see (2.24)), we need to consider $9$ regions of $\Omega$ (see figure 2\textit{a}). 
We also impose that the line $\xi=0$ maps to the line $x=0$. Hence,
\scriptsize
\begin{equation}\label{eq:piecewisexi3blobs}
	\xi = \begin{cases}
		\delta x \qquad & \text{in R1},
		\\
		 x- \left(\frac{(x-6)^3}{3}+x(y-2)^2\right)(1-\delta)+ \frac{2}{3}(1-\delta)(y-1)(3-y)(\sqrt{(y-1)(3-y)}-9)  \qquad & \text{in R2},
		\\
		\delta x +\frac{4}{3}(1-\delta)((y-1)(3-y)^{3/2} \qquad & \text{in R3},
		\\
		x- \frac{1}{4}\left(\frac{(x-10)^3}{3}+x(y-5)^2\right)(1-\delta) + \frac{1}{6}(1-\delta)(y-3)(7-y)(\sqrt{(y-3)(7-y)}-15)  \qquad & \text{in R4},
		\\
		\delta x +\frac{1}{3}(1-\delta)((y-3)(7-y))^{3/2} \qquad & \text{in R5},
		\\
		2 x- \frac{1}{9}\left(\frac{(x-4)^3}{3}+x(y-7)^2\right)(2-\delta) + 
		\frac{2}{27}(2-\delta)(y-4)(10-y)(\sqrt{(y-4)(10-y)}-6)  \qquad & \text{in R6},
		\\
		\delta x +\frac{4}{27}(2-\delta)((y-4)(10-y))^{3/2}  \qquad & \text{in R7},
		\\
		  x- \frac{1}{4}\left(\frac{(x-10)^3}{3}+x(y-5)^2\right)(1-\delta) 
		\\
		\quad + 
		\frac{1}{6}(1-\delta)(y-3)(7-y)(\sqrt{(y-3)(7-y)}-15)
		+\frac{4}{27}(2-\delta)((y-4)(10-y))^{3/2}  \qquad & \text{in R8},
		\\
		\delta x +\frac{1}{3}(1-\delta)((y-3)(7-y))^{3/2}+\frac{4}{27}(2-\delta)((y-4)(10-y))^{3/2}  \qquad & \text{in R9}.
	\end{cases}
\end{equation} \normalsize
We now describe the coordinates of the boundaries and edges of circular deposits in the deformed Lagrangian coordinates. Three of the four Lagrangian boundaries are $0\leq \xi \leq 13 \delta $ for both $\eta=0$ and $\eta =11$, and $0\leq \eta \leq 11$ for $\xi=0$. The fourth boundary $\partial \Omega_R$ (the right hand side boundary of the deformed domain shown in figure 2\textit{b}) is given by 
\begin{subequations}\label{eq:RHEdge1}
\begin{align}
	\xi &= 13\delta,  \quad &0\leq \eta \leq 1 \ \text{for R1}, \\
	\xi &= 13 \delta  +\tfrac{4}{3}(1-\delta)((\eta-1)(3-\eta))^{3/2} \quad &1< \eta \leq 3 \ \text{for R3}, \\
	\xi &= 13\delta  +\tfrac{1}{3}(1-\delta)((\eta-3)(7-\eta))^{3/2} \quad &3< \eta \leq 4 \ \text{for R5},  \\
	\xi &= 13\delta  +\tfrac{1}{3}(1-\delta)((\eta-3)(7-\eta))^{3/2} +\tfrac{4}{27}(2-\delta)((\eta-4)(10-\eta))^{3/2} \quad &4< \eta \leq 7 \ \text{for R9},  \\
	\xi &= 13\delta +\tfrac{4}{27}(2-\delta)((\eta-4)(10-\eta))^{3/2} \quad &7< \eta \leq 10 \ \text{for R7}, \\
	\xi &= 13 \delta  \quad& 10< \eta \leq 11 \ \text{for R1}.
 \label{eq:RHEdgeend}
\end{align} 
\end{subequations} 
The boundaries of the three deposits are given by
\scriptsize
\begin{equation}
	\xi = \delta(6-\sqrt{1-(\eta-2)^2}) \quad \text{and}  \quad \delta(6+\sqrt{1-(\eta-2)^2}) +\frac{4}{3}(1-\delta)((\eta-1)(3-\eta))^{3/2} \quad \text{with} \quad
	1\leq \eta \leq 3,
\end{equation}
 \normalsize for R2, 
 \scriptsize
 \begin{equation}
\xi = \delta(10-\sqrt{4-(\eta-5)^2}) \quad \text{and}  \quad \delta(10+\sqrt{4-(\eta-5)^2}) +\frac{1}{3}(1-\delta)((\eta-3)(7-\eta))^{3/2} \quad \text{with}  \quad
3\leq \eta \leq 4,
\end{equation}
 \normalsize for R4, 
 \scriptsize
 \begin{multline}
\delta(10-\sqrt{4-(\eta-5)^2})  +\frac{4}{27}(2-\delta)((\eta-4)(10-\eta))^{3/2}   
\\ 
 \text{and}  \quad \delta(10+\sqrt{4-(\eta-5)^2}) +\frac{1}{3}(1-\delta)((\eta-3)(7-\eta))^{3/2}+\frac{4}{27}(2-\delta)((\eta-4)(10-\eta))^{3/2}\quad
 \text{with}  \quad
4\leq \eta \leq 7,
\end{multline}
 \normalsize for R8, 
 \scriptsize
 \begin{equation}
\xi = \delta(4-\sqrt{9-(\eta-7)^2}) \quad \text{and}  \quad \delta(4+\sqrt{9-(\eta-7)^2}) +\frac{4}{27}(2-\delta)((\eta-4)(10-\eta))^{3/2} \quad \text{with}  \quad
4\leq \eta \leq 10.
\end{equation} 
 \normalsize for R6.
 
\section{Eulerian and Lagrangian comparison}
\label{sec:AppEulLagComp}

\begin{figure}
	\centering
	\begin{subfigure}{0.49\textwidth}
		\includegraphics[width=1.1\textwidth]{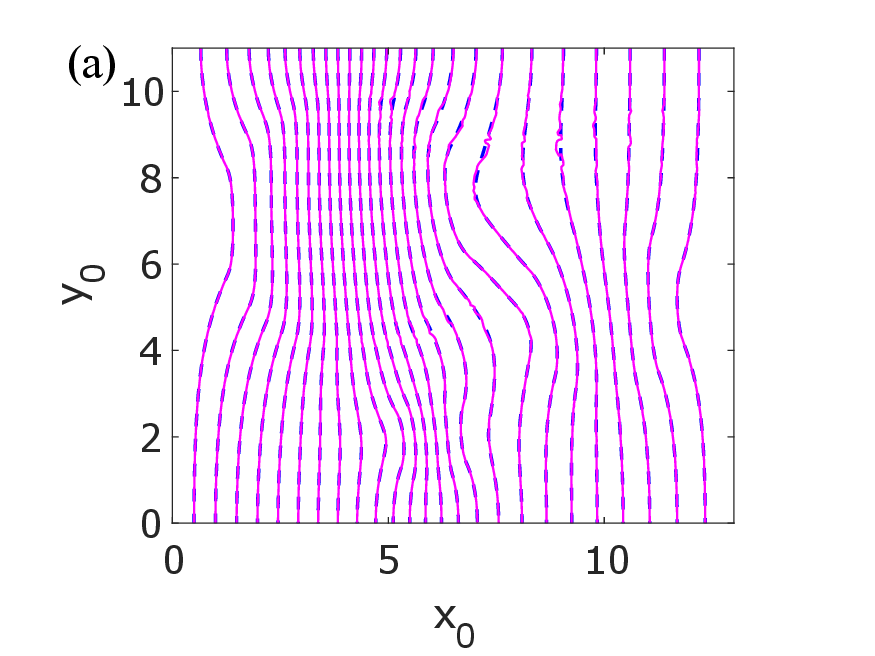}
	\end{subfigure}
	\begin{subfigure}{0.49\textwidth}
		\includegraphics[width=1.1\textwidth]{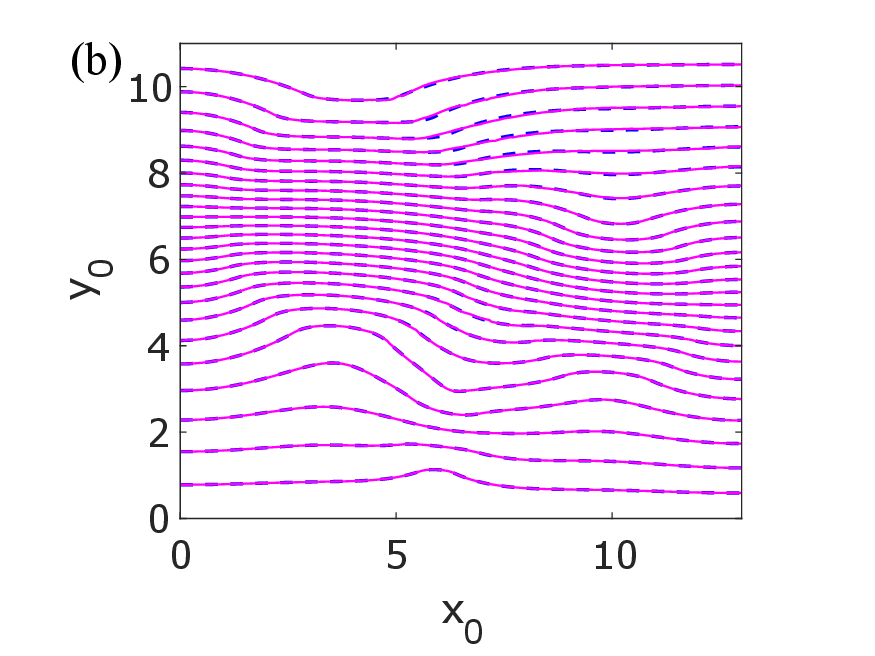}
	\end{subfigure}
	\caption{ Graphs (a) and (b) show the overlay of the $X$ and $Y$ components, respectively, of the maps taken from Eul[0.25] and Lag[0.25]. The solution from Eul[0.25] is given by the dashed blue contours, and the solution from Lag[0.25] is given by the solid magenta contours. }
	\label{fig:EulLagComp}
\end{figure}

Figure \ref{fig:EulLagComp} overlays the contours of both components of the map found from Eul[0.25] and Lag[0.25]. Both solutions are at $t=t_f$, a time found from consideration of (B12) implemented during the calculation of Eul[0.25], such that both $X_{tol}$ and $Y_{tol}$ were set to $10^{-3}$. Both sets of contours in figure \ref{fig:EulLagComp} coincide with each other, apart from small regions in figure \ref{fig:EulLagComp}(\textit{a}) where small oscillations occur on the magenta contours (see, e.g., for $7\leq x_0\leq 11$ and $8\leq y_0\leq 10$). The fact that these oscillations occur for $X$, and not for $Y$ is evidence that they are an artefact caused by the deformation of the Lagrangian domain (figure 2), which is implemented in the $x_0$ coordinate direction only.

\section{The evolution of the concentration field in time}

\begin{figure}
	\centering
 \begin{subfigure}{0.89\textwidth}
    \begin{subfigure}{0.32\textwidth}
    \centering
\includegraphics[keepaspectratio=true,width=1.1\textwidth]{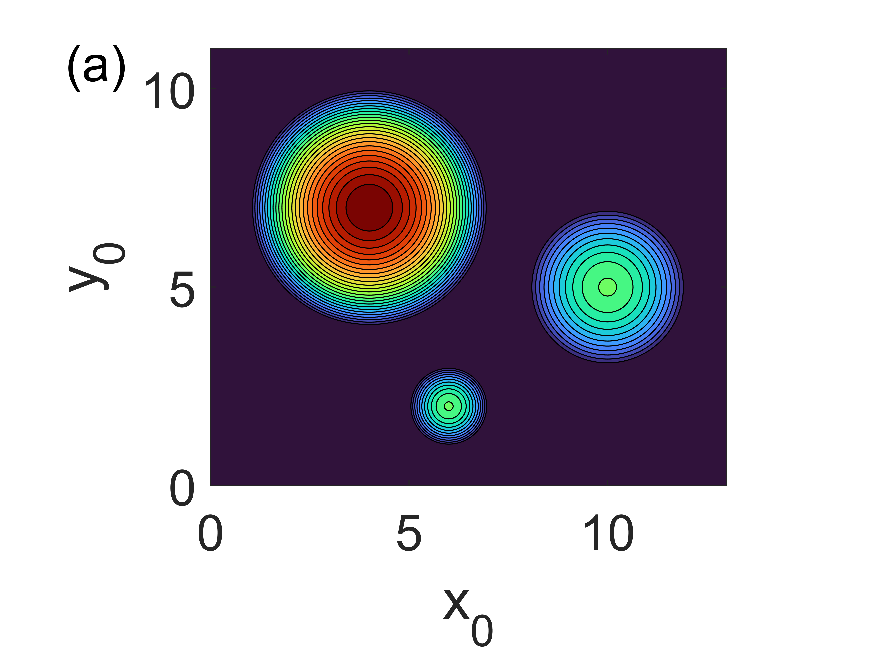}
    \end{subfigure}
     \begin{subfigure}{0.32\textwidth}
    \centering
    \hspace{-10mm}
\includegraphics[keepaspectratio=true,width=1.1\textwidth]{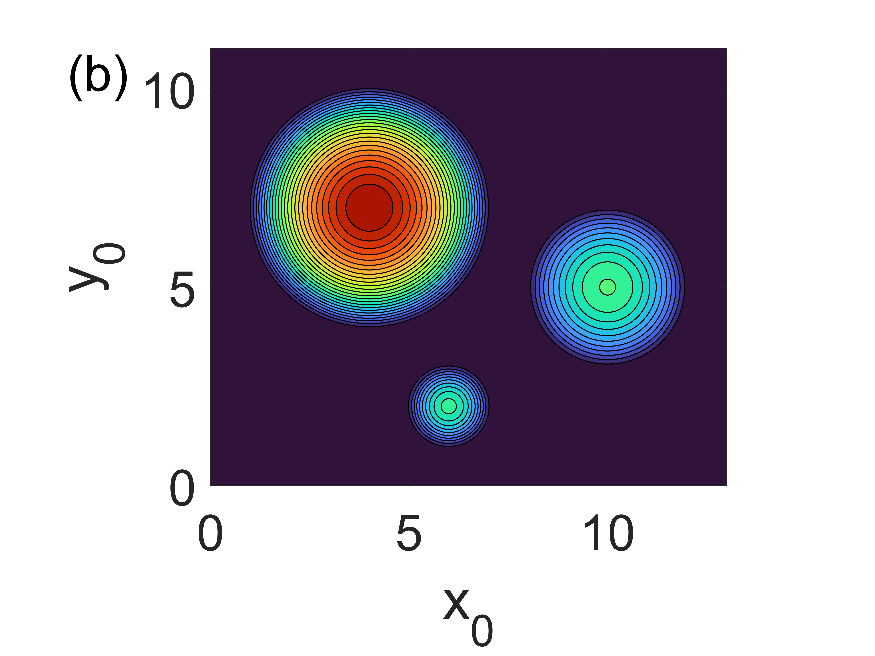}
    \end{subfigure}
    \begin{subfigure}{0.32\textwidth}
    \centering
    \hspace{-20mm}
\includegraphics[keepaspectratio=true,width=1.1\textwidth]{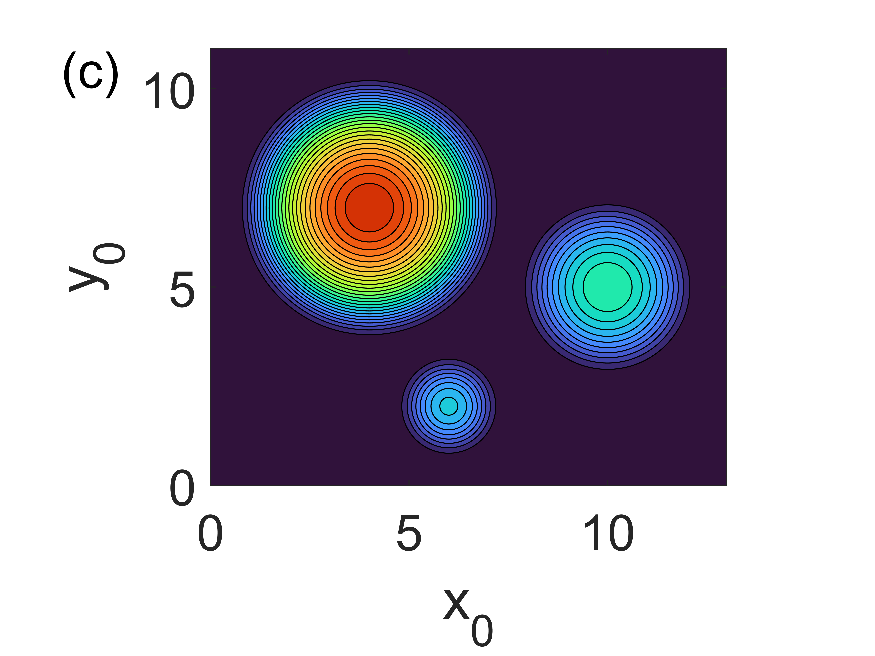}
    \end{subfigure}
     \begin{subfigure}{0.32\textwidth}
    \centering
\includegraphics[keepaspectratio=true,width=1.1\textwidth]{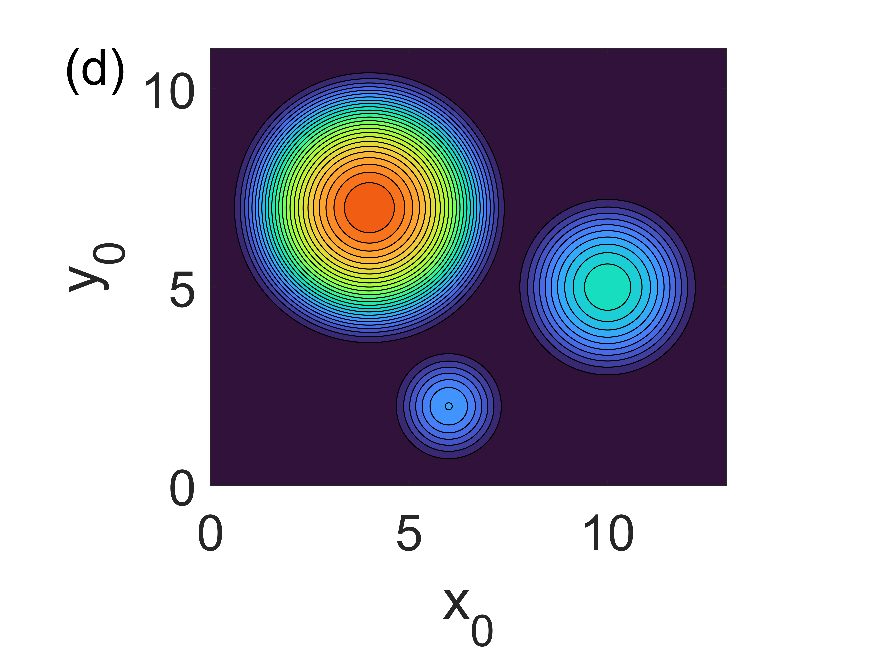}
    \end{subfigure}
    \begin{subfigure}{0.32\textwidth}
    \centering
    \hspace{-10mm}
\includegraphics[keepaspectratio=true,width=1.1\textwidth]{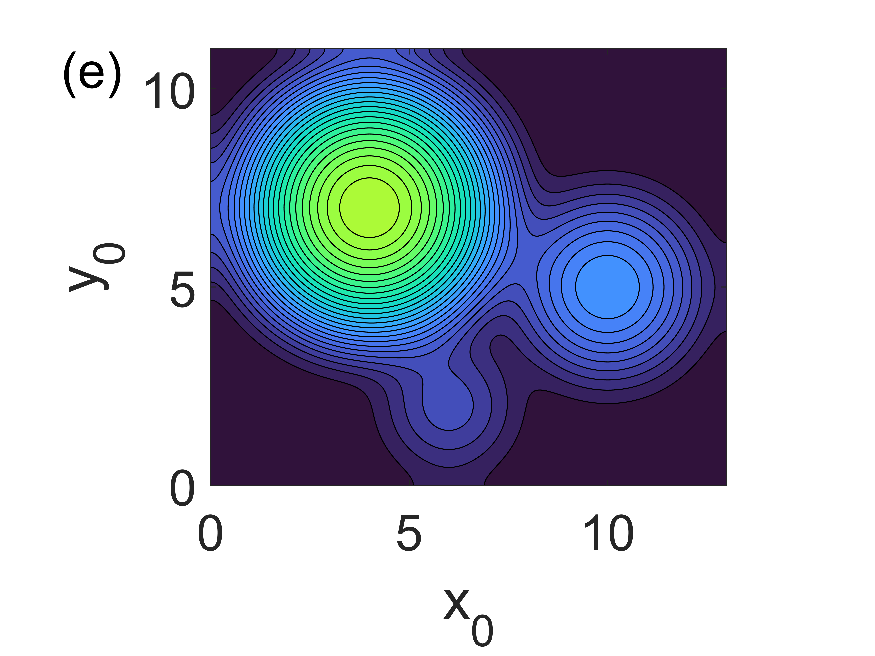}
    \end{subfigure}
     \begin{subfigure}{0.32\textwidth}
    \centering
    \hspace{-20mm}
\includegraphics[keepaspectratio=true,width=1.1\textwidth]{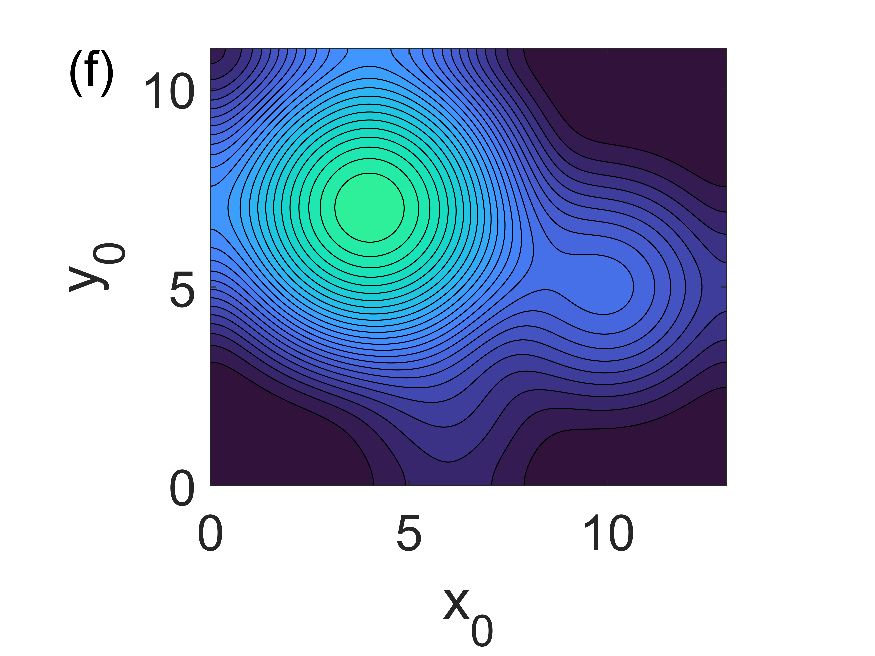}
    \end{subfigure}
    \begin{subfigure}{0.32\textwidth}
    \centering
\includegraphics[keepaspectratio=true,width=1.1\textwidth]{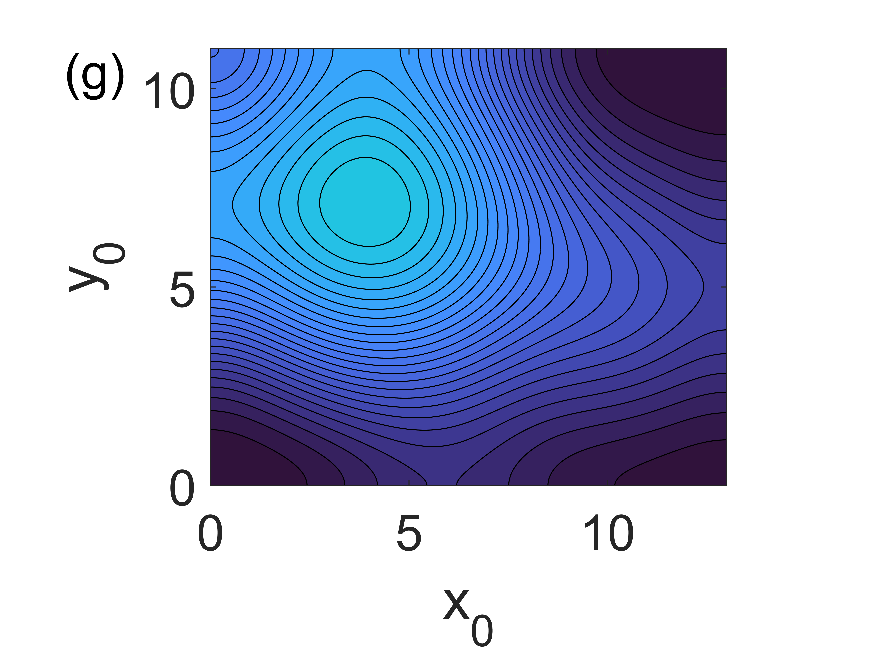}
    \end{subfigure}
    \begin{subfigure}{0.32\textwidth}
    \centering
    \hspace{-6mm}
\includegraphics[keepaspectratio=true,width=1.1\textwidth]{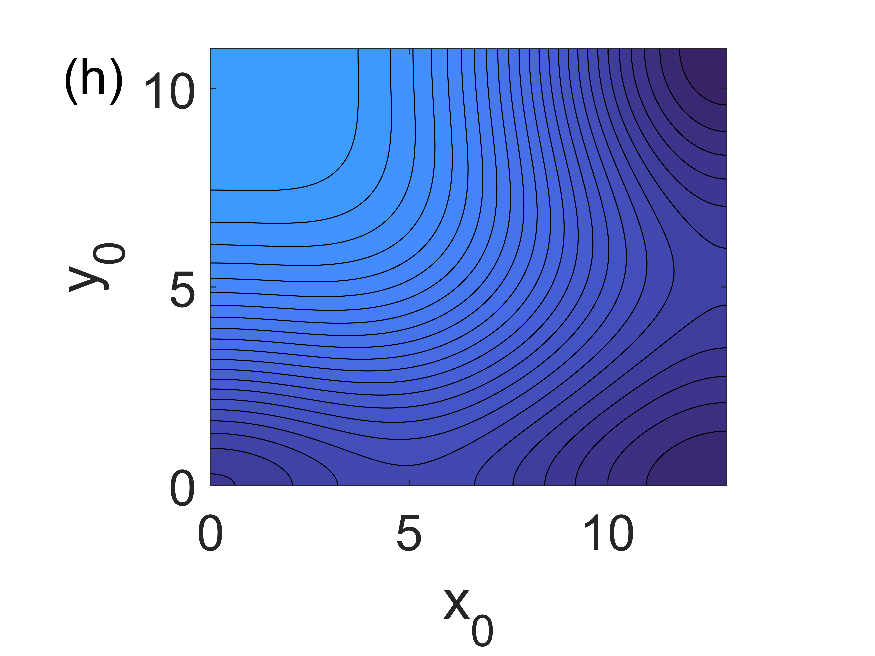}
    \end{subfigure}
    \begin{subfigure}{0.32\textwidth}
    \centering
    \hspace{-16mm}
\includegraphics[keepaspectratio=true,width=1.1\textwidth]{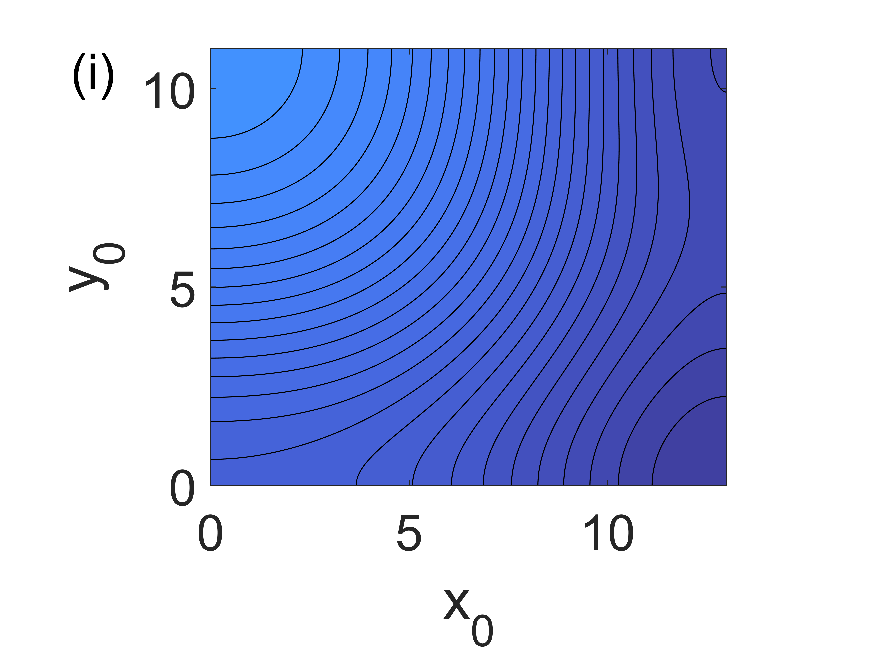}
    \end{subfigure}
     \vspace{-15mm}
    \end{subfigure}
    \begin{subfigure}{0.1\textwidth}
    \centering
    \hspace{-20mm}
\includegraphics[width=1.2\textwidth]{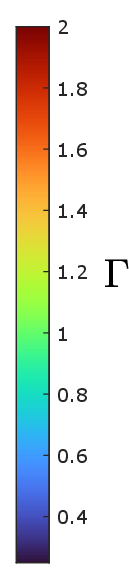}
    \end{subfigure}
\vspace{10mm}
	\caption{The concentration evolution over time given as a colour map for $\Gamma$ taken from Eul[0.25], shown at $t=0$ in (\textit{a}), $t=0.1$ in (\textit{b}), $t=0.5$ in (\textit{c}), $t=1$ in (\textit{d}), $t=5$ in (\textit{e}), $t=10$ in (\textit{f}), $t=20$ in (\textit{g}), $t=50$ in (\textit{h}), and $t=100$ in (\textit{i}).}
	\label{fig:conc}
\end{figure}

In figure \ref{fig:conc} we present a time series of colour maps of the concentration field $\Gamma$ taken from Eul[0.25]. This concentration evolution is calculated from (2.1), from initial conditions (2.7, 2.9) for $\delta=0.25$, $(x_1,y_1) = (6,2)$, $(x_2,y_2) = (10,5)$, $(x_3,y_3)=(4,7)$, $r_2=2$, $r_3=3$, $\Gamma_2=1$ and $\Gamma_3=2$. This solution terminates at $t_f=1047.8$ as stated in table 1, but here we see the concentration field is smoothed to a shallow gradient between top-left and bottom-right corners by $t=100$ (panel \textit{i}).

\section{Comparison between the Eulerian particle-tracking solution in the steady-state and the Monge--Amp\`ere  solution}

\begin{figure}
	\centering
    \begin{subfigure}{0.49\textwidth}
    \centering
\includegraphics[keepaspectratio=true,width=1\textwidth]{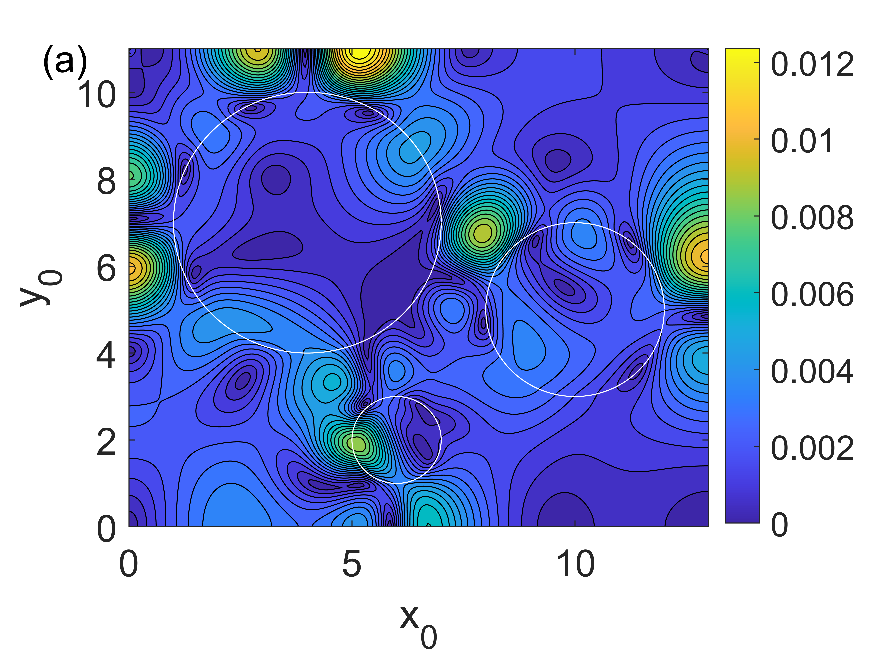}
    \end{subfigure}
     \begin{subfigure}{0.49\textwidth}
    \centering
\includegraphics[keepaspectratio=true,width=1\textwidth]{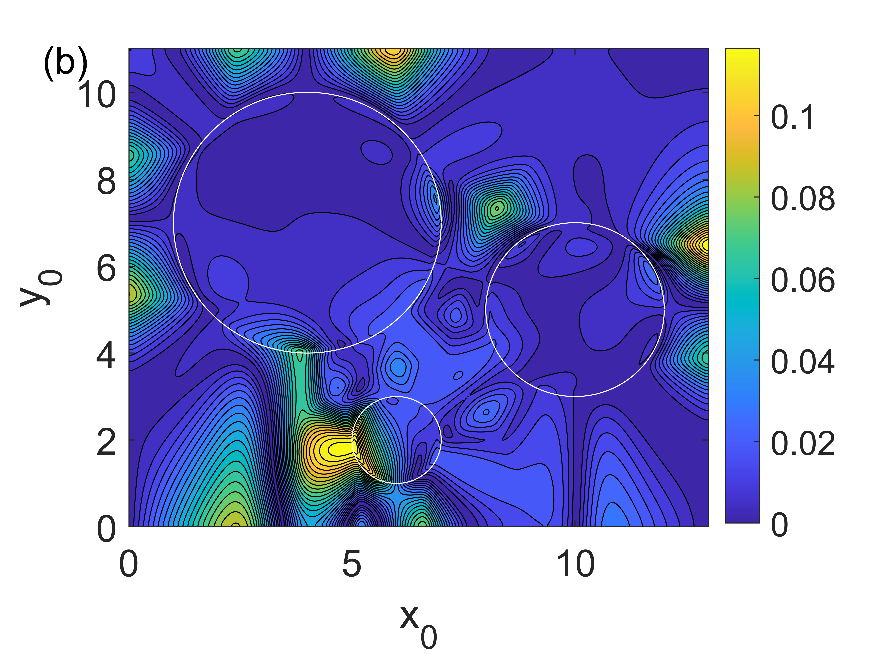}
    \end{subfigure}
    \begin{subfigure}{0.49\textwidth}
    \centering
\includegraphics[keepaspectratio=true,width=1\textwidth]{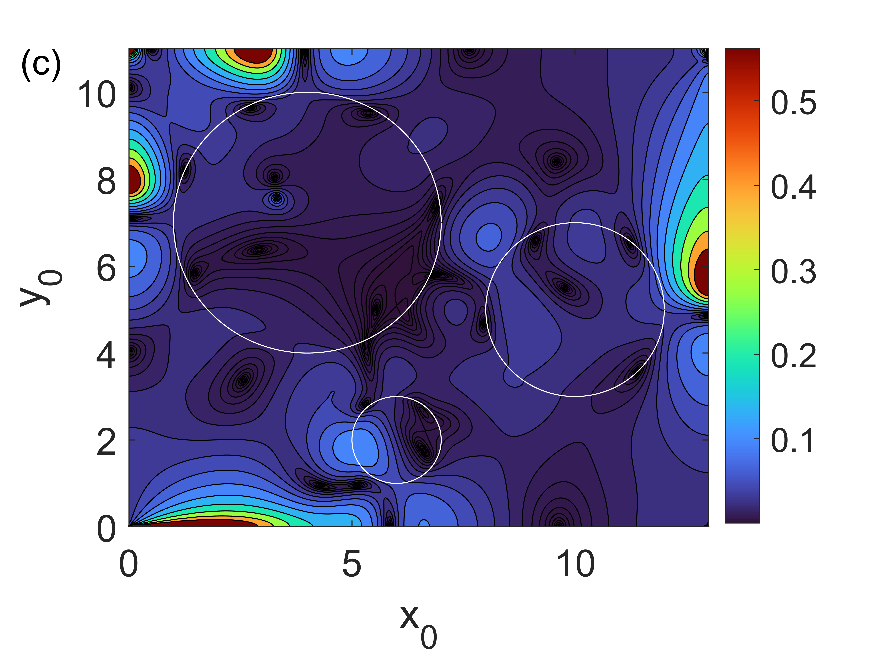}
    \end{subfigure}
     \begin{subfigure}{0.49\textwidth}
    \centering
\includegraphics[keepaspectratio=true,width=1\textwidth]{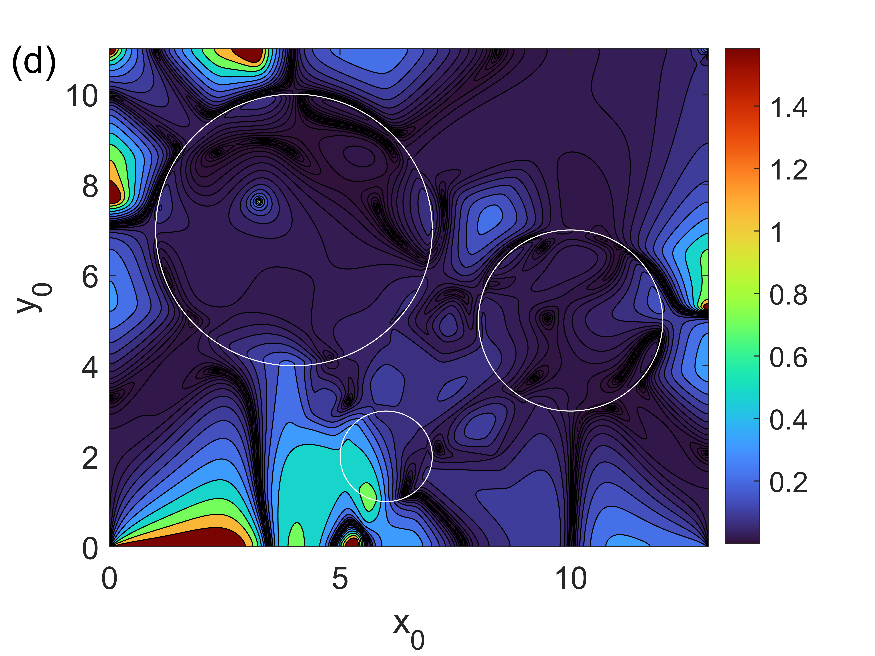}
    \end{subfigure}
	\caption{Contour plots of the Euclidean distance between predictions of final particle location from the Eulerian particle-tracking and the Monge--Amp\`ere methods, as functions of Lagrangian coordinates for the three-deposits problem (see \S2.1.2). (\textit{a}) Shows the normalized absolute error $|X_{EU}-X_{MA}|/L_1$ for Eul[0.25] and MA[0.25], (\textit{b}) shows the same for Eul[0.002] and MA[0.002]. (\textit{c}) Shows the relative error $||X_{EU}-X_{MA}||$ (see \eqref{eq:EucDistnormed}) for Eul[0.25] and MA[0.25] with contours spaced logarithmically, and (\textit{d}) shows the same for Eul[0.002] and MA[0.002]. White circles show the initial deposit locations (see figure 2\textit{a}).}
	\label{fig:AbsRelErr}
\end{figure}

Figure \ref{fig:AbsRelErr} shows a comparison between the final particle locations found by both the finite-difference Eulerian solution and the Monge--Amp\`ere approximation for the three-deposit problem (see \S2.1.2). Figures \ref{fig:AbsRelErr}(\textit{a}) and \ref{fig:AbsRelErr}(\textit{b}) show the normalized absolute error defined as (2.34) in the main text, while figures \ref{fig:AbsRelErr}(\textit{c}) and \ref{fig:AbsRelErr}(\textit{d}) show a relative error defined as
\begin{equation}\label{eq:EucDistnormed}
||{X}_{EU}-{X}_{MA}|| = \frac{\sqrt{(X_{EU}-X_{MA})^2+(Y_{EU}-Y_{MA})^2}}{\sqrt{(X_{EU}-x_0)^2+(Y_{EU}-y_0)^2}}.
\end{equation}
The relative error normalizes the absolute error with respect to the distance travelled by each particle using the Eulerian method. Note that the relative error diverges near fixed points where $\mathbf{X}_{EU}(t_f)\approx(x_0,y_0)$.

For moderate $\delta$, the error is very small everywhere except for a few places in the region initially occupied by endogenous surfactant where a significant rotational part of the map is acquired due to deposits interacting with each other and with boundaries. A larger error exists near fixed points of Eul[0.25]. For $\delta=0.002$ the error is larger than for $\delta=0.25$, but even for small $\delta$, the difference between the two solutions is close to zero in most of the domain, and especially inside the exogenous deposits. 

\begin{figure}
	\centering
    \begin{subfigure}{0.49\textwidth}
    \centering
\includegraphics[keepaspectratio=true,width=1\textwidth]{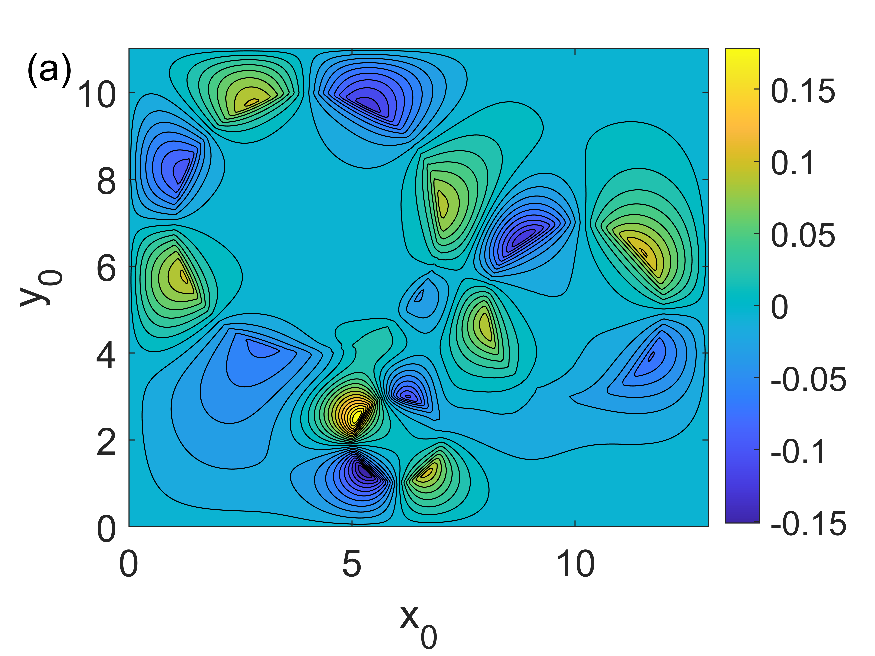}
    \end{subfigure}
     \begin{subfigure}{0.49\textwidth}
    \centering
\includegraphics[keepaspectratio=true,width=1\textwidth]{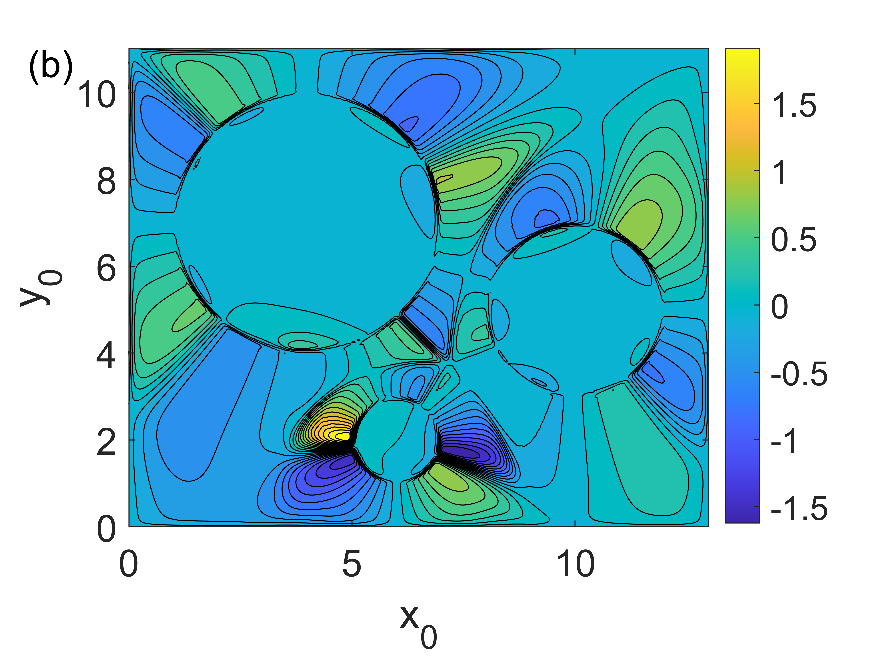}
    \end{subfigure}
	\caption{ Contour plots showing the ratio $\nabla_{x_0}^2\psi/\nabla_{x_0}^2\phi$ for Eul[0.25] in (\textit{a}) and Eul[0.002] in (\textit{b}).}
	\label{fig:lappsiovlapphi}
\end{figure}

Figure \ref{fig:lappsiovlapphi} shows $\nabla^2_{x_0}\psi/\nabla^2_{x_0}\phi$ for Eul[0.25] and Eul[0.002]; the data provide evidence supporting the claim (2.30) from the main text which implies that $\nabla^2_{x_0}\psi \ll \nabla^2_{x_0}\phi$, given the boundary conditions (2.29).  The evidence is weaker for $\delta=0.002$, although the inequality holds in most parts of the domain.

\section{Statistics for the difference in particle location for different drop configurations}

\begin{table}
\centering
\begin{tabular}{c c c c c c c c }
  Method   & $\delta$ & No. of & Dep. centres & Dep. radii & Max conc. & Profile &  Key  \\
     &  & deposits  & $(x_c,y_c)$ $1/2/3$ & r $1/2/3$ & $\Gamma_{0,c}$ $1/2/3$ & shape &    \\\hline
   EU  & $0.25$ & 1 & Various/NA/NA & 1/NA/NA  & 1/NA/NA & Cosine & NA \\ 
   MA   & $0.25$  & 1 & Various/NA/NA & 1/NA/NA & 1/NA/NA & Cosine & NA
   \\
  EU   & $0.25$ & 2 & Various/Various/NA & 1/1/NA & 1/1/NA & Cosine & NA \\ 
   MA   & $0.25$  & 2 & Various/Various/NA & 1/1/NA & 1/1/NA & Cosine & NA
   \\
   EU   & $0.25$  & 3 & (6,2)/(10,5)/(4,7) & 1/2/3 & 1/1/2 & Cosine & Eul\textsubscript{c}[0.25] \\ 
   MA   & $0.25$  & 3 & (6,2)/(10,5)/(4,7) & 1/2/3 & 1/1/2 & Cosine & MA\textsubscript{c}[0.25] 
   \\
   EU  & $0.25$  & 3 & (6,2)/(10,5)/(4,7) & 1/2/3 & 1/1/2 & Quadratic & Eul[0.25] \\ 
   MA   & $0.25$  & 3 & (6,2)/(10,5)/(4,7) & 1/2/3 & 1/1/2 & Quadratic & MA[0.25] 
   \\
   EU   & $0.25$  & 3 & (6,2)/(10,5)/(4,7) & 1/2/3 & $\frac{11}{8}$/$\frac{11}{8}$/$\frac{23}{8}$ & Linear & Eul\textsubscript{L}[0.25] \\ 
   MA   & $0.25$  & 3 & (6,2)/(10,5)/(4,7) & 1/2/3 & $\frac{11}{8}$/$\frac{11}{8}$/$\frac{23}{8}$ & Linear & MA\textsubscript{L}[0.25] 
   \\
   EU   & $0.25$  & 3 & (6,2)/(10,5)/(4,7) & 1/2/3 & $\frac{5}{8}$/$\frac{5}{8}$/$\frac{9}{8}$ & Top Hat & Eul\textsubscript{T}[0.25] \\ 
   MA  & $0.25$  & 3 & (6,2)/(10,5)/(4,7) & 1/2/3 & $\frac{5}{8}$/$\frac{5}{8}$/$\frac{9}{8}$ & Top Hat & MA\textsubscript{T}[0.25] 
   \\
   \hline
\end{tabular}
\caption{Table showing the various simulations studied in \S{S5}. The only methods considered here are Eulerian particle tracking (EU) and Monge--Amp\`ere (MA). We consider a variety of one- and two-deposit configurations with centres of the deposits reported in figures \ref{fig:1dep} and \ref{fig:2dep}. We consider the three-deposit problem for a variety of different initial concentration profile shapes, with the maximum concentration $\Gamma_{0,c}$ within each deposit chosen such that the mass of surfactant is the same in each deposit for each method.}
\label{tab:SimulationSummarySupp}
\end{table}

\begin{figure}
	\centering
    \begin{subfigure}{0.49\textwidth}
    \centering
\includegraphics[keepaspectratio=true,width=1\textwidth]{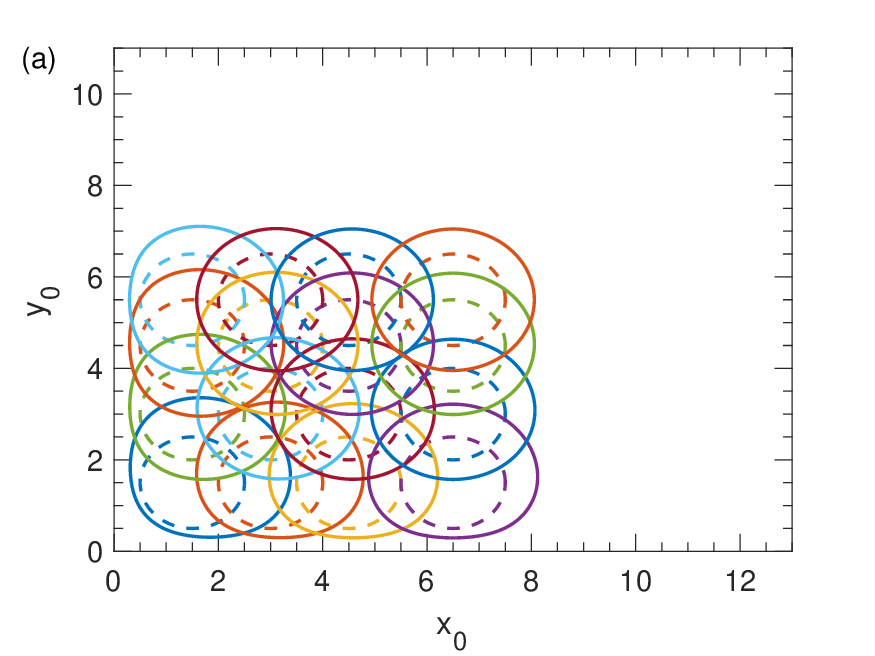}
    \end{subfigure}
    \begin{subfigure}{0.49\textwidth}
    \centering
\includegraphics[keepaspectratio=true,width=1\textwidth]{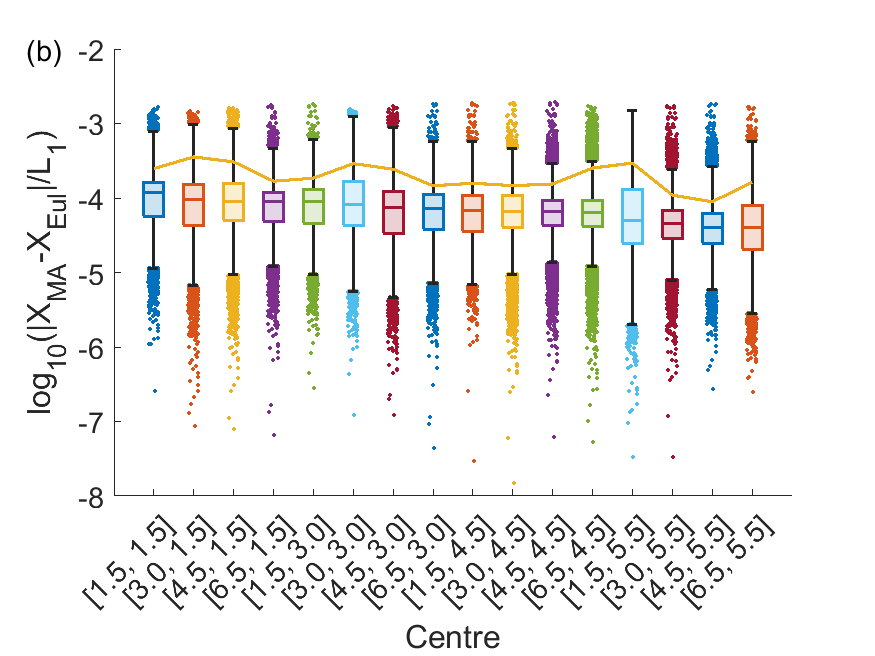}
    \end{subfigure}
	\caption{Graphs showing results for the one-deposit problem which consists of a single deposit of unit radius and initial concentration profile $\delta+\mathcal{C}_c(x_c,y_c,1,1-\delta)$ spreading to equilibrium, with $[x_c, y_c]$ shown on the x-axis and for $\delta=0.25$ ($\mathcal{C}_c$ is defined in (F1) in the main text). (\textit{a}) Shows the 16 different initial boundaries of the single deposits as dashed lines, with final locations of the boundaries found by the Eulerian particle-tracking method as solid lines. (\textit{b}) Shows box and whisker plots of the logarithm of $|X_{EU}-X_{MA}|/L_1$, where each data point corresponds to a unique initial particle location calculated from a uniformly-spaced grid, and each plot is colour-coded to match (\textit{a}). The orange curve denotes the 90th percentile of each solution. Owing to symmetry considerations, we have showed only initial deposit locations in one quadrant of the domain. The plots in (\textit{b}) are ordered by descending median. Outliers are plotted as a cloud of points above and below the whiskers.}
	\label{fig:1dep}
\end{figure}

\begin{figure}
	\centering
    \begin{subfigure}{0.49\textwidth}
    \centering
\includegraphics[keepaspectratio=true,width=1\textwidth]{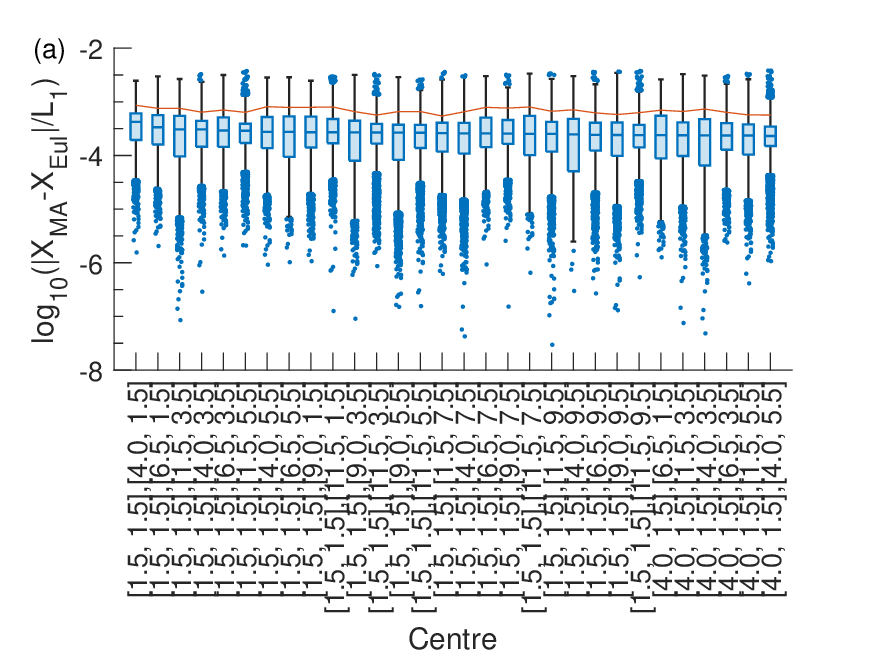}
    \end{subfigure}
     \begin{subfigure}{0.49\textwidth}
    \centering
\includegraphics[keepaspectratio=true,width=1\textwidth]{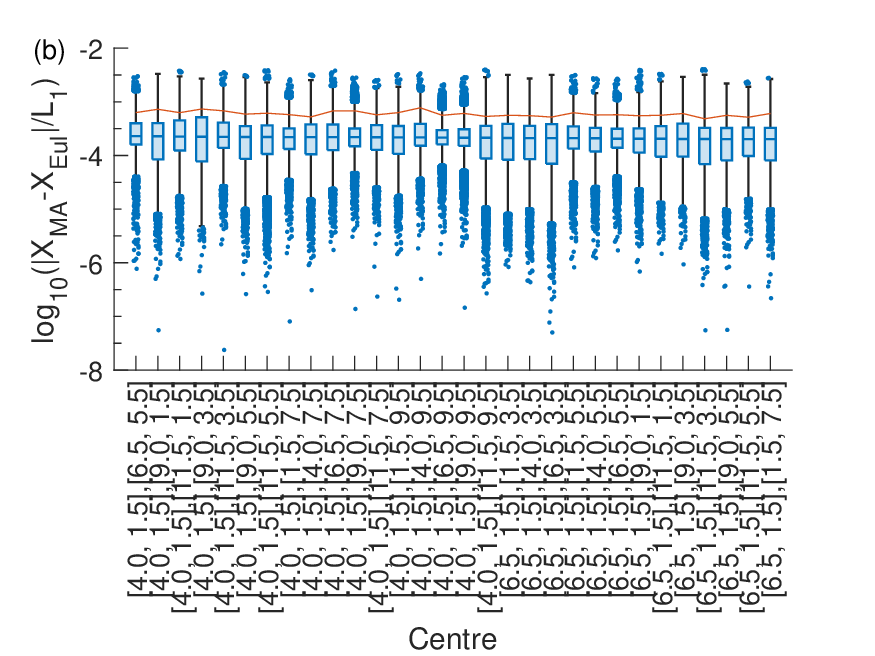}
    \end{subfigure}
    \begin{subfigure}{0.49\textwidth}
    \centering
\includegraphics[keepaspectratio=true,width=1\textwidth]{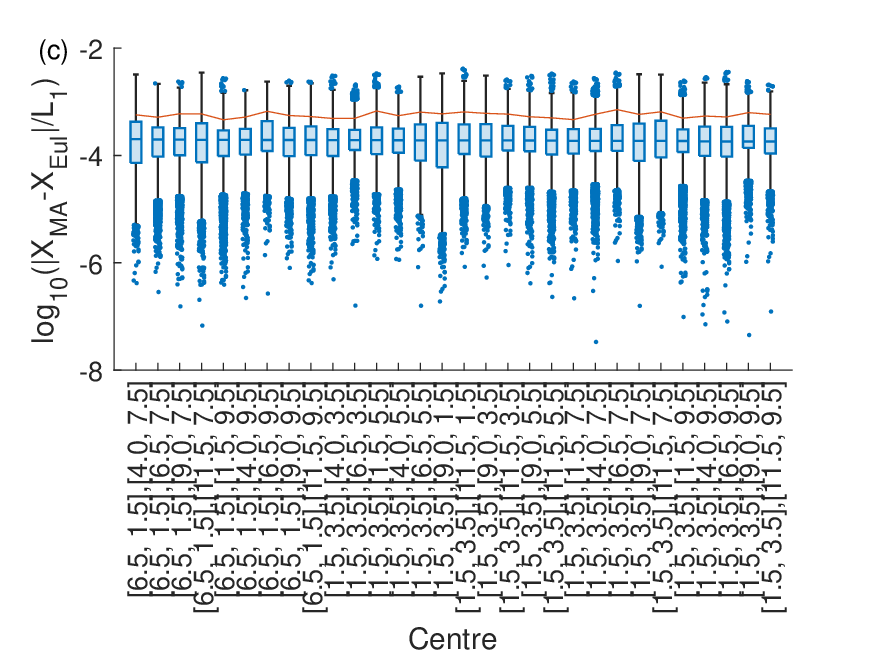}
    \end{subfigure}
     \begin{subfigure}{0.49\textwidth}
    \centering
\includegraphics[keepaspectratio=true,width=1\textwidth]{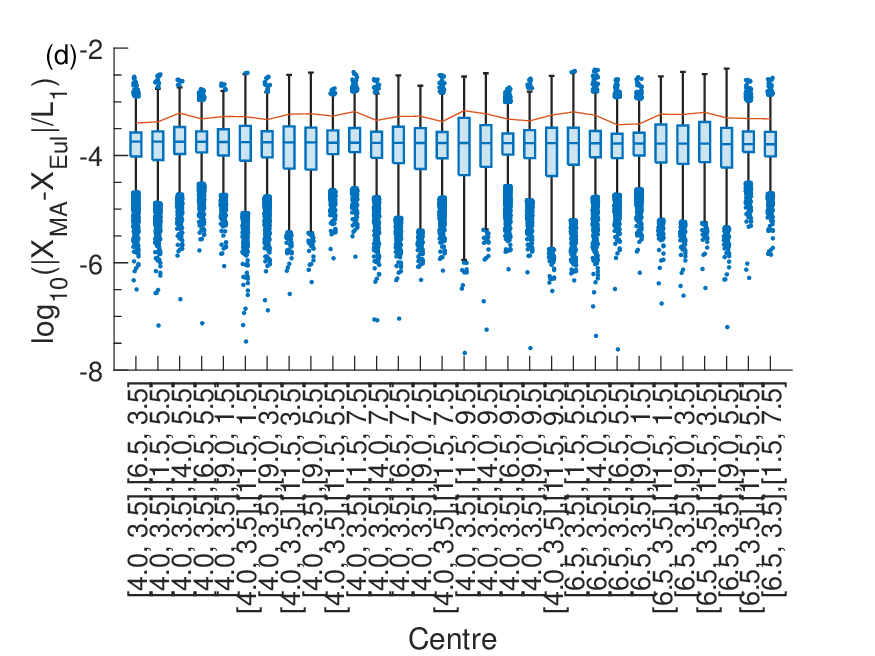}
    \end{subfigure}
    \begin{subfigure}{0.49\textwidth}
    \centering
\includegraphics[keepaspectratio=true,width=1\textwidth]{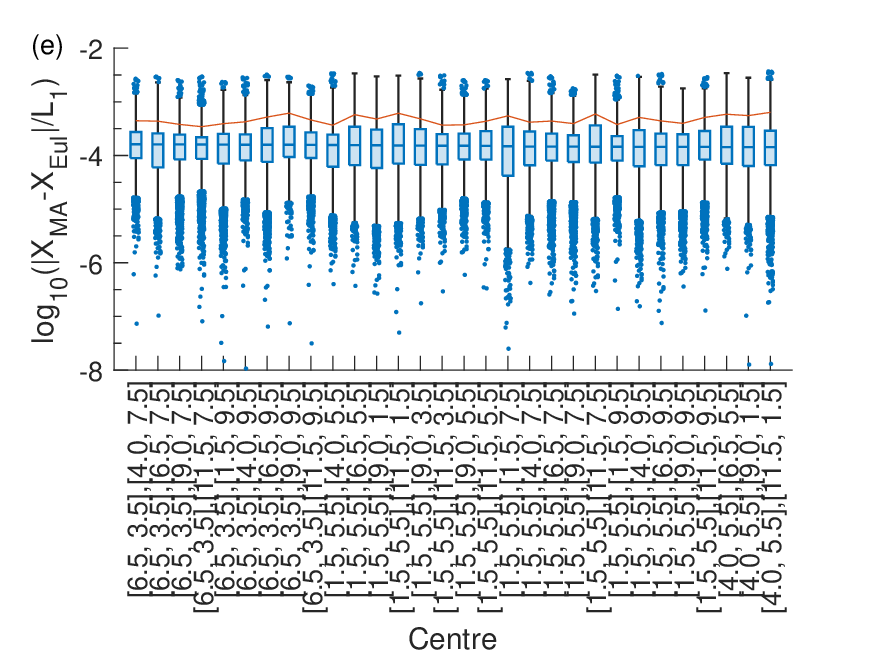}
    \end{subfigure}
     \begin{subfigure}{0.49\textwidth}
    \centering
\includegraphics[keepaspectratio=true,width=1\textwidth]{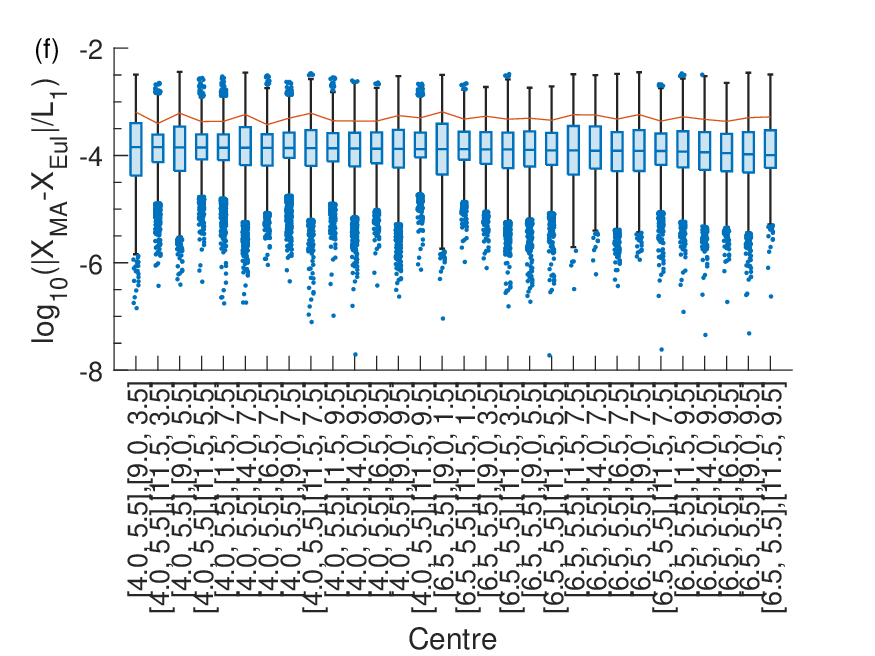}
    \end{subfigure}
	\caption{Box and whisker plots of the logarithm of the normalized absolute error $|X_{EU}-X_{MA}|/L_1$ between the predictions of final particle location for the Eulerian particle tracking and Monge--Amp\`ere methods for two deposits of unit radius. The initial concentration profile is $\delta+\mathcal{C}_c(x_{c1},y_{c1},1,1-\delta)+\mathcal{C}_c(x_{c2},y_{c2},1,1-\delta)$, with $[x_{c1}, y_{c1}],[x_{c2}, y_{c2}]$ indicated on the x-axis, and $\delta=0.25$. Each box plot is for a unique two-deposit configuration, and each data point corresponds to a unique initial particle location taken from an evenly spaced grid. Graphs are ordered by descending median. The orange curve shows the 90th percentiles. Outliers are plotted as a cloud of points above and below the whiskers.}
	\label{fig:2dep}
\end{figure}

\begin{figure}
	\centering
    \centering
\includegraphics[keepaspectratio=true,width=0.8\textwidth]{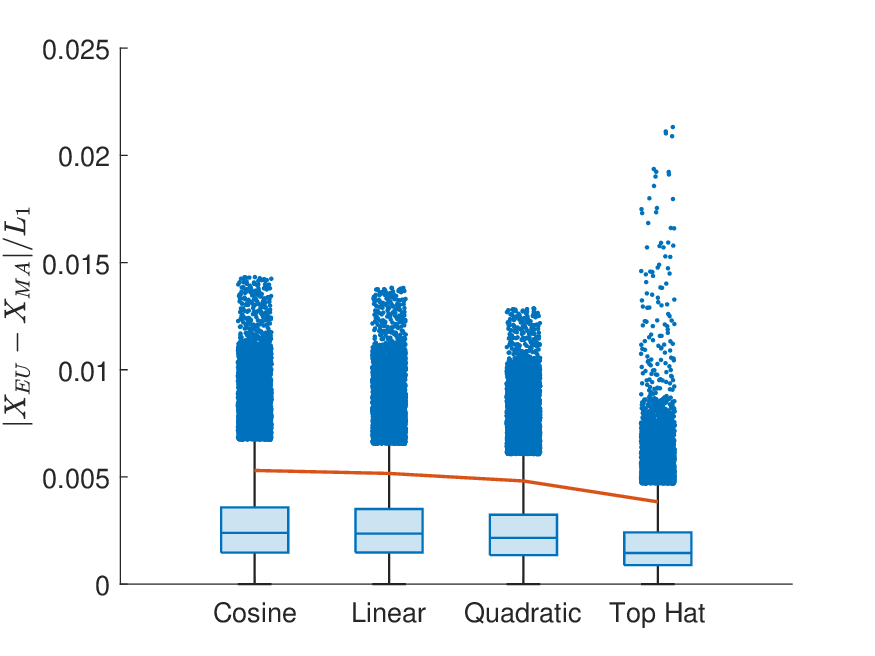}
	\caption{Graph showing box and whisker plots for the normalised absolute error between predictions of final particle location from the Eulerian particle-tracking and Monge--Amp\`ere methods, where each data point is a unique starting location taken from an evenly spaced grid.  The graph plots $|X_{EU}-X_{MA}|/L_1$ for the three-deposit problem considered in the main paper (see \S2.2.2), but where the exogenous deposits concentration profiles have different initial shapes as indicated on the x-axis. For the cosine profiles, $\mathcal{C}_q$ in (2.9) of the main text is replaced by $\mathcal{C}_c$ (equation (F1)) for a cosine profile with $\Gamma_{0,c}=1$ for the smallest two deposits and $\Gamma_{0,c}=2$ for the largest deposit. For the linear profile, $\mathcal{C}_q$ is replaced by $\mathcal{C}_l$ (equation (\ref{eq:circlenotationtri})) with $\Gamma_{0,c}=11/8$ for the smallest two deposits and $\Gamma_{0,c}=23/8$ for the largest deposit. Lastly, for the top hat profile, $\mathcal{C}_q$ is replaced by $\mathcal{C}_{t}$ (equation (\ref{eq:circlenotationTH})) with $\Gamma_{0,c}=5/8$ for the smallest two deposits and $\Gamma_{0,c}=11/8$ for the largest. These choices of $\Gamma_{0,c}$ yield the same total mass of exogenous surfactant between all the simulations. The orange curve denotes the 90th percentile of each solution. Outliers are plotted as a cloud of points above the top whisker.}
	\label{fig:3depmanyprof}
\end{figure}

Figures \ref{fig:1dep}, \ref{fig:2dep} and \ref{fig:3depmanyprof}  show box-and-whisker plots for a variety of different exogenous deposit configurations (see table~\ref{tab:SimulationSummarySupp}) not considered in the main paper to provide further quantitative comparison between the Monge--Amp\`ere method and the Eulerian solution (assumed to be close to the exact solution). While quadratic concentration profiles are considered in the main paper, and cosine profiles in Appendix F, here we also consider linear and top-hat shaped initial concentration profiles within circular deposits. Cosine shaped circular deposits with centres at $\mathbf{x}_c=(x_c, y_c)$, radii $r$ and maximum concentration $\Gamma_{0,c}$ are given by (F1) in the main text, linear concentration profiles within the circular deposits are given by
\begin{equation}\label{eq:circlenotationtri}
 \mathcal{C}_{l}(\mathbf{x}_0;\mathbf{x}_c,r,\Gamma_{0,c}-\delta) = \begin{cases}
   (\Gamma_{0,c}-\delta)\left(1- \frac{|\mathbf{x}_0-\mathbf{x}_c|}{r}\right)&\quad |\mathbf{x}_0-\mathbf{x}_c|\leq r \\
0 &\quad |\mathbf{x}_0-\mathbf{x}_c| > r\end{cases},
\end{equation}
and top-hat shaped profiles are given by
\begin{equation}\label{eq:circlenotationTH}
 \mathcal{C}_{t}(\mathbf{x}_0;\mathbf{x}_c,r,\Gamma_{0,c}-\delta) = \begin{cases}
  \Gamma_{0,c}-\delta &\quad |\mathbf{x}_0-\mathbf{x}_c|\leq r \\
0 &\quad |\mathbf{x}_0-\mathbf{x}_c| > r\end{cases}.
\end{equation} 

In the case of a one-deposit problem, with initial conditions described in table~\ref{tab:SimulationSummarySupp}, figure \ref{fig:1dep} shows that a single deposit configuration (shown in figure \ref{fig:1dep}\textit{a}) produces a larger median error (shown in figure \ref{fig:1dep}\textit{b}) when the deposit is initially placed closer to a corner of the domain, and the median error decreases as the deposit approaches the centre of the domain. Nevertheless, we find that overall the maximum distance error is largely insensitive to drop location. The error is bounded above by 0.002 for $\delta=0.25$, which corresponds to 0.2\% of the domain length $L_1$.   

Figure \ref{fig:2dep} shows a variety of independent two-deposit configurations, with initial conditions described in table~\ref{tab:SimulationSummarySupp}. As in figure \ref{fig:1dep} the most asymmetric configurations have the largest median error, and the error decreases slightly with increasing symmetry of the initial configuration.   All distance errors in these examples are bounded above by 0.005, corresponding to 0.5\% of the domain length.

Figure \ref{fig:3depmanyprof} shows particle data for the three-deposits problem considered in \S(2.1.2), where the initial concentration distribution has the profile shown on the x-axis, varying from a cosine (F1), a linear \eqref{eq:circlenotationtri}, a quadratic (2.8) and a top-hat \eqref{eq:circlenotationTH} profile. The largest median discrepancy between the Monge--Amp\`ere and the Eulerian solution is found for the profiles with the more complicated first derivative (the cosine profile), despite the cosine profile being the smoothest of the shown profiles. In all the examples, the maximum distance error is bounded above  by 0.02, corresponding to 2\% of the domain length.

Figures \ref{fig:1dep},   \ref{fig:2dep} and \ref{fig:3depmanyprof} show  little variation in the error between the two methods. The specific  configuration of a fixed number of deposits,  the initial shape of the profiles within the deposits do not seem to affect much the error between MA and EU methods. Overall, we find that the parameters which tend to increase the most significantly the error between the two methods are the endogenous surfactant concentration $\delta$, and the number of deposits.

\bibliographystyle{jfm}
